\documentclass[12pt,preprint]{aastex}

\usepackage{amsmath}
\usepackage{amsfonts}
\usepackage{longtable}

\def\M#1{{\mathbf{#1}}}                          
\def\T#1{\M{#1}^{T}}
\def\T#1{{{#1}^{\top}}}                             
\def\R#1{{\mathrm{#1}}}
\def\ltsima{$\; \buildrel < \over \sim \;$}
\def\simlt{\lower.5ex\hbox{\ltsima}}
\def\gtsima{$\; \buildrel > \over \sim \;$}
\def\simgt{\lower.5ex\hbox{\gtsima}}
\def\d{{\R{d}}}                                           
\def\mdot{\!\cdot\!}                                    
\def\feh{\mathrm{[Fe/H]}}
\def\MH{\mathrm{[M/H]}}
\def\mh{\mathrm{[m/H]}}
\def\alphaenh{\mathrm{[\alpha/Fe]}}
\def\logg{\log g}
\def\vrot{{V_\mathrm{rot}}}
\def\teff{{T_\mathrm{eff}}}
\def\la{\lambda}

\def\deg{^{\mathrm o}}
\def\kms{\mbox{km~s$^{-1}$}}

\def\ljubljana{            1}
\def\potsdam{              2}
\def\strasbourg{           3}
\def\asiago{               4}
\def\canberra{             5}
\def\aao{                  6}
\def\baltimore{            7}
\def\macquarie{            8}
\def\cambridge{            9}
\def\brunel{              10}
\def\heidelberg{          11}
\def\groningen{           12}
\def\victoria{            13}
\def\swinburne{           14}
\def\oxford{              15}
\def\usydney{             16}
\def\leiden{              17}
\def\leicester{           18}
\def\mpe{                 19}
\def\preston{             20}
\def\rochester{           21}
\def\edinburgh{           22}

\shorttitle{RAVE 2nd data release}
\shortauthors{Zwitter et al.}

\begin{document}


\title{The Radial Velocity Experiment (RAVE): second data release}
\author{
T. Zwitter\altaffilmark{\ljubljana},
A. Siebert\altaffilmark{\potsdam,\strasbourg},
U. Munari\altaffilmark{\asiago},
K.~C. Freeman\altaffilmark{\canberra},
A. Siviero\altaffilmark{\asiago},
F.~G. Watson\altaffilmark{\aao},
J.~P. Fulbright\altaffilmark{\baltimore},
R.~F.~G. Wyse\altaffilmark{\baltimore},
R. Campbell\altaffilmark{\macquarie,\potsdam},
G.~M. Seabroke\altaffilmark{\cambridge,\brunel},
M. Williams\altaffilmark{\canberra,\potsdam},
M. Steinmetz\altaffilmark{\potsdam},
O. Bienaym\'e\altaffilmark{\strasbourg},
G. Gilmore\altaffilmark{\cambridge},
E.~K. Grebel\altaffilmark{\heidelberg},
A. Helmi\altaffilmark{\groningen},
J.~F. Navarro\altaffilmark{\victoria},
B. Anguiano\altaffilmark{\potsdam},
C. Boeche\altaffilmark{\potsdam},
D. Burton\altaffilmark{\aao},
P. Cass\altaffilmark{\aao},
J. Dawe\altaffilmark{\dagger,\aao},
K. Fiegert\altaffilmark{\aao},
M. Hartley\altaffilmark{\aao},
K. Russell\altaffilmark{\aao},
L. Veltz\altaffilmark{\strasbourg,\potsdam},
J. Bailin\altaffilmark{\swinburne},
J. Binney\altaffilmark{\oxford}, 
J. Bland-Hawthorn\altaffilmark{\usydney},
A. Brown\altaffilmark{\leiden},
W. Dehnen\altaffilmark{\leicester},
N.~W. Evans\altaffilmark{\cambridge},
P. Re Fiorentin\altaffilmark{\ljubljana},
M. Fiorucci\altaffilmark{\asiago},
O. Gerhard\altaffilmark{\mpe},
B. Gibson\altaffilmark{\preston}, 
A. Kelz\altaffilmark{\potsdam},
K. Kujken\altaffilmark{\groningen},
G. Matijevi\v{c}\altaffilmark{\ljubljana},
I. Minchev\altaffilmark{\rochester},
Q.~A. Parker\altaffilmark{\macquarie},
J. Pe\~narrubia\altaffilmark{\victoria},
A. Quillen\altaffilmark{\rochester},
M.~A. Read\altaffilmark{\edinburgh},
W. Reid\altaffilmark{\macquarie},
S. Roeser\altaffilmark{\heidelberg},
G. Ruchti\altaffilmark{\baltimore},
R.-D. Scholz\altaffilmark{\potsdam},
M.~C. Smith\altaffilmark{\cambridge},
R. Sordo\altaffilmark{\asiago},
E. Tolstoi\altaffilmark{\groningen},
L. Tomasella\altaffilmark{\asiago}, 
S. Vidrih \altaffilmark{\heidelberg,\cambridge,\ljubljana},
E. Wylie de Boer \altaffilmark{\canberra}
}
%
\altaffiltext{\ljubljana}{University of Ljubljana, Faculty of Mathematics and Physics, Ljubljana, Slovenia}
\altaffiltext{\potsdam}{Astrophysikalisches Institut Potsdam, Potsdam, Germany}
\altaffiltext{\strasbourg}{Observatoire de Strasbourg, Strasbourg, France}
\altaffiltext{\asiago}{INAF, Osservatorio Astronomico di Padova, Sede di Asiago, Italy}
\altaffiltext{\canberra}{RSAA, Australian national University, Canberra, Australia}
\altaffiltext{\aao}{Anglo Australian Observatory, Sydney, Australia}
\altaffiltext{\baltimore}{Johns Hopkins University, Baltimore MD, USA}
\altaffiltext{\macquarie}{Macquarie University, Sydney, Australia}
\altaffiltext{\cambridge}{Institute of Astronomy, University of Cambridge, UK}
\altaffiltext{\brunel}{e2v Centre for Electronic Imaging, School of Engineering and Design,
Brunel University, Uxbridge, UK}
\altaffiltext{\heidelberg}{Astronomisches Rechen-Institut, Center for Astronomy of the University of 
Heidelberg, Heidelberg, Germany}
\altaffiltext{\groningen}{Kapteyn Astronomical Institute, University of Groningen, Groningen, the Netherlands}
\altaffiltext{\victoria}{University of Victoria, Victoria, Canada}
\altaffiltext{\swinburne}{Centre for Astrophysics and Supercomputing, Swinburne University
	of Technology, Hawthorn, Australia}
\altaffiltext{\oxford}{Rudolf Pierls Center for Theoretical Physics, University of Oxford, UK}
\altaffiltext{\usydney}{Institute of Astronomy, School of Physics,  
University of Sydney, NSW 2006, Australia}
\altaffiltext{\leiden}{Sterrewacht Leiden, University of Leiden, Leiden, the Netherlands}
\altaffiltext{\leicester}{University of Leicester, Leicester, UK}
\altaffiltext{\mpe}{MPI fuer extraterrestrische Physik, Garching, Germany}
\altaffiltext{\preston}{University of Central Lancashire, Preston, UK}
\altaffiltext{\rochester}{University of Rochester, Rochester NY, USA}
\altaffiltext{\edinburgh}{University of Edinburgh, Edinburgh, UK}
\altaffiltext{$\dagger$}{deceased}




\begin{abstract}
We present the second data  release of the Radial Velocity Experiment (RAVE),
an ambitious  spectroscopic survey to measure radial  velocities and stellar
atmosphere parameters  (temperature, metallicity, surface gravity, and 
rotational velocity) of up to
one million  stars using the 6dF  multi-object spectrograph on  the 1.2-m UK
Schmidt  Telescope  of the  Anglo-Australian  Observatory  (AAO).  The  RAVE
program  started  in  2003,  obtaining  medium  resolution  spectra  (median
R=7,500) in  the Ca-triplet  region ($\lambda\lambda$ 8,410--8,795  \AA) for
southern hemisphere stars drawn from the Tycho-2 and SuperCOSMOS catalogues,
in the magnitude range  $9\!<\!I\!<\!12$.  Following the first data release 
 \citep{dr1} the current release doubles the sample of published radial 
velocities, now containing 51,829 radial velocities for 49,327 individual 
stars observed on 141 nights between April 11 2003 and  
March 31 2005. Comparison with external data sets shows that the new data
collected since April 3 2004 show a standard deviation of 1.3~\kms, about 
twice better than for the first data release.
For the first time this data release contains values 
of stellar parameters from 22,407 spectra of 21,121 individual stars. They
were derived by a penalized $\chi^2$ method using an extensive grid of synthetic
spectra calculated from the latest version of Kurucz stellar atmosphere 
models. From comparison with external data sets, our conservative estimates of
errors of the stellar parameters for a spectrum with an average signal 
to noise ratio of $\sim\!\!40$ are 400~K in temperature, 0.5~dex in gravity, 
and 0.2 dex in metallicity. We note however that, for all three stellar 
parameters, the internal errors estimated from repeat RAVE observations 
of 822 stars are at least a factor 2 smaller. We demonstrate that the 
results show no systematic offsets if compared to values derived from 
photometry or complementary spectroscopic analyses.
The data release includes proper motions from Starnet2, Tycho2, 
and UCAC2 catalogs and photometric measurements from Tycho-2
USNO-B, DENIS  and 2MASS.  The data release can be accessed via the RAVE
webpage: http://www.rave-survey.org and through CDS.
\end{abstract}



\section{Introduction}
\label{s:introduction}

This paper presents the second data release from the Radial Velocity 
Experiment (RAVE), an ambitious spectroscopic survey of the southern sky 
which has already observed over 200,000 stars away from the 
plane of the Milky Way ($|b| > 25\deg$) and with apparent magnitudes 
$9<I_\mathrm{DENIS}<13$. The paper follows the first data 
release, described in \citet{dr1}, hereafter Paper~I. It doubles the 
number of published radial velocities. For the first time it also 
uses spectroscopic analysis to provide information on values of stellar 
parameters: temperature, gravity, and metallicity. Note that the latter 
in general differs from iron abundance, because metallicity is the proportion
of matter made up of all chemical elements
other than hydrogen and helium in the stellar atmosphere. 
Stellar parameters are given for the majority of the newly published stars. 
This information is supplemented by additional data from the 
literature: stellar position, proper motion, and photometric measurements 
from DENIS, 2MASS and Tycho surveys.  

Scientific uses of such a data set were described in \citet{steinmetz2003}. They 
include the identification and study of the current structure of the Galaxy and 
of remnants of its formation, recent accretion events, as well as discovery 
of individual peculiar objects and spectroscopic binary stars. 
Kinematic  information derived from the RAVE dataset has been used 
\citep{escape_speed} to constrain the Galactic escape speed at the Solar 
radius to $v_\mathrm{esc} = 536^{+58}_{-44} \ \kms$ (90 percent confidence). The 
fact that $v_\mathrm{esc}^2$ is significantly greater than $2 v_\mathrm{circ}^2$ (where 
$v_\mathrm{circ} = 220~\kms$ is the local circular velocity) is a model-independent 
confirmation that there must be a significant amount of mass exterior to the 
Solar circle, i.e.\ it convincingly demonstrates the presence of a dark 
halo in the Galaxy. A model-dependent estimate yields the virial mass of the 
Galaxy of $1.31^{+0.97}_{-0.49} \times 10^{12}$~M$_\odot$ and 
the virial radius of $297^{+60}_{-44}$~kpc (90 per cent confidence).
\citet{veltz} discussed kinematics towards the Galactic poles and identified 
discontinuities that separate thin disk, thick disk and a hotter 
component. \citet{seabroke2008} searched for in-falling stellar streams
on to the local Milky Way disc and found that it is devoid of any 
vertically coherent streams containing hundreds of stars. The passage  
of the disrupting Sagittarius dwarf galaxy leading tidal stream through 
the Solar neighborhood is therefore ruled out. 
Additional ongoing studies have been listed in Paper~I. 

The structure of this paper is as follows: Section 2 is a description of the  
observations, which is followed by a section on data reduction and processing. 
Data quality is discussed in Section 4, with a particular 
emphasis on a comparison of the derived values of stellar parameters with 
results from an analysis of external data sets. Section 5 is a presentation 
of the data product, followed by concluding remarks on the results in 
the context of current large spectroscopic surveys. 

\section{Observations}

\begin{figure}[hbtp]
\centering
\includegraphics[width=8.7cm,angle=270]{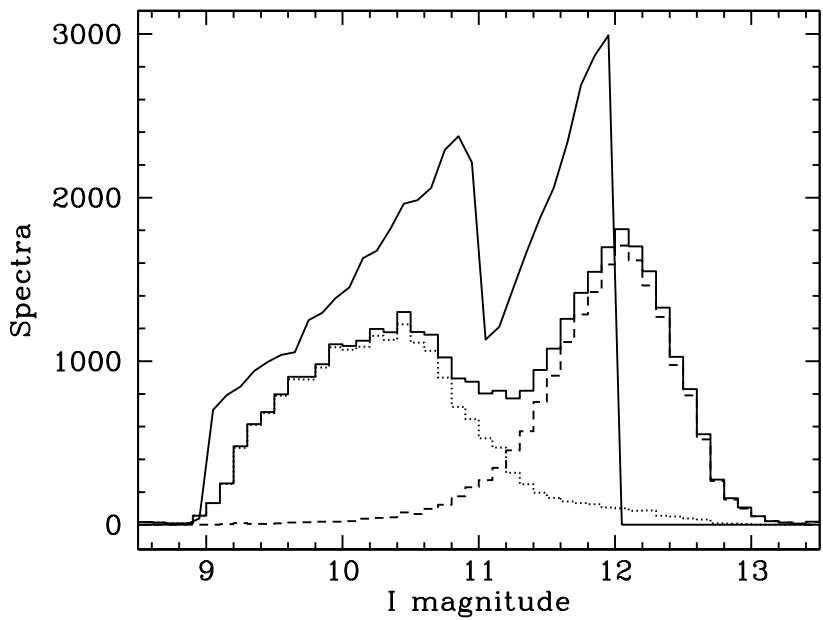}
\caption{
Cousins $I$-band magnitudes of RAVE spectra in the 2nd data release. 
The smooth line denotes magnitudes 
derived from Tycho-2 and SSS survey photometry which were used 
as an input catalog for RAVE. The solid line histogram depicts  
DENIS $I$ magnitudes for the 77\%\ of stars which are also 
in the 2nd release of the DENIS catalog. Short and long dashed
lines are histograms of DENIS $I$ magnitudes for stars from the Tycho-2 
and SSS surveys, respectively. Test fields close to the Galactic plane 
($|b|<25\deg$) are not plotted.
}
\label{f01}
\end{figure}

RAVE is a magnitude limited spectroscopic survey. For this reason it 
avoids any kinematic bias in the target selection. The wavelength range 
of 8410 to 8795~\AA\ overlaps with the photometric Cousins $I$ band. 
However the DENIS and 2MASS catalogs were not yet available at the 
time of planning of the observations we present here. 
So this data release uses the same input catalog as Paper~I: the 
bright stars were selected using $I$ magnitudes estimated from the 
Tycho-2 $V_\mathrm{T}$ and $B_\mathrm{T}$ magnitudes \citep{hog2000}, and the faint ones 
were chosen by their $I$ magnitudes in the SuperCOSMOS Sky Survey 
\citep{hambly2001}, 
hereafter SSS. Transformations to derive the $I$ magnitude and its 
relation to the DENIS $I$ magnitude values 
are discussed in Paper~I. There we also comment on the fact that 
SuperCOSMOS photographic $I$ magnitudes show an offset with respect 
to DENIS $I$ magnitudes (Fig.~\ref{f01}). So, although the initial 
magnitude limit of the survey was planned to be 12.0, the actual 
limit is up to one magnitude fainter.

The survey spans a limited range in apparent magnitude, still it 
probes both the nearby and more distant Galaxy. Typical distances 
for K0 dwarfs are between 50 and 250~pc, while the K0 giants are 
located at distances of 0.7 to 3~kpc.

The instrumental setup is similar to the one used in Paper~I. 
Two field plates with robotically positioned fibers are used in turn in the 
focus of the UK Schmidt telescope at the Anglo-Australian Observatory.
A field plate covers a $5.7\deg$ field of view  and feeds light to
up to 150 fibers each with an angular diameter of 6.7" on the sky. 
One should be careful to avoid chance superpositions with target stars
when using such wide fibers. As a precaution we avoid regions close to the 
Galactic plane ($|b|<25\deg$) or dense stellar clusters. Also, all candidate 
stars are visually checked for possible contamination prior to observing using
the 1-arcmin SSS thumbnails from the on-line SSS R-band data.

\begin{figure}[hbtp]
\centering
\includegraphics[width=12.7cm,angle=270]{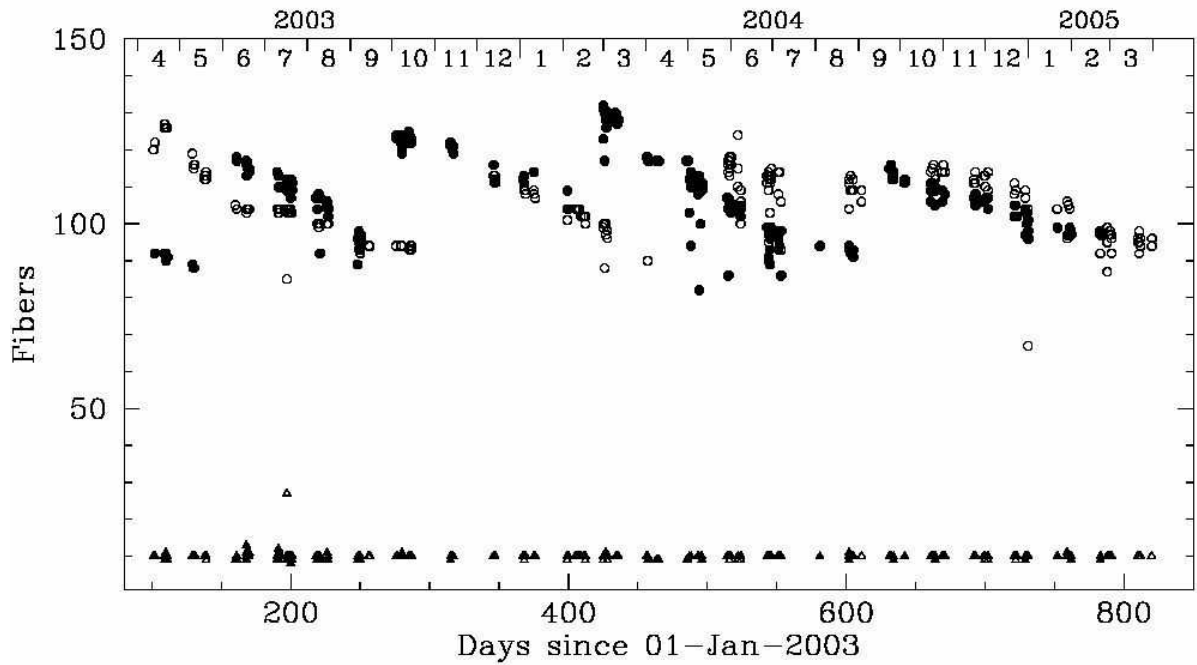}
\caption{
Number of fibers observing stars (circles) and sky background (triangles)
for fields in the 2nd data release.
Filled symbols mark observations obtained with fiber plate 1 and open 
symbols those with plate 2. Test fields close to the Galactic plane 
($|b|<25\deg$) have been omitted from the graph.
}
\label{f02}
\end{figure}

Each field plate contains 150 science fibers, with additional bundles 
used for guiding. A robot positioner configures the plate for each 
field by moving each fiber end to the desired position. The associated 
mechanical stress occasionally causes the fiber to break, so it needs 
to be repaired. A typical fiber is broken after every 2~years of use on 
average, and is repaired in the next 8 months. Figure \ref{f02} shows 
the number of fibers which were used successfully to collect star light 
for each of the 517 pointings. The number varies with time. 
A period of decline is followed by a sharp rise after the repair of 
broken fibers on the corresponding field plate. 
Each pointing was typically used to successfully observe 106 stars. 
An additional 9 or 10 fibers were used to monitor the sky background. 

The light is dispersed by a bench-mounted Schmidt-type spectrograph to 
produce spectra  with a 
resolving power of $R \sim 7500$. The main improvement introduced since 
the first data release is the use of a blue light blocking filter (Schott OG531) 
which blocks the second order spectrum. This allows for an unambiguous 
placement of the continuum level and so permits the derivation of values 
of stellar parameters, in addition to the radial velocity. 
The introduction of the blocking filter lowers the number of 
collected photons by only $\sim 25$\%, so we decided to keep the same 
observing routine as described in Paper~I. The observation of a given field 
consists of 5 consecutive 10-minute exposures, which are accompanied by 
flat-field and Neon arc calibration frames. 

Note that we use two field plates on an alternating basis
(fibers from one fiber plate are being configured while we observe 
with the other field plate). So fibers from a given field 
plate are mounted to the spectrograph slit prior to observation 
of each field. To do this the cover of the spectrograph needs to be 
removed, so its temperature may change abruptly. The associated thermal 
stress implies that it is best to use the flatfield and Neon arc lamp 
exposures obtained immediately after the set of scientific exposures 
when the spectrograph is largely thermally stabilized. For all data new 
to this data release we ensured that such flatfield and arc lamp exposures 
have been obtained and used in the data reduction.

Observations were obtained between April 11 2003 and March 31 
2005. The observations obtained since April 3 2004 yielded data which 
were not published in Paper~I, so they are new to this data release. 
Statistics on the number of useful nights, of field centers and of 
stellar spectra are given in Table~\ref{t:obslog}. These numbers 
make the present, second data release about twice as large as the 
one presented in Paper~I. Stars were mostly observed only once, but 
75 stars from the field centered on 
R.A.~=~16$^\mathrm{h}$~07$^\mathrm{m}$, Dec.~$ = -49^\mathrm{o}$ 
were deliberately observed 8 times to study their variability. 

Observations are limited 
to the southern hemisphere and have a distance of at least 
25 degrees from the Galactic plane (except for a few test fields).
Their distribution is plotted in Figure~\ref{aitoffRVs}. The unvisited area 
is concentrated around the Galactic plane and in the direction of the 
Magellanic clouds.

\begin{deluxetable}{lrr}
\tablecaption{Observing statistics
\label{t:obslog}}
\tablewidth{0pt}
\tablecolumns{3}
\tablehead{
\colhead{ }                        & All     & New Data\\
                                   & Data    & in this DR\\
}
\startdata
Number of nights of observation    &     141 &     72 \\
Number of fields (incl.\ repeats)  &     517 &    266 \\
Sky area covered (square degrees)  &   7,200 &  2,440 \\
Stellar spectra                    &  51,829 & 25,850 \\
Number of different stars          &  49,327 & 24,010 \\
Number of stars observed once      &  47,492 & 22,676 \\
Number of stars observed twice     &   1,618 &  1,232 \\
Number of stars observed 3 times   &     124 &     25 \\
Number of stars observed 4 times   &       2 &      1 \\
Number of stars observed 5 times   &       2 &      0 \\
Number of stars observed 6 times   &       0 &      0 \\
Number of stars observed 7 times   &       1 &      1 \\
Number of stars observed 8 times   &      88 &     75 \\
\enddata
\tablecomments{The middle column counts all data in the present 
data release, the right one only data obtained 
after April 3 2004, i.e.\ new in this data release.}
\end{deluxetable}


\section{Data reduction and processing}

The data reduction is performed in several steps: 
\begin{enumerate}
\item Quality control of the acquired data.
\item Spectra reduction.
\item Radial velocity determination and estimation of physical stellar parameters. 
\end{enumerate}
In the first step the RAVEdr software package and plotting tools are used to make a 
preliminary estimate of data quality in terms of signal levels, focus 
quality and of possible interference patterns. This serves two goals: to 
quickly determine which observations need to be repeated because 
of unsatisfactory data quality, and to exclude any problematic data 
from further reduction steps. For the first data release 17\%\ of 
all pointings were classified as problematic, while in this data release 
the overall dropout rate fell to 13\%. Problematic data are kept 
separately and are not part of this data release. The next two steps of the 
data reduction process are described below. 


\subsection{Spectra reduction}

\label{spectra_reduction}
We use a custom set of IRAF routines which have been described in 
detail in Paper~I. Here we highlight only the improvements introduced
for reduction of data new to this data release. 

\begin{figure}[hbtp]
\centering
\includegraphics[width=12.7cm,angle=270]{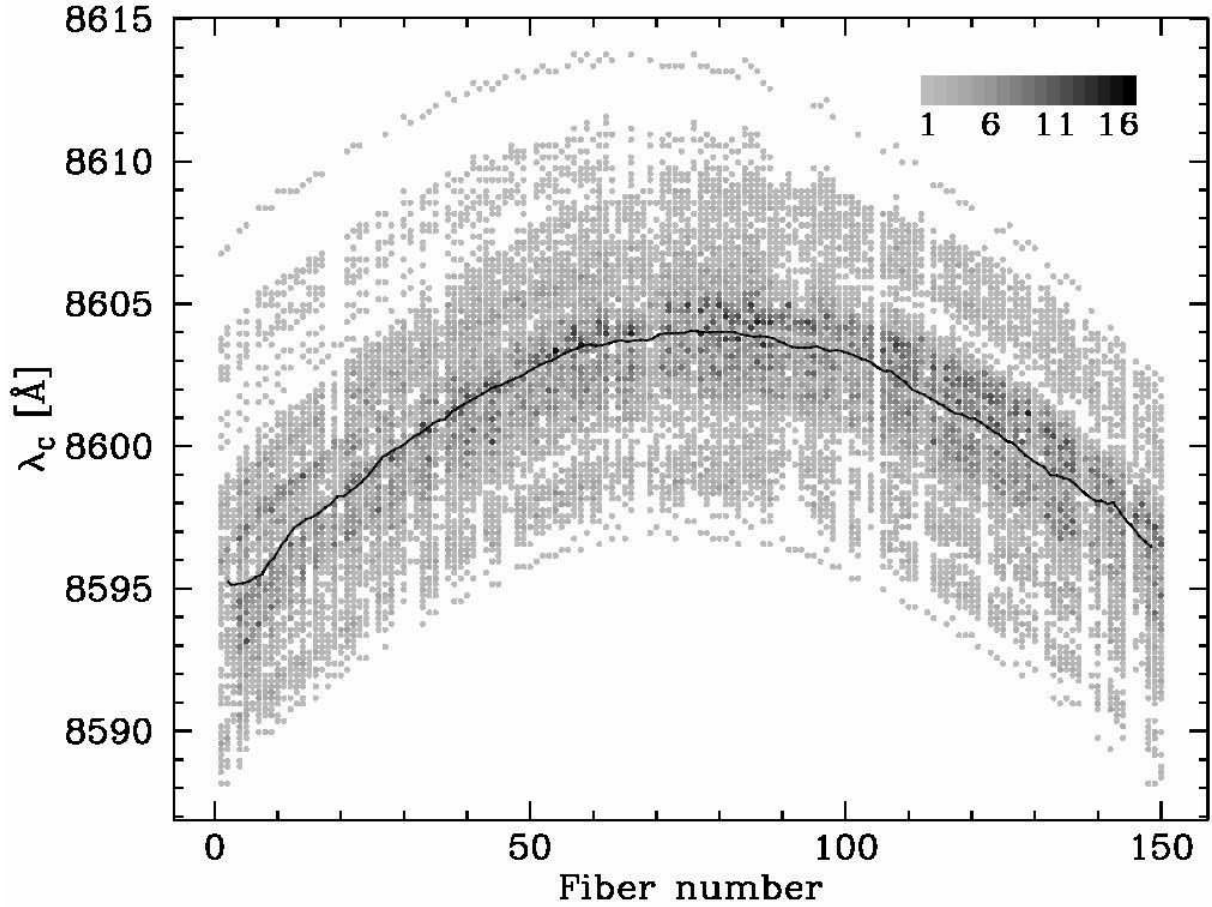}
\caption{Variation of central wavelength as a function of fiber number 
for data new to this release. Shades of gray code the number of spectra 
in a certain bin, as given in the key. The line follows half-point central 
wavelengths as a function of fiber number. 
}
\label{f04}
\end{figure}

The use of the blue light blocking filter permits a more accurate 
flatfielding of the data. The spectra have a length of 1031 pixels, 
and are found to cover a  
wavelength interval of $384.6 \pm 1.7$~\AA. The resolving power is the 
same as estimated in Paper~I, we use the value of $R \simeq 7,500$ 
throughout. The camera of the spectrograph has a very fast focal ratio (F/1). 
The associated optical aberrations at large off-axis angles 
imply that the central wavelength of the spectrograph is not constant, 
but depends on the fiber number (Figure\ \ref{f04}). This means that 
the wavelengths covered by a spectrum depend on its fiber number. Also any 
residual cross--talk between the spectra in adjacent fibers is generally 
shifted in wavelength. This makes an iterative procedure to remove illumination 
from adjacent fibers even more important (see Paper~I for details). The peak 
of central wavelengths around the half-point of their distribution shows that 
our instrumental setup remained quite stable for one year when the data new 
to this data release were obtained. 

The determination of radial velocity and stellar parameters is based 
on the 788 pixels of the central part of the wavelength range only
($8449.77$~\AA~$< \lambda < 8746.84$~\AA). This avoids 
telluric absorption lines and a ghost image caused by internal 
reflections of non-dispersed light at the borders of the wavelength range 
which are occasionally present and could jeopardize the results, as 
described in Paper~I. The edges of the spectral interval are avoided also 
because of a poorer focus, lower resolving power and a lower quality of the 
wavelength calibration. 

\begin{figure}[hbtp]
\centering
\includegraphics[width=12.7cm,angle=270]{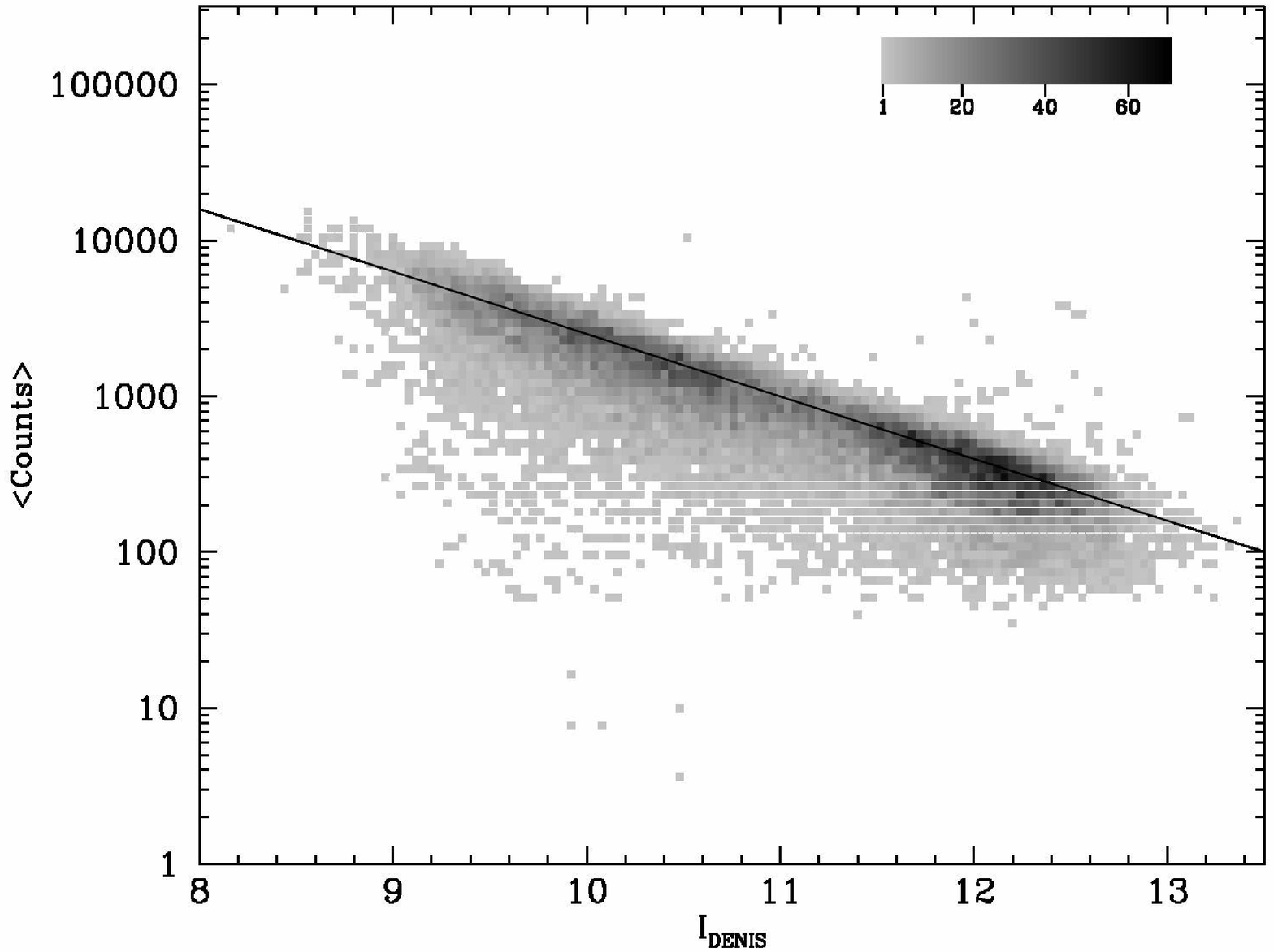}
\caption{Average number of counts per pixel per hour of exposure time 
as a function of DENIS $I$ magnitude. Shades of grey code the number of 
spectra in a certain bin, as given in the key. The average count level 
is calculated from the central part of the spectrum only 
($8449.77\ \mathrm{\AA} \le \lambda \le 8746.84\ \mathrm{\AA}$). 
The inclined line follows eq.~\ref{e:countsvsI}.
}
\label{f:countsIdenis}
\end{figure}

Figure~\ref{f:countsIdenis} plots the average ADU count level of the 
central part of the final 1-D spectrum, and per one hour of exposure time, 
as a function of Denis $I$ magnitude. Only data new to this data release 
are plotted. The line follows the relation 
\begin{equation}
N_{\mathrm{counts}} = 10^{-0.4 (I_\mathrm{DENIS} - 20.25)}
\label{e:countsvsI}
\end{equation} 
where the constant term is the mode of the magnitude corrected count 
distribution. These count levels are 0.25~mag below those in Paper~I.
The difference is due to the 2nd order blocking filter. Note however that 
the filter allowed for a more accurate flatfielding, and so better 
determined count levels. This information has been used in data quality 
control.

%

\subsection{Radial velocity determination}

\label{radialvelocitydetermination}

The general routine stayed the same as described in detail 
in Paper~I. Radial velocities are computed from sky-subtracted 
normalized spectra, while sky unsubtracted spectra are used to 
compute the zero-point correction. The latter is needed because of 
thermal variations of the spectrograph which cause a shift of the 
order of one tenth of a pixel or 1.5~\kms. Radial velocities are computed 
from cross-correlation with an extensive library of synthetic spectra. 
A set of 57,943 spectra degraded to the resolving power of RAVE 
from \citet{munari2005} is used. It is based on the latest generation 
of Kurucz models. It covers all loci of non-degenerate 
stars  in the H-R diagram, with metallicities in the range of  
$-2.5 \le [\mathrm{M}/\mathrm{H}] \le +0.5$. Most spectra have a microturbulent velocity 
of 2~\kms\ (with additional entries for 1 and 4 \kms), while the $\alpha$ 
enhancements of $\alphaenh = 0.0$ and $+0.4$ are used. The use 
of the blue blocking filter simplifies the computations, as 
no contribution from the 2nd order spectrum needs to be considered. 
Both the observed spectra and theoretical templates are normalized prior to the  
radial velocity measurement. We use IRAF's task {\it continuum } with 
a two-piece cubic spline. The rejection criteria used in 10 consecutive 
iterations of the continuum level are asymmetric (1.5-$\sigma$ low and 
3-$\sigma$ high). 

Kurucz synthetic spectra used in cross-correlation do not 
include corrections of radial velocity due to convective motions 
in the stellar atmosphere or due to a gravitational redshift of light 
leaving the star (F.\ Castelli, private 
communication). The combined shift is in the range of --0.4~\kms\ for F 
dwarfs to +0.4~\kms\ for K dwarfs \citep{gullberg2002}, while the 
near absence of gravitational redshift in giants causes a $\sim 0.4$~\kms\ 
shift between giants and dwarfs. The exact value of 
these corrections is difficult to calculate, so we follow the Resolution 
C1 of the IAU General Assembly in Manchester \citep{rickman2002} and 
report the heliocentric radial velocities without corrections for 
gravitational or convective shifts in the stellar atmosphere. Note 
however that these values may be different from the line-of-sight 
component of the velocity of the stellar center of mass 
\citep{Lindegren1999,Latham2001}. 

In the final data product we report the heliocentric 
radial velocity and its error, together with the value of the applied 
zero-point velocity correction, the radial velocity of sky lines and 
their correlation properties. A detailed description of the data release 
is given in Section~\ref{secondrelease}.

\subsection{Stellar parameter determination}

\label{estimationofstellarparameters}

The name of the survey suggests that RAVE is predominantly a radial velocity 
survey. However, the spectral type of the survey stars is generally not 
known and the input catalog does not use any color criterion, so RAVE 
stars are expected to include all evolutionary stages and a wide range of 
masses in the H-R diagram. The properties of the stellar spectra in the 
wavelength interval used by RAVE strongly 
depend on the values of the stellar parameters \citep{munari2001}. While the 
Ca~II IR triplet is almost always present, the occurrence and strength of 
Paschen, metallic and molecular lines depends on temperature, 
gravity and metallicity (see e.g.\ Figure~4 in \citet{zwitter2004}). 
So we cannot adopt the common practice of using a small number of
spectral templates to derive the radial velocity alone, as it has been  
commonly done at, e.g., the ELODIE spectrograph at OHP. We therefore construct 
the best matching template from a large library of synthetic Kurucz spectra
(see Sec.~\ref{radialvelocitydetermination}). The parameters of the 
best matching spectrum are assumed to present the true physical parameters
in the stellar atmosphere. 

\begin{figure}[hbtp]
\centering
\includegraphics[width=17.2cm,angle=0]{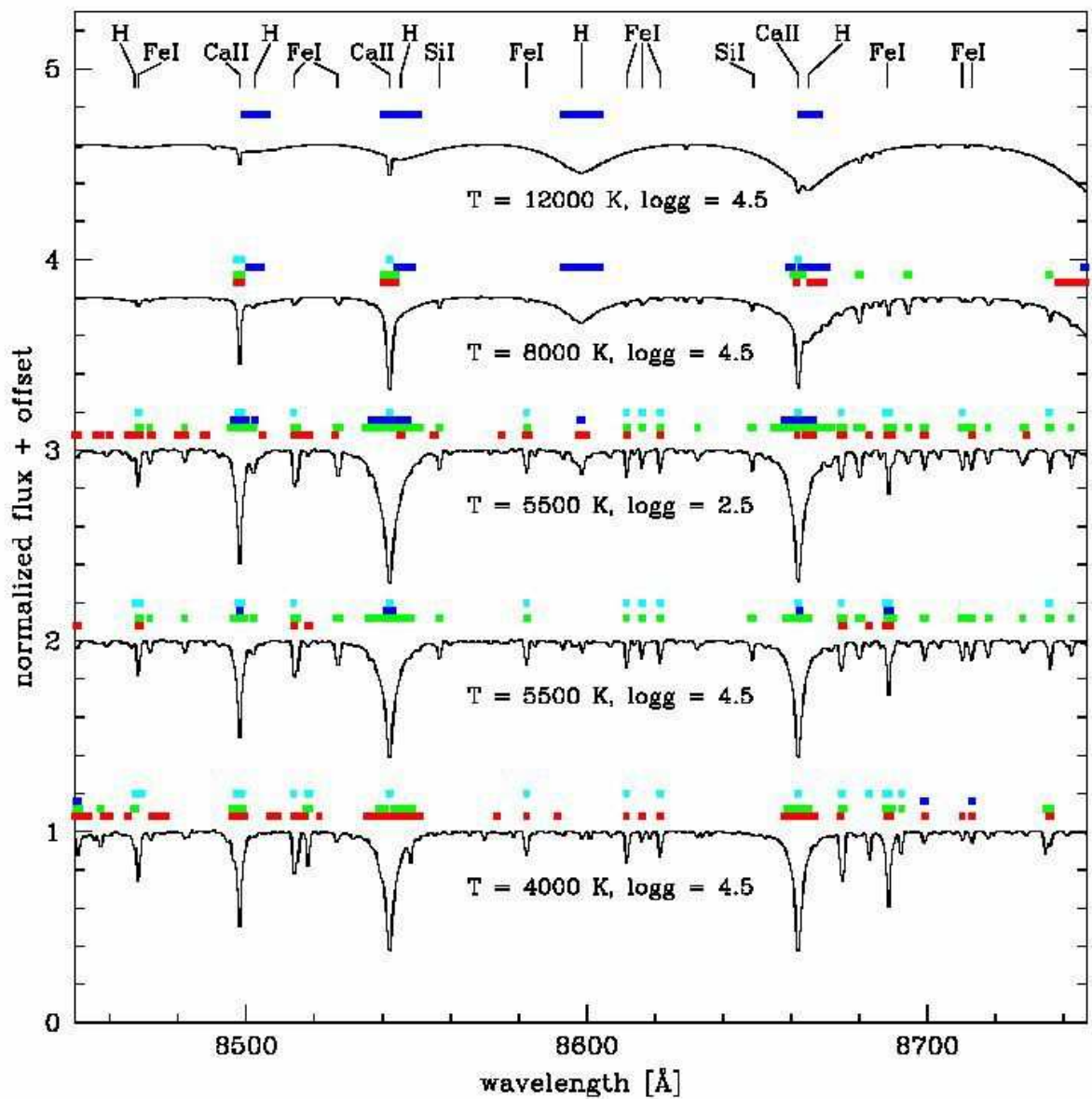}
\caption{Sensitivity of synthetic spectra to stellar parameters.
These are synthetic spectra of non-rotating stars with Solar metallicity 
and microturbulent velocity of  2~\kms. Intermittent lines mark regions 
where a change in one of the parameters causes a change of at least 
3\%\ in normalized flux. From bottom up  
the intermittent lines mark flux changes for: a 500~K decrease 
in temperature (red), a 0.5~dex decrease in metallicity (green), 
a 0.5~dex decrease in gravity (blue),
a 30~\kms\ increase in rotational velocity (cyan). 
The wavelength range of the spectra is the one actually used for the 
determination of stellar parameters.
}
\label{f06}
\end{figure}

Two comments are in order before we outline the template spectrum construction method. 
First, the template library only covers normal stars. So peculiar objects 
cannot be classified correctly. Such objects include double lined 
spectroscopic binaries and emission line objects. Sometimes a peculiar 
nature of the spectrum can be inferred from a poor match of the templates, 
despite a high S/N ratio of the observed spectrum.

The second important point concerns the non-orthogonality of the physical parameters 
we use. This is demonstrated in Figure~\ref{f06}: the wavelength ranges with 
flux levels sensitive to a change in temperature overlap with those sensitive 
to metallicity and the rotational velocity. On the other hand sensitivity to 
changes in both gravity and temperature depend on spectral type and class. 
The intermittent lines in Figure~\ref{f06} mark wavelengths where the normalized flux 
level changes for at least $3$\%\ if the value of one of the parameters is 
modified by a given amount (temperature by 500~K, gravity or metallicity 
by 0.5~dex, or rotational velocity by 30~\kms). We note that a 3\%\ change 
is marginally detectable in a typical RAVE spectrum with S/N~$=40$, 
but the non-orthogonality of 
individual parameters can present a serious problem (see also Figure~1 in
\citet{zwitter2002}). If the temperature or gravity would be known 
{\it a priori}, the ambiguities would be largely resolved. An obvious idea 
is to use photometric colors to constrain the value of stellar temperature. 
Unfortunately the errors of current photometric surveys are too large: a 
change of 0.03~mag in $J-K$ corresponds to a shift of 230~K in temperature in a 
mid-G main sequence star. Also, stellar colors may be seriously compromised 
by interstellar extinction or by stellar binarity. We therefore decided not 
to use any outside information but to base our estimates of stellar parameters 
exclusively on spectral matching. This may change in the future when results of 
multicolor and multi-epoch all-sky photometric surveys such as SkyMapper 
\citep{skymapper} will become available. 

Our parameter estimation procedure makes use of a full set of theoretical templates.
They span a grid in 6 parameters: temperature, gravity, metallicity, $\alpha$ enhancement,
microturbulent and rotational velocity. The sampling in gravity, metallicity, and temperature is 
very good, with $\simgt 9$ tabulated values for the former two and even more for the temperature.
On the other hand the current synthetic library contains only one non-Solar $\alpha$ 
enhancement value ($\alphaenh = +0.4$) and only up to 3 values of microturbulent
velocity (1, 2, 4 \kms, but only 2~\kms\ is available for the whole grid). So we decided 
to publish values of temperature, gravity and metallicity. The alpha enhancement values are 
also listed but they should be interpreted with caution, as they are derived from 2 grid 
values only. These two values may not span the whole range of $\alpha$ enhancement which 
is present in nature. Also the error of $\alpha$ enhancement can be comparable 
to the whole range of the grid in this parameter (see Sec.~\ref{methodvalidation}). 
Microturbulent velocity values are not published, because their errors are typically much 
larger than the range of microturbulent velocities in the grid. Similarly, the rather low 
resolving power of RAVE spectra does not allow the determination of rotational velocities ($\vrot$)  
for slow rotators which represent the vast majority of RAVE stars. Hence the rotational velocity is not 
published, but fast rotators will be discussed in a separate paper. So we aim at the estimation 
of three stellar parameters: effective temperature ($\teff$), gravity ($\logg$), 
and metallicity ($\MH$). The adopted reference system of these parameters is the latest 
set of Kurucz template spectra. Next we describe the inverse method used to derive 
values of stellar parameters.

\subsubsection{Method}

To derive the  stellar parameters, we use a  penalized $\chi^2$ technique to
construct a synthetic spectrum matching the observed spectrum (for other uses 
of  similar methods  see for  example  \citet{pichon02}, \citet{ocvirk06a}).
The observed  spectrum is  modelled as a  weighted sum of template spectra
with known parameters  and it is assumed that  the stellar parameters follow
the same weight relation.  The continuous problem is therefore written as
\begin{equation}
\begin{cases}
{\cal F}_\M{P'}(\la)= \int \tilde{w}(\M{P}) {\cal S}(\la,\M{P})\d^6\M{P}\\
\M{P'}=\int \tilde{w}(\M{P})\M{P}\d^6\M{P}\, ,
\end{cases}
\end{equation}
where ${\cal F}$ is the spectrum  we want the stellar parameters for, ${\cal
S}(\la,\M{P})$  are  the  template  spectra with  known  stellar  parameters
$\M{P}=\T{(\teff,\logg,\MH,\alphaenh,\vrot,\mu)}$,   $\M{P'}$  is  the
stellar  parameter set  we want  to  measure and  $\tilde{w}(\M{P})$ is  the
weight function  we try to  recover. In the  perfect case, where we  have an
infinite number of  template spectra and the observed  spectrum depends only
on  the stellar  parameters (perfect  match between  the observed  and model
spectra),  $\tilde{w}(\M{P})=\delta(\M{P}-\M{P'})$.  In  a  real case  where
noise  plays an  important role  and a  real spectrum  can not  be perfectly
reproduced, $\tilde{w}(\M{P})$ is not a Dirac function but a smooth function
which is non-zero on a limited range. Also, we have the additional 
constraint $\int \tilde{w}(\M{P})\d^6\M{P}=1$.\\
 
In the  more general case, we have  access to a limited  number of templates
and the problem becomes discrete. The problem then can be rewritten as 
\begin{equation}
\begin{cases} {\cal S}_{\M{P}}(\la)=\sum_{i}\, w_{i} . {\cal S}_{\M{P}_{i}}(\la)\\
\M{P} = \sum_{i}\, w_{i} . \M{P}_{i}\, ,
\end{cases}
\label{e:discrete}
\end{equation}
where $w_i$ is the discrete form of $\tilde{w}(\M{P})$.

This problem is ill-conditioned, the number of template spectra being larger
than the number of pixels, and  the information contained in a spectrum being
largely  redundant.   Therefore,  we  make  use  of  penalization  terms  to
regularize the  solution.  Also, the  recovered weights must be  positive to
have a physical meaning, which changes the problem from linear to non-linear.
The following paragraphs will present briefly the linear problem which has a
well defined solution  before entering the realm of  the non-linear problem.
For a full discussion and description  of the method, the reader is referred
to \citet{pichon02}, \citet{ocvirk06a,ocvirk06b} and references therein.\\

\subsubsection{Linear inverse problem}

The  discrete problem  of Eq.~\ref{e:discrete}  can be  written in  a matrix
form.   Calling $\M{\tilde{y}}=\T{({\cal  F}(\la_1),\dots,{\cal F}(\la_n))}$
the  observed   spectrum,  $\M{x}=\T{(w_{1},\ldots,w_{m})}$  the   array  of
weights, $\M{a}=\T{({\cal  S}_{i=1},\ldots,{\cal S}_{i=m})}$ the  library of
template spectra  and $\M{b}=\T{(\M{P}_{i=1},\ldots,\M{P}_{i=m})}$ the array
of parameters, the problem then reads
\begin{equation}
\begin{cases}
\tilde{\M{y}}=\M{a} \mdot \M{x} +\M{e}\\
\M{P}=\M{b} \mdot \M{x}\, ,\\
\end{cases}
\label{e:matrix_form}
\end{equation}
where $\M{e}$  accounts for the noise  in the observed  spectrum. $\M{a}$ is
also referred to as the model matrix or kernel.\\

Using Bayes theorem, solving equation~\ref{e:discrete} or \ref{e:matrix_form}
is  equivalent to  maximizing  the a posteriori  conditional probability  
density $f_\mathrm{post}(\M{x}|\tilde{\M{y}})$ defined as
\begin{equation}
f_\mathrm{post}(\M{x}|\tilde{\M{y}})={\cal L}(\tilde{\M{y}}|\M{x})
f_\mathrm{prior}(\M{x})\, .
\end{equation}
Here $f_\mathrm{prior}(\M{x})$ is our prior on the stellar parameters and ${\cal
  L}(\tilde{\M{y}}|\M{x})$ is the likelihood of the data given the model.

In the case of Gaussian errors, the likelihood is
\begin{equation}
{\cal L}(\tilde{\M{y}}|\M{x})\propto \exp \bigg(-\frac{1}{2}
\T{(\tilde{\M{y}}-\M{a}\mdot\M{x})}\mdot \M{W} \mdot
(\tilde{\M{y}}-\M{a}\mdot\M{x}) \bigg) \, ,
\label{e:likelihood}
\end{equation}
where the expression in the exponent is the $\chi^2$ operator
\begin{equation}
\chi^2(\tilde{\M{y}}|\M{x})=\T{(\tilde{y}-\M{a}\mdot\M{x})}\mdot \M{W} \mdot
(\tilde{y}-\M{a}\mdot\M{x})\, ,
\label{e:chi2}
\end{equation}
$\M{W}$  is  the   inverse  of  the  covariance  matrix   of  the  noise;
$\M{W}=cov(\M{e})^{-1}$. Maximizing  $f_\mathrm{post}(\M{x}|\tilde{\M{y}})$ is
equivalent to minimizing the penalty operator ${\cal Q}(\M{x})$ given by
\begin{eqnarray}
{\cal Q}(\M{x})&=&\chi^2(\tilde{\M{y}}|\M{x})-2 \log (f_\mathrm{prior}(\M{x}))\\ 
&=&\chi^2(\tilde{\M{y}}|\M{x})+\la\,\, \M{R}(\M{x})\, ,
\label{e:Q}
\end{eqnarray}
where  in  the  second form,  the  a  priori  probability density  has  been
rewritten as a penalization or  regularization operator $\M{R}$ and $\la$ is
a Lagrange multiplier.  

When        $\M{R}(\M{x})$        is        a       quadratic        function,
e.g.             $\M{R}(\M{x})=\T{\M{x}}\mdot\M{K}\mdot\M{x}$            and
$\M{K}=\T{\M{L}}\mdot\M{L}$, the problem has a well defined solution
\begin{equation}
\M{x}=(\T{\M{a}}\mdot \M{W} \mdot \M{a} + \la \,\,
\M{K})^{-1} \mdot \T{\M{a}} \mdot \M{W} \mdot \tilde{\M{y}}\, ,
\label{e:linear_solution}
\end{equation}
and the optimal $\la$ is given by the generalized cross validation
(GCV):           $\la_0=\textrm{GCV}(\la)=\textrm{min}_\la           \bigg\{
\frac{\|(\M{1}-\tilde{\M{a}})                                           \mdot
\tilde{\M{y}}\|^2}{[\textrm{trace}(\M{1}-\tilde{\M{a}})]^2} \bigg\}$ , where
$\tilde{\M{a}}=\M{a}\mdot(\T{\M{a}}\mdot  \M{W}  \mdot   \M{a}  +  \la  \,\,
\M{K})^{-1} \mdot \T{\M{a}} \mdot \M{W}$.\\

Using equation~\ref{e:linear_solution}, $\M{x}  \in \mathbb{R}^m$ and the weights
$\M{x}_i$  can have  negative  values.  Negative  weights  have no  physical
meaning  and will result  in non-physical  solutions.  We  therefore require
that $\M{x}\in\mathbb{R}^{+m}$ which leads to the non linear problem discussed
below.

\subsubsection{Non-linear extension}

Unfortunately, there is no simple extension from the analytic linear problem
to the non linear case, and there is no analytic solution for the minimum of
${\cal Q}$.   In the non  linear regime, the  minimum of ${\cal Q}$  must be
obtained  using  efficient  minimization  algorithms  and  can  be  computer
intensive.\\

Nevertheless, as stressed by  \citet{ocvirk06a}, solving the non linear case
has also advantages.  First, we  will obtain a physically motivated solution
(with positive or null weights everywhere) then, imposing positivity reduces
significantly the  allowed parameter  space and reduces  the level  of Gibbs
phenomenon (or ringing artifacts) in  the solution.  This comes at the price
of a higher computing time and asymmetric (non Gaussian) errors.\\

To ensure  that the weights  are positive, we  pose $\M{x}=\exp(\M{\alpha})$
and solve  Eq.~\ref{e:discrete} for $\M{\alpha}$.  The exponential transform
has  the   property  that  while   $\M{\alpha}\in\mathbb{R}^m$,  $\M{x}  \in
\mathbb{R}^{+*m}$  which  ensures  that  the weights  $\M{x}$  are  strictly
positive. Equation~\ref{e:Q} can be rewritten as
\begin{eqnarray}
{\cal Q}(\M{\alpha})=&\T{(\tilde{y}-\M{a}\mdot\exp\M{\alpha})}\mdot
\M{W} \mdot (\tilde{y}-\M{a}\mdot \exp\M{\alpha})\nonumber\\
&+\la_1
\M{P_1}(\M{\alpha})+\la_2
\M{P_2}(\M{\alpha})\ldots \, ,
\label{e:Qnonlinear}
\end{eqnarray}
and the  problem now is  to find the  minimum of ${\cal  Q}(\M{\alpha})$ for
$\M{\alpha}$. Note  that in the  last equation, the  regularization operator
$\M{R}$ has  been split in a  set of regularization operators each with its
own Lagrange parameter.  The penalization operators will be discussed in the
next section.

We mentioned that in the linear  case, the GCV provides an optimal value for
the Lagrange parameter  however, in the non-linear case,  this definition is
no longer valid.   Also, no method is known that allows  a quick estimate of
the optimal $\la$s for the non linear problem.  In our case, we estimate the
proper  Lagrange parameter values  by means  of numerical  simulations using
synthetic spectra and  Gaussian noise. The $\la$s used  in the pipeline were
chosen to optimize  the computation time and the  accuracy (highest possible
accuracy in a minimal computation time).   It must be stressed here that the
Lagrange parameters, obtained from numerical simulations, may not be optimal
as  the  simulations can  not  cover  all the  parameter  space  and as  the
idealized  simulations do  not incorporate  all  the ingredients  of a  real
spectrum.  Nevertheless, the simulations allow us to find a solution for the
Lagrange parameters matching predefined requirements.\\

Finally,  using the  exponential  transform  can cause  the  solution to  be
unbound. For  example, we expect  the weights of  spectra far away  from the
true solution to be zero.  In this case, for $\M{x}_i=0$, $\M{\alpha}_i
\rightarrow  -\infty$ and  the solution  is  unbound.  This  problem can  be
solved using an additional  term in the regularization, penalizing solutions
where  $\M{\alpha}_i$  becomes  lower  than  a  predefined  threshold.   For
example, in the  case of continuum subtracted spectra,  the threshold can be
set  from  $\M{x}_i=\frac{10^{-3}}{N_\mathrm{lib}}$  to  $\frac{10^{-5}}{N_\mathrm{lib}}$,
$N_\mathrm{lib}$  being the number  of spectra  in the  library, ensuring  that the
contribution of a template spectrum away from the solution is negligible.

\subsubsection{Penalization}

The problem  of determining the stellar  parameters from a  RAVE spectrum is
ill-conditioned and requires regularization in order to recover a physically
meaningful solution.  Also, the size of the synthetic spectra library we are
using  is too  large  to  enable us  to  process a  RAVE  spectrum within  a
realistic time frame considering the number of spectra to process.

Our first  operation reduces  the size of  the parameter space  by selecting
templates according to a $\chi^2$ criterion. We use the transform
\begin{equation}
\exp(\alpha'_i)=\exp(\alpha_i) \theta(\M{P}_i)\, ,
\end{equation}
and     solve    Eq.~\ref{e:Qnonlinear}     replacing     $\M{\alpha}$    by
$\M{\alpha'}$. $\theta(\M{P}_i)$ is  a gate function in the 6D stellar
parameter space. In  the 1D case it reads
\begin{eqnarray}
\theta(\M{P}_i) =&
\begin{cases}
1\,\, \textrm{if\,\,}-\frac{1}{2}<i-i'\leq\frac{1}{2} \textrm{\,\, and\,\,} \chi^2(\M{P}_i')<\chi^2_\mathrm{lim}\\
0\,\, \textrm{otherwise}
\end{cases}
&\, .
\end{eqnarray}
At  each  point  on  the   grid  defined  by  the  library,  the  derivative
$\theta(\M{P})$  is   0.   Therefore,  solving   Eq.~\ref{e:Qnonlinear}  for
$\M{\alpha'}$ is  equivalent to solving  the same equation  for $\M{\alpha}$
but on a reduced subset  of $\M{\alpha}$ matching the $\chi^2$ condition, and
we shall drop the prime in the following.

We  choose to use  a $\chi^2$  criterion to  select the  subset in  order to
include local minima with a  $\chi^2$ value close to the minimum $\chi^2$.
This selection criterion avoids potential problems where the noise, ghost or
cosmic  rays  create spurious  minima  which could  lead  to  biases in  the
estimated  stellar  parameters.   Care  must  be taken  when  selecting  the
$\chi^2$ limit  as, if the number of  spectra in the subset  of templates is
not  large enough,  biases  can be  introduced  in the  solution. The  limit
$\chi^2_\mathrm{lim}$  was  chosen  according  to numerical  simulation  using  the
synthetic   template   library.    Simulations   have  shown   that,   using
Eq.~\ref{e:Qnonlinear}, at  least the 150  template spectra from  the lowest
$\chi^2$  must be  used  to  minimize the  reconstruction  errors and  avoid
biases. As those  simulations were run using idealized  spectra, in practice
$\chi^2_\mathrm{lim}$  is  set  to  the  300$^{th}$ lowest  $\chi^2$  for  a  given
spectrum. This leads to a subsample  of the library containing between 2 and
4  values  per  parameter,  depending  on  the  location  in  the  parameter
space. This number is lower than  $3^6$ which would be the number of spectra
used for  a quadratic interpolation on  a complete 6D  grid, and is due  to the
fact  that  the  stellar  parameter  space  is not  evenly  covered  by  the
library. The average  number of direct neighbors (on a grid  point next to a
given parameter) is 85, varying between 1 and 314.\\

Reducing the number  of template spectra does not  solve the ill-conditioned
nature of  the problem, even if  the number of templates  becomes lower than
the number of pixels. This is due  to the fact that the pixel values are not
independent and  the information on  effective temperature, gravity  etc. is
redundant in a spectrum. To regularize the problem we use the property that,
in the idealized continuous case,  the solution $w(\M{P})$ is expected to be
close to a  Gaussian function centered on the  true solution.  Therefore, we
expect the discrete solution to follow the same behaviour and we require the
solution $\M{x}$  to be smooth in  the parameter space.  Nevertheless, as in
the real case  the solution might have local minima because  of the noise or
features in  the spectrum,  we do  not impose any  particular shape  for the
solution and  we keep the  method non parametric\footnote{The method  is non
parametric in the sense that no  functional form is imposed for the array of
parameters.}.   We only require  that the  variation of  the weights  in the
parameter space is locally smooth. We define the penalization operator $P_2$
as
\begin{equation}
P_2(\M{\alpha})=\T{\exp(\M{\alpha})}\mdot\T{\M{L}}\mdot \M{L}\mdot
\exp(\M{\alpha})
\end{equation}
where 
\begin{eqnarray}
\M{L}_{i,j} \propto 
\begin{cases}
\frac{-1}{<d>_{{\cal N}i}}\,\,\, i \neq j\,\,, i \in {\cal N}_i\\
1 \,\,\, i=j\\
0 \,\,\, \textrm{otherwise.}
\end{cases}
\end{eqnarray}
$d$ is the distance in the parameter space defined as
\begin{equation}
d(\M{P_i,P_j}) = \sqrt{ \sum_k \frac{(P_{i,k}-P_{j,k})^2}{\sigma_k^2}}\, ,
\end{equation}
$k$ being an  index over the dimensions of the  stellar parameter space, 
$<d>_{{\cal N}i}$  the mean  distance over a  fixed neighborhood $\cal N$
of  the point defined  by  the  index  $i$  in  the parameter  space, and 
$\sigma_k$ the dispersion in the stellar parameter $k$.  In 
practice, the neighborhood is set to the 40 closest points in $\M{P}$ which is
approximately half  the average number of  neighbors. The fact  that not the
entire  neighborhood  is used  to  compute  the  average distance  does  not
introduce  errors  as the  operator  is local  and  all  the templates  will
contribute as the operator is applied over the entire set of templates. Note
here that  $<d>_{{\cal N}i}$ is always  lower than 1.   With this definition,
$\T{\M{L}}\mdot \M{L}$ will be large  for $i=j$, negative in the surrounding
of $i$  in the parameter space  and 0 outside.   $P_2$ is then large  when a
large value  for a given template  $i$ is not balanced  by its neighborhood,
penalizing  strong local  variations  like  peaks of  width  lower than  the
$\sigma_k$ of the library.\\

To derive the  stellar parameters, the method presented  above is applied on
continuum  subtracted spectra  and to recover the proper continuum level 
we have the additional constraint that 
$\sum_i \exp(\alpha_i)=1$.  Therefore, we add  a third penalization  term to
ensure that the  sum of the weights $\sum \exp (\M{\alpha})$  is one. 
For clean spectra this last  penalization can  be omitted. 
But in the case of RAVE a ghost can affect the  blue part of some spectra and 
there may be residuals of cosmic ray strikes. So imposing the continuum level 
enables us to  avoid potential  problems in automatic processing. The
operator $P_3$ is defined as
\begin{equation}
P_3(\M{\alpha})=1.-\exp \bigg(
-\frac{(1-\sum\exp(\M{\alpha}))^2}{2\sigma_3^2}\bigg) \, .
\end{equation}
This    operator    has     an   inverted    Gaussian    behaviour    around
$\sum\exp(\M{\alpha})=1$,  with  $P_3=0$  for  $\sum\exp(\M{\alpha})=1$  and
$P_3=1$ away  from this  value. To estimate  the stellar parameters,  we use
$\sigma_3=0.01$ or 1\% of the continuum value.

\subsubsection{Validation of the method}

\label{methodvalidation}
To establish the validity of  our approach to recover the stellar parameters
in the RAVE  regime, we tested the algorithm on a  series of 20,000 synthetic
spectra built using the same template library. As the accuracy of the method
depends on  the resolution and wavelength interval  (the Lagrange parameters
must be defined  separately for each instrument and library),  we do not try
to validate the method outside  of our observational regime and this section
will  focus on  an idealized  case mimicking  the RAVE  spectra.   A complete
discussion  of  error  estimates  and  zero  point  offsets,  comparing  our
measurements to other sources, is presented in Sec.~\ref{zeropointoffsets}.

The synthetic spectra are randomly  generated from a linear interpolation of
the library and three ingredients are added :
\begin{itemize}
\item[1)] a Gaussian white noise with SNR in the range [10..40]
\item[2)] a RV mismatch up to 5~km/s
\item[3)] continuum structures amounting to up to 5\% of the continuum level. 
\end{itemize}
 
These three ingredients  were added to mimic observed  and expected features
in the RAVE spectra: random noise  level typical of the RAVE spectra, a mean
internal    RV   error    of    $\sim$5~km/s   after    the   1$^{st}$    RV
estimation\footnote{The  spectra  used   for  parameter  estimation  are  RV
corrected after a first RV estimation  using a reduced set of templates (see
paper I), a  better template with proper parameters  is then generated using
the algorithm and is then used  for the final RV calculation.}  and residual
continuum  features that  can be  left after  data reduction.   The residual
continuum  features are included  using 1  to 5  cosine functions  each with
arbitrary  phase, frequency  (between 0.5  and 5  periods on  the wavelength
interval) and normalization (within 0 and 5\% of the continuum level).

\begin{figure*}[hbtp]
\centering
\includegraphics[width=9cm,angle=270]{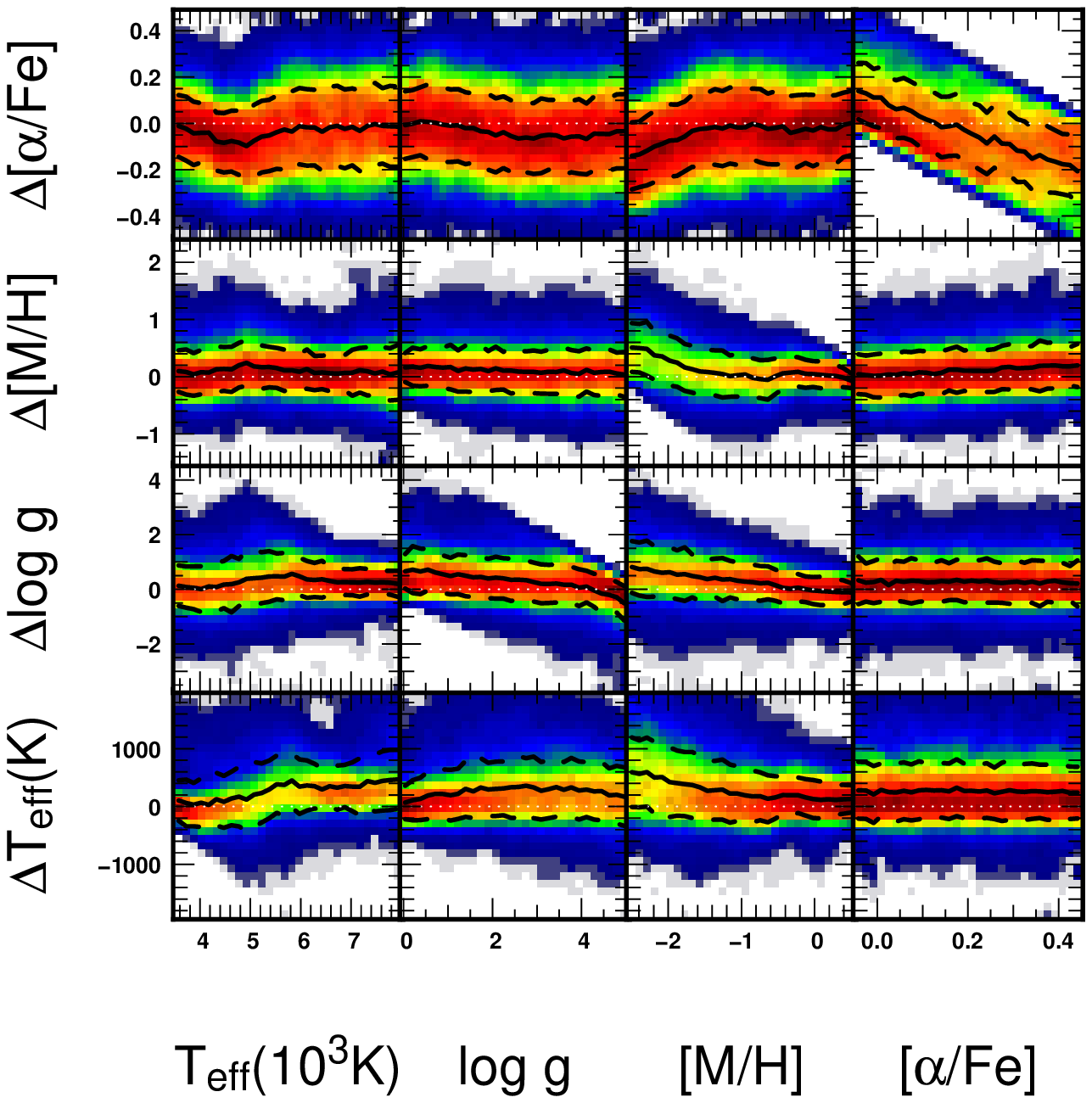}
\includegraphics[width=9cm,angle=270]{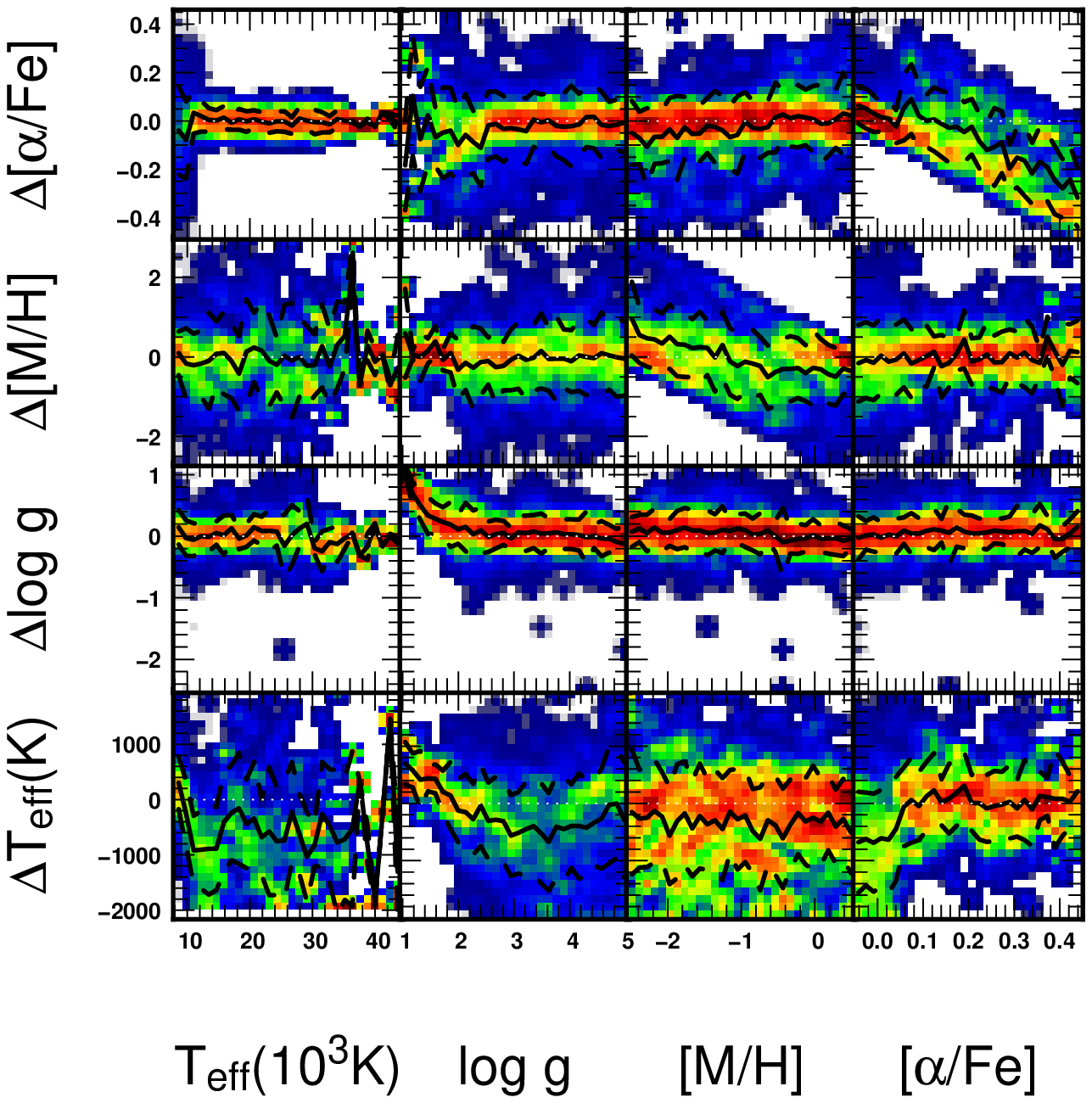}
\caption{2D histograms of the reconstruction error (RAVE-true) as a function of
the true parameters for the  four parameters reported in this release. The
color gradients  follow the  number density per  bin. Thick black  lines and
dashed  lines are  the mean  error and  RMS per column. White  dotted lines
indicate  a zero reconstruction  error. Left  panel: spectra  with $\teff$
below 8000~K. Right panel: spectra with $\teff \geq 8000$~K.}
\label{f:reconserror}
\end{figure*}

Figure~\ref{f:reconserror} presents the  reconstruction error (RAVE-true) as a
function of the  various parameters released in DR2  for the 20000 simulated
spectra. As mentioned  before, the rotational velocity will  be discussed in
another paper, and microturbulence can not be recovered  in the RAVE regime.
Therefore, these  two parameters are  not presented here.   Nevertheless, we
stress  that all  6  parameters were  used  in the  simulations, the same 
as was done in the standard pipeline on observed spectra. 
The  left panel  represents the  spectra  with effective
temperature  below  8000~K (CaII  lines  dominated)  while  the right  panel
presents  the hotter  spectra that  are  dominated by  the hydrogen  Paschen
lines. The  number of  simulated spectra in  the left panel  is $\sim$17,500
while the  right panel  contains $\sim2,500$ simulated  spectra. This  is an
effect of  the template library, the  cool part of the  library having a denser
grid of spectra than the hot side. Also, a smoothing  was applied to the 
right panel for the visualization, to lower the effect of the noise.

These  simulations enable  us to  assess  the expected  dependencies of  our
errors  as  a   function  of  the  various  stellar   parameters.  The  main
characteristics we observe are:
\begin{itemize}
\item[-] Below 8000~K, there is little dependence of the recovered
parameters on $\teff$ but for $\teff$ itself with an overestimation that
increases as the effective temperature becomes larger.
\item[-] [M/H] is the main driver for the errors in the low metallicity
regime ([M/H]$\simlt-1.0$)  with all  parameters but  $\alphaenh$ being
overestimated, while $\alphaenh$ is underestimated. This indicates that for
the  metallicity, the true  metal content  can only  be recovered  when both
[M/H]     and    $\alphaenh$     are     considered    (see     discussion
Sec.~\ref{zeropointoffsets}).
\item[-] $\alphaenh$, as expected, is not properly recovered in the RAVE
regime as shown by the upper right panels.
\item[-] $\logg$ is better constrained in hot spectra than in cool spectra.
\item[-] $\teff$ is systematically underestimated  for hot spectra.
\end{itemize}

The overall  accuracy we can expect  for the stellar parameters  in the RAVE
regime ranges  then from 200~K  to 500~K for  $\teff$, 0.2 to  0.5~dex for
$\log  g$  and  0.1  to  0.4~dex   for  [M/H]  (depending  on  the  value  of
$\alpha$ enhancement) while $\alphaenh$ alone is not recovered.

\begin{figure}[hbtp]
\centering
\includegraphics[width=14.7cm,angle=270]{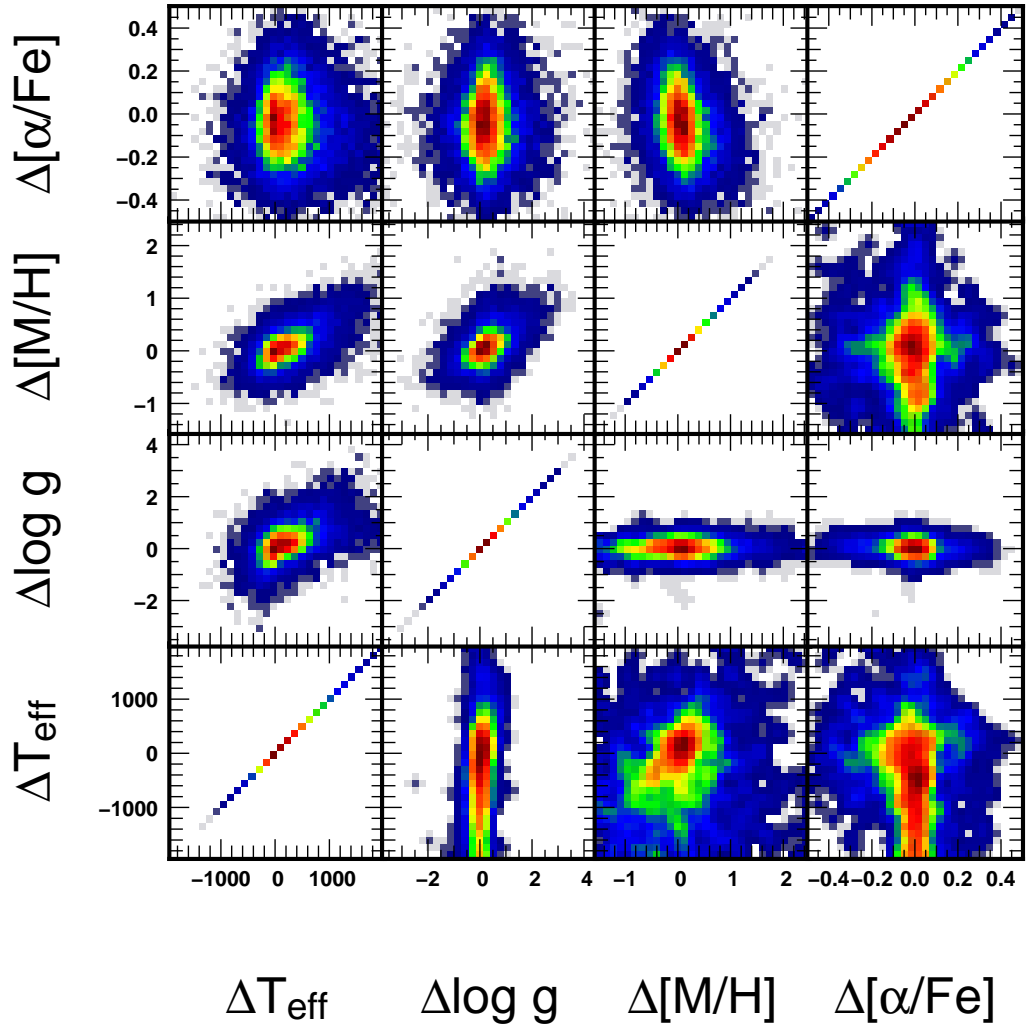}
\caption{2D histograms of the reconstruction errors vs reconstruction
error. The  upper triangle is  for spectra with effective  temperature below
8000~K,  the lower  triangle for  spectra with  $\teff$ above  8000~K. The
color coding follows the number density per bin.}
\label{errorcorrel}
\end{figure}

A  better understanding  of the relations and mutual influences of the 
errors on the
stellar parameters is gained from the correlations between the reconstruction
errors.  These  are presented in Fig.~\ref{errorcorrel}  where the different
behavior of the  hot stars and of the cool stars  is apparent. The upper
triangle presents  the correlations  between the errors  for the  cool stars
while  the lower triangle  shows the  correlation for  the hot  spectra (the
lower triangle has been smoothed for visual rendering).

It is clear  in this figure that  in the cool spectra regime,  the errors on
the parameter reconstruction are strongly correlated which indicates that an
error on one  parameter results in errors on the  other parameters. There is
however an exception for $\alphaenh$ which is only anti-correlated to [M/H]
and not correlated to the other parameters which further indicates that only
a combination of [M/H] and $\alphaenh$  is recovered, and that these two 
quantities cannot be uniquely separated. 

The  situation  is different  for  the hot  stars,  where  the only  visible
correlation  is between $\teff$  and [M/H]  and only  for large  errors on
[M/H].  Otherwise, no  correlation is  seen  indicating that  the system  is
better constrained. Nevertheless, typical errors for hot stars are larger
than for cool stars with similar noise levels. 

Overall, the  method presented allows  us to recover the  stellar parameters
with a good  accuracy knowing that our wavelength interval  is small and our
resolution is  limited (R$\sim$7500). The expected  correlations between the
reconstruction errors for the  different parameters are well behaved (simple
one mode  correlations) if one is  able to distinguish a  posteriori the two
cases, hot and cool stars.

\subsection{Estimate of the ratio of signal to noise}

\label{S:S2N}
The initial estimate of the signal to noise (S/N) comes from comparison of 1-D 
spectra derived from typically 5 subexposures of a given field 
(see Paper~I for details). This estimate is model independent 
and readily available for the calculation of $\chi^2$ for the radial velocity 
and stellar parameter determination routines. However any change of observing 
conditions during the observing run may contribute to differences of 
subexposure spectra and therefore render the value of S/N too low. 
We therefore wrote a procedure which calculates the S/N from the 
final spectrum only. We refer to it as the S2N value in the data release, 
while the one calculated from subexposure variation is labeled SNR. 

Line-free regions in observed spectra are very scarce. Moreover, the spectra 
are quite noisy, so one does not know {\it a priori\/} if an apparently
line-free region does not hide weak absorption lines. So it seems obvious 
that suitable regions should be chosen by comparison of the observed spectrum 
to the best matching template. 

The procedure is as follows:
\begin{enumerate}
\item
The {\it normalized\/} final observed spectrum (shifted and 
resampled to the rest frame) is compared with the synthetic library 
template with the best correlation.
The two spectra are not identical for two reasons: noise in 
the observed spectrum and systematic deviations (due to 
observational or theoretical computation deficiencies). We want to avoid 
the latter. The difference between the observed and theoretical spectrum 
often alternates in sign between consecutive wavelength pixels if it is 
due to noise. But systematics usually affect several adjacent 
wavelength bins, so the sign of the difference does not vary so frequently. 
We therefore decided to use only those pixels for which the difference 
changes sign from the previous {\it or} towards the next adjacent pixel.
This selection scheme retains 75\%\ of all pixels if the reason for variation
is just noise. This seems a reasonable price to pay in order to avoid 
systematics. Note that we impose restrictions 
only on the {\it sign} of the difference, not on its absolute value, so 
noise properties are not affected.
\item
Regions of strong spectral lines are prone to systematic errors. So we 
discard any pixel for which the flux of the template would 
be less than 0.9 of the continuum flux. Strong spectral lines span a 
small fraction of the entire spectral range, except in high 
temperature objects. The derived S/N estimate is representative of the 
continuum S/N, but the value generally does not differ by more than 5-10\%\ 
from the S/N averaged over the whole spectrum. 
\item
Next we calculate the difference between the observed and theoretical 
spectrum, and divide it by the theoretical spectrum flux. 
The final S2N estimate is an inverse of its standard deviation, of course 
only using the pixels retained in the steps above.
\item
The observed spectra we used for the three steps above are 
shifted to the rest frame and resampled with  
respect to the original ones given in observed wavelengths.
This is important, as resampling damps the pixel to pixel variation and 
therefore artificially increases the measured value of the signal to 
noise. So we need to take it into account. The resampled and the original 
spectrum have the same number of pixels, so resampling can be 
characterized by its average fractional-pixel shift. 
A zero shift obviously does nothing, but a shift of half a pixel means 
that the S/N estimate measured in previous steps needs to be 
multiplied by a factor of $1/\sqrt{2}$. If the shift is a fraction $x$ of the 
pixel separation, the expression for the damping factor $f_\mathrm{SN} =
\frac{1-1/\sqrt{2}}{0.5^2} (x-0.5)^2 + 1/\sqrt{2}$. The S/N 
value calculated by the 3 steps above needs to be multiplied by this 
damping factor to obtain the final S2N estimate of the observed spectrum. 
In the case of RAVE data the fractional
shift of pixels at both edges of the spectrum is zero, while the pixels 
in-between are resampled from an observed non-linear to a linear increase of 
wavelength with pixel number. Because this non-linearity is always very 
similar also the resulting damping factor turns out to be well constrained: 
$f_\mathrm{SN} = 0.78 \pm 0.01$. This value is actually very close to 
0.805 obtained for a uniform distribution of fractional pixel shifts $x$ 
in the $[0,1]$ interval.
\end{enumerate}

The first two points limit the fraction of pixels used in the S2N estimate 
to $46\% \pm 6\%$. This is true also for hot stars, so the selection 
outlined above does not seem to be too constraining. 
Note that step 4 means that the S2N values are lower than the 
ones calculated by e.g.\ the {\it splot} package of IRAF, because the latter 
does not take into account the effects of resampling. The SNR estimate is 
very sensitive to variations of atmosphere transparency and instrumental 
effects during the observing sequence while the S2N is not. So S2N values are 
similar to the SNR ones, with the average value of S2N being 
$\sim 33$\%\ higher. We propose to use the S2N as the final S/N estimate 
for the spectrum. So the quantity S/N below always refers to the S2N value.

\begin{figure}[hbtp]
\centering
\includegraphics[width=9.7cm,angle=270]{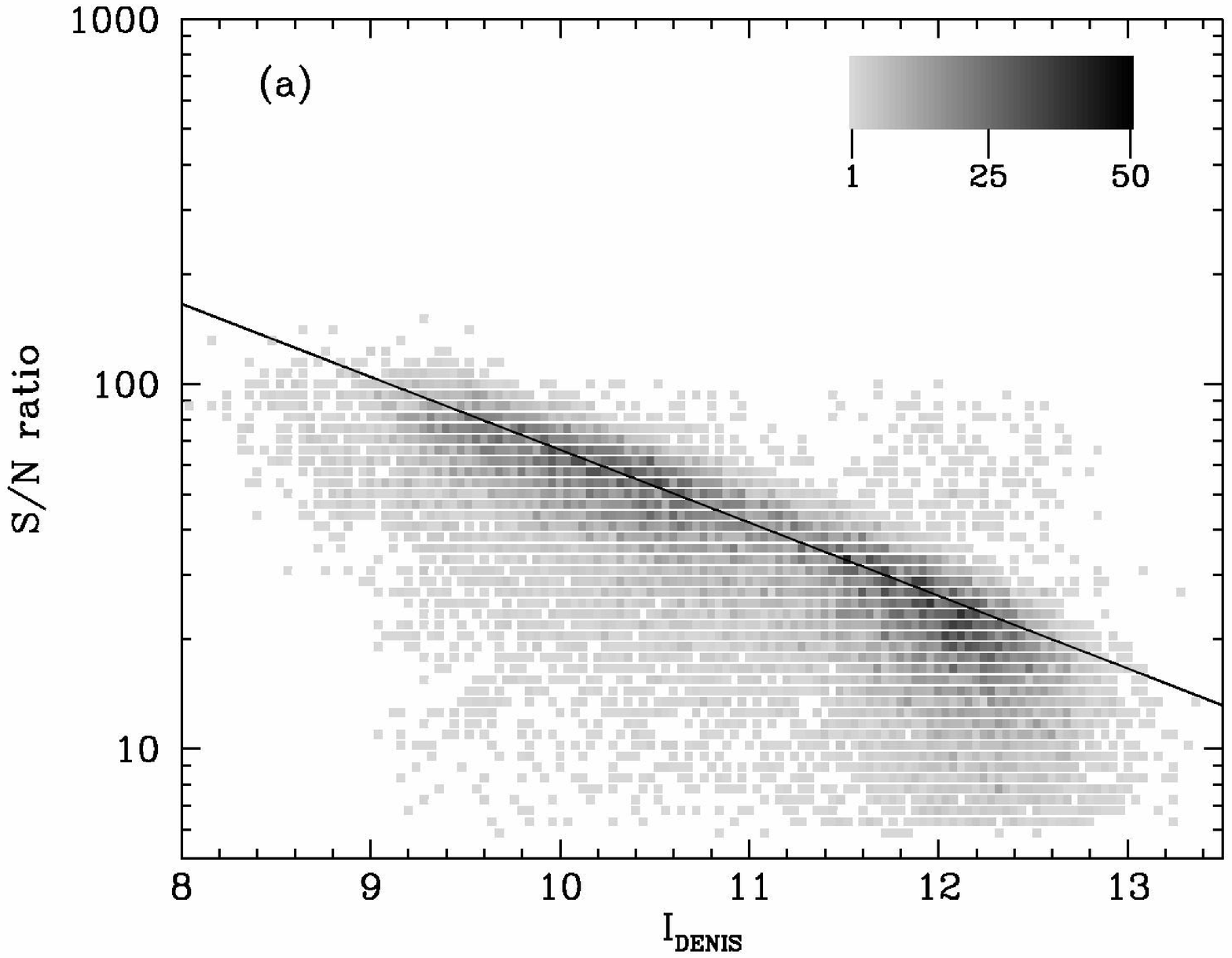}
\includegraphics[width=9.7cm,angle=270]{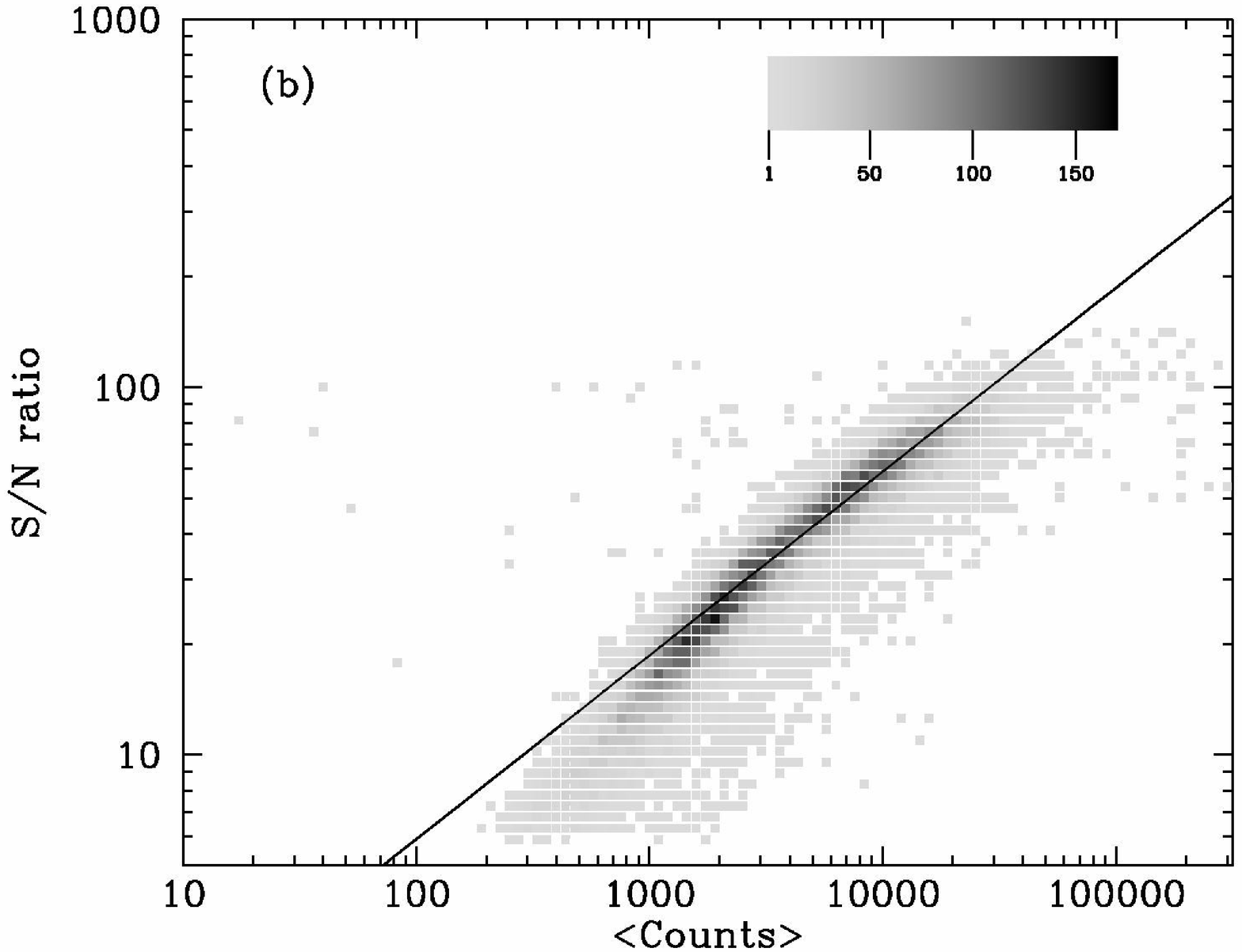}
\caption{Signal to noise ratio (S2N) as a function of Denis $I$
magnitude (a), and average level of counts per wavelength bin 
per hour of exposure time (b). Shades of 
grey mark the number of spectra in a given bin, as explained in 
the key. The straight line in the top panel follows the relation 
(\ref{e:SNvsI}), while the one in the bottom panel follows from 
equations (\ref{e:countsvsI}) and (\ref{e:SNvsI}).
}
\label{f07}
\end{figure}

Fig.~\ref{f07} plots the signal to noise ratio (S2N) as a function of the 
Denis $I$ magnitude and average number of counts per pixel.
The latter was calculated in the central part of the spectrum 
($8449.77\ \mathrm{\AA} \le \lambda \le 8746.84\ \mathrm{\AA}$). 
The straight line in the magnitude graph (Fig.~\ref{f07}a) follows the 
relation
\begin{equation}
S/N = 10^{-0.2 (I_\mathrm{DENIS} - 19.1)}
\label{e:SNvsI}
\end{equation}
while the one in Fig.~\ref{f07}b is obtained from combining it with
equation (\ref{e:countsvsI}). The constant term in eq.~\ref{e:SNvsI} is 
the mode of the magnitude corrected S/N distribution. The magnitude 
graph shows that the signal to noise can be predicted from DENIS $I$ magnitude
with an average error of $\approx \pm 50$\%. The dependence of S/N on the count 
level is much better determined, with a dispersion of the central ridge of 
only $\pm 15$\%. The difference is due to an uneven transparency of the Earth's 
atmosphere and of optical fibers which have a stronger effect on the magnitude 
graph. Sky background as well as light scattered within the spectrograph 
are of increasing relative importance for faint objects. They cause 
the deviation from a straight line seen in both panels at faint  
count or magnitude levels.

\section{Data quality}

\subsection{Radial velocity accuracy}

\label{radialvelocityaccuracy}
\begin{figure}[hbtp]
\centering
\includegraphics[width=10.7cm,angle=0]{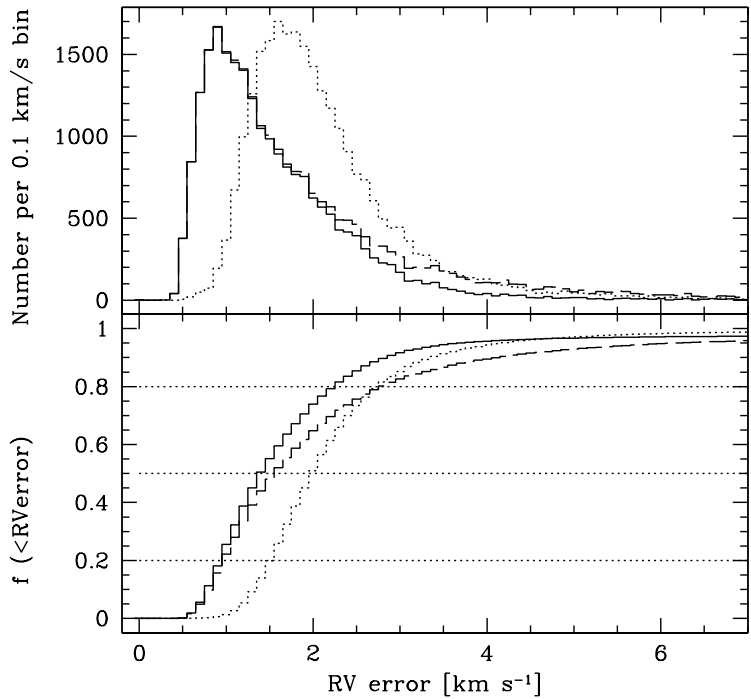}
\caption{Top panel: distribution of the internal radial velocity error. 
Solid line is for normal stars which also have their values of stellar parameters 
published. The dashed line is for all stars new in the present data release. The 
dotted histogram are stars from the first data release. 
Bottom panel:
fraction of RAVE targets with a radial velocity error lower than a given value. 
The dotted lines indicate limits of 20\%, 50\% and 80\%. The line types are the 
same as in the top panel.
}
\label{f_rv_internal_error_histo}
\end{figure}

The distribution of the internal radial velocity errors is presented in 
Fig.~\ref{f_rv_internal_error_histo}. These are the estimated uncertainties
of fitting a parabola to the top of the correlation peak (Paper~I). The top
panel shows the histogram of the radial velocity error in 0.1 \kms\ bins, while the bottom
panel is the cumulative distribution. Results for the spectra new to this data release
and for the ones from Paper~I are shown separately. In the former case we also add 
the results for spectra for which we are publishing the values of stellar parameters 
(see below). These are spectra of sufficiently high quality and without peculiarities 
(binarity, emission lines etc.). Table~\ref{t:internalrverror} summarizes the values of 
the most probable and average internal velocity errors. 

The blue light blocking filter, which cuts the second order light and was used for data 
new to this release, clearly improves the match between theoretical templates and observed 
spectra. This is mostly a consequence of the more accurate flatfielding of a rather 
narrow spectral range of the first order light, compared to a mix of relative contributions 
of the first and the second order spectra which emphasizes any differences in the color 
temperature between the star and the calibration lamp. Also the level of the continuum 
is much easier to determine if the blocking filter is used. The most probable value of 
the internal velocity error is 0.9~\kms\ for the data new to this data release, compared 
to 1.7~\kms\ in Paper~I. On the other hand the possibility of a better match also 
increases the chances to identify any types of peculiarities. So there is a rather notable 
tail of large internal velocity errors if we consider all data new to this data release 
(dashed line in Fig.~\ref{f_rv_internal_error_histo}). If only normal stars are 
plotted (solid line in Fig.~\ref{f_rv_internal_error_histo}) large velocity errors 
are much less common. This is reflected also in the average errors reported in 
Table~\ref{t:internalrverror}.

\begin{figure}[hbtp]
\centering
\includegraphics[width=10.7cm,angle=270]{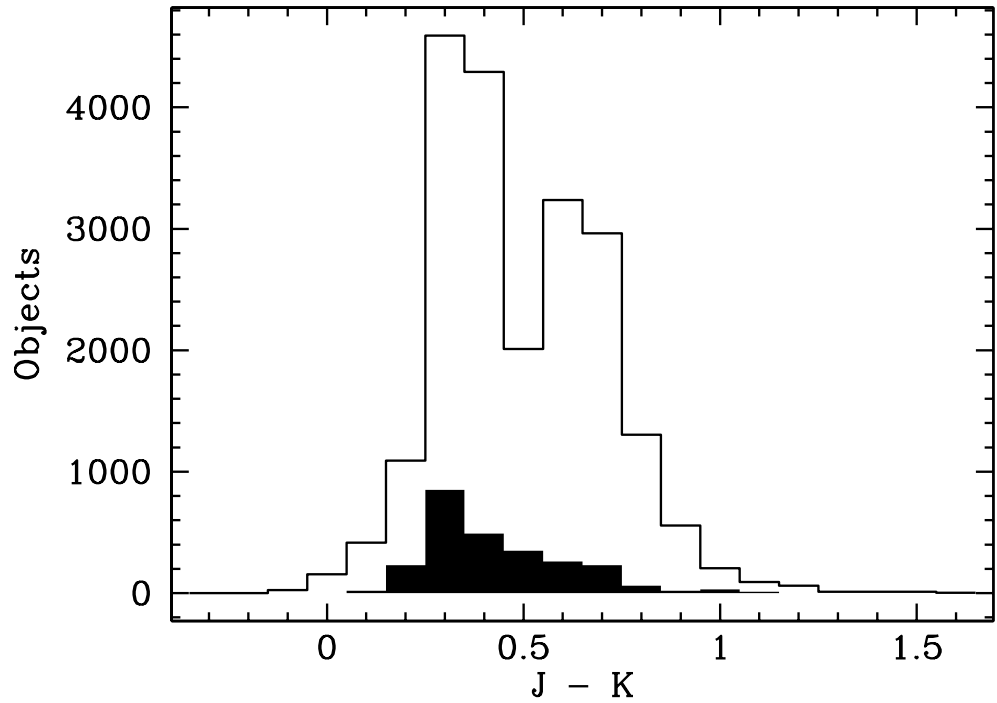}
\caption{$(J-K)_\mathrm{2MASS}$  colors for the second data release (solid line) and 
for reference stars used to check the radial velocity accuracy (filled 
histogram, its values multiplied by 10). Most reference stars, especially 
those from the Geneva Copenhagen Survey, are yellow dwarfs with only a 
small fraction of red giants. 
}
\label{f:JK_histo_rvreference_rave}
\end{figure}

Internal velocity errors are useful, but they do not include possible systematic 
effects. As mentioned in Sec.~\ref{radialvelocitydetermination} 
the reported radial velocities do not allow for shifts due to non-vanishing 
convective motions in the stellar atmosphere and for gravitational redshift 
of the light leaving the stellar surface. This is the case also with other 
spectroscopically determined radial velocities.
We compared RAVE radial velocities with those obtained from external datasets.
255 stars from 4 different datasets were used to assess the accuracy of 
radial velocities of stars new to this data release. From these, 213 stars 
turn out to have normal spectra
without emission lines, strong stellar activity or stellar multiplicity
and have radial velocity errors smaller than 5 \kms, so that they are retained 
for further analysis. They include 144 stars from the Geneva Copenhagen 
Survey (GCS), and three datasets observed specifically to check RAVE 
radial velocities: 33 stars were observed 
with the Sophie and 15 with the Elodie spectrograph at the Observatoire de 
Haute Provence, and 21 stars with the echelle spectrograph at the Asiago 
observatory. Stars observed in Asiago span the whole range of colors, 
while most other datasets and especially GCS focus on yellow 
dwarfs. The whole survey includes a larger number of red stars 
(Fig.~\ref{f:JK_histo_rvreference_rave}) which are mostly giants, as can be 
seen from temperature-gravity distributions derived by RAVE for the 
whole survey (Fig.~\ref{paramscalib}). A smaller fraction 
of giants in the reference datasets does not present a real problem, as 
radial velocities for giants tend to be more accurate than for dwarfs. 

\begin{figure}[hbtp]
\centering
\includegraphics[width=10.7cm,angle=270]{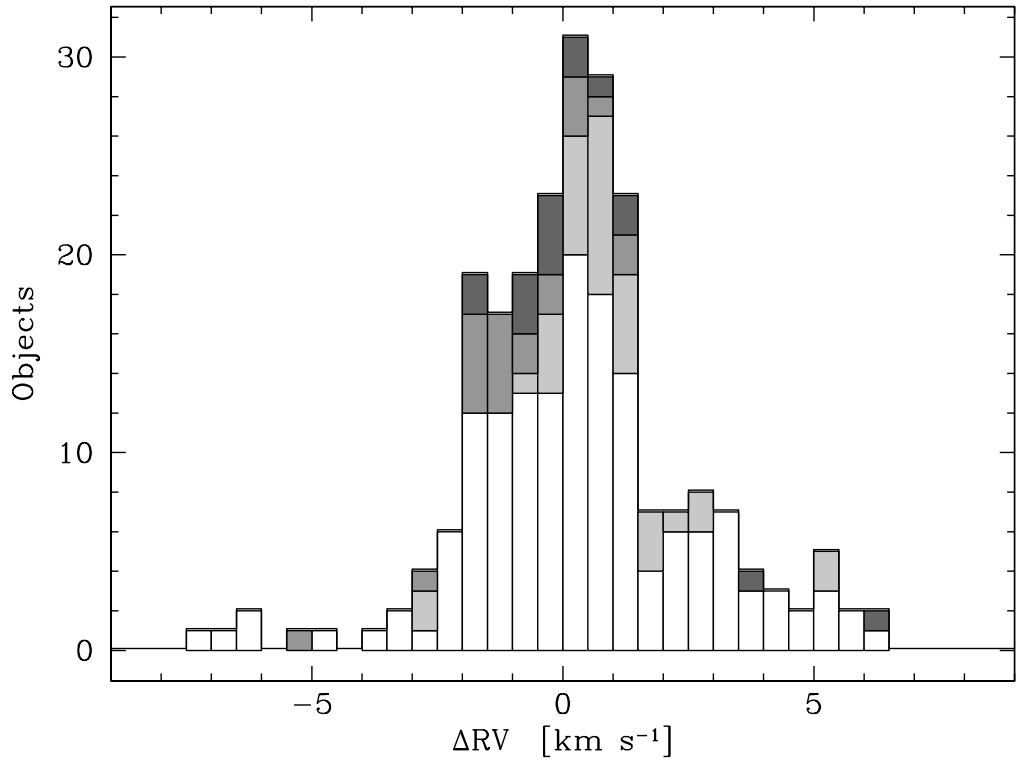}
\caption{
Difference of radial velocities as derived by RAVE and by the reference 
instruments. The solid histogram is for all objects with colored rectangles 
belonging to individual external datasets: the Geneva Copenhagen 
Survey (white), Sophie (light grey), Asiago (dark grey), and the Elodie (black).
}
\label{f08}
\end{figure}

\begin{figure}[hbtp]
\centering
\includegraphics[width=10.7cm,angle=270]{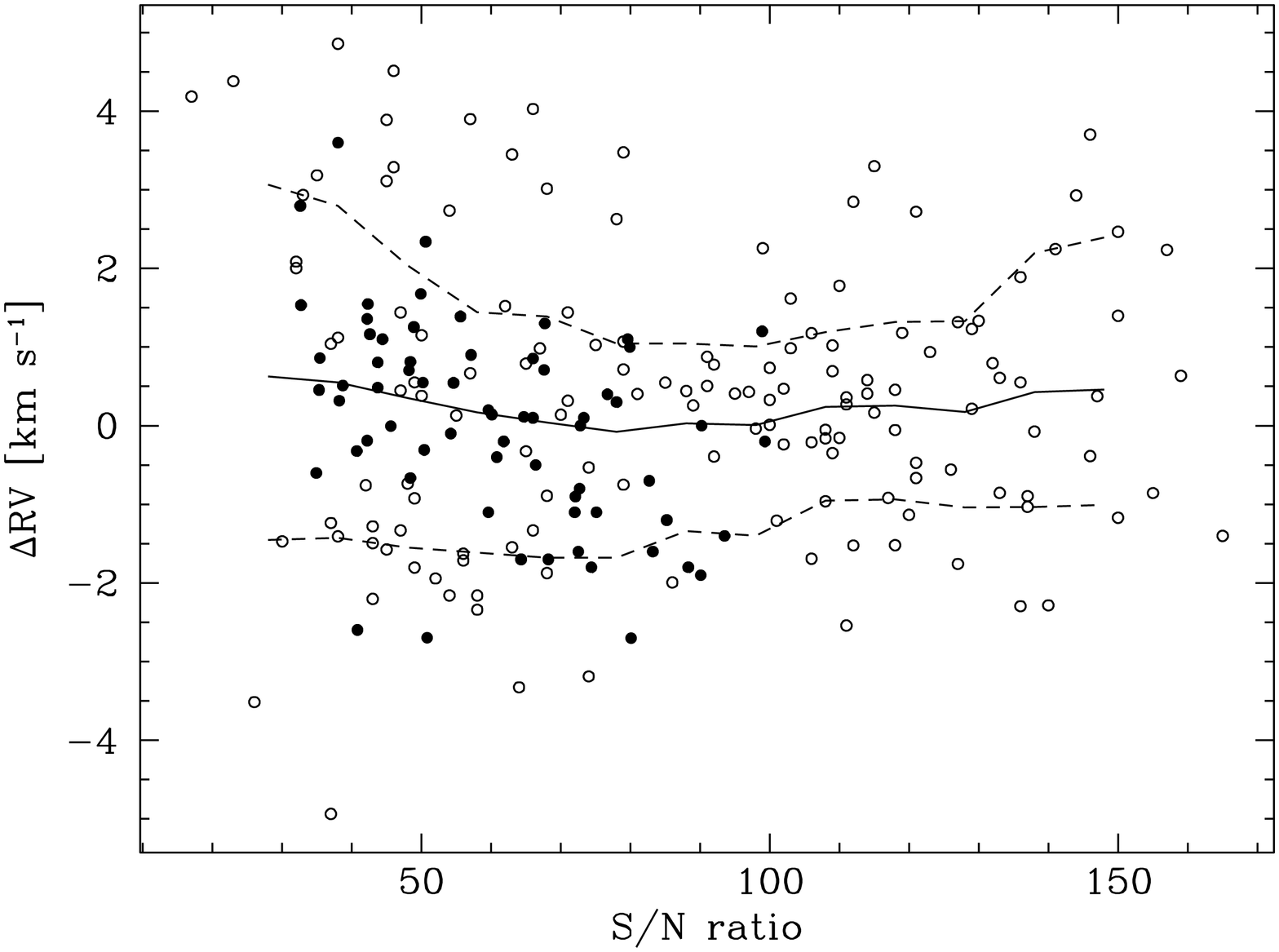}
\caption{Difference of radial velocities as derived by RAVE and by the reference 
instruments, as a function of the S/N ratio of the RAVE observation. Dots depict 
individual measurements, the solid line is a running mean (with a boxcar 
smoothing of $\pm 15$ in S/N). Similarly the area between the dashed lines 
includes 68.2\%\ of the measurements, i.e.\ $\pm \sigma$. The open symbols are 
measurements of the GCS stars, while the filled ones mark the other 3 reference 
datasets.
}
\label{f10}
\end{figure}

A comparison of radial velocities obtained by RAVE and by the reference 
datasets is presented in Figure~\ref{f08} and summarized in 
Table~\ref{t:rvaccuracy}. $N$ is the number of objects in each dataset, 
$<\!\!\Delta RV\!\!>$ is the mean of differences between RAVE and reference 
measurements and $\sigma(\Delta RV)$ is their standard deviation. We note 
that mean zero point offsets are non-zero and of different size and sign for 
separate datasets. So the difference in zero point is likely due to a different 
zero point calibration of each instrument. Most of the reference stars 
are taken from the GCS. So the large number of dwarfs from the GCS
drive also the final value of the zero point offset and its dispersion. 
If one omits those the mean zero point difference is only 0.1~\kms\ 
and the dispersion ($\sigma(\Delta RV)$) is 1.30~\kms. These estimates neglect 
the intrinsic measurement errors of each reference dataset. A typical error 
of 0.7~\kms\ and a zero point offset of 0.3~\kms\ for the GCS 
suggest that the RV error of RAVE is $\sim 1.3$~\kms. This is also the 
value derived from the other datasets. Figure~\ref{f10} shows that 
the standard deviation stays within $\sim 2$~\kms\ even at low ratios 
of signal to noise. This value decreases to $\sim 1.5$~\kms\ if one 
omits the stars from the GCS dataset. 

Most of the stars in external datasets are dwarfs with a metallicity close to 
the Solar one. The midpoint of $ | \Delta RV | $ stays at $\sim 1.2$~\kms\ \ for temperatures
lower than 5800~K, increasing to $\sim 2$~\kms\ for stars with 6800~K. There is 
no significant variation of the radial velocity difference with metallicity in the 
range $-0.5 < \MH < 0.3$ covered by the external datasets. 

One can also use repeated observations of RAVE stars to assess the internal 
consistency of the measurements. Section \ref{repeatedobservations} shows 
that radial velocity from a pair of measurements of a single star 
differ by  $\le 1.80$~\kms\ in 68.2\%\ of the cases. This corresponds 
to an error of 1.3~\kms\ for a single measurement.

We conclude that the typical RV error for data new to this data release is 
$\simlt 2$~\kms. For the measurements with a high 
value of S/N the error is only 1.3~\kms\  with a negligible zero point error.

\begin{deluxetable}{lcc}
\tablecaption{Internal radial velocity errors 
\label{t:internalrverror}}
\tablewidth{0pt}
\tablecolumns{3}
\tablehead{
\colhead{Dataset} &\multicolumn{1}{c}{Peak}&\multicolumn{1}{c}{Average}\\
                  & (\kms)& (\kms) \\
}
\startdata
Spectra of normal stars new to DR2 & 0.9 & 2.0 \\
Spectra of all stars new to DR2    & 0.9 & 2.5 \\
Spectra of all stars in DR1        & 1.7 & 2.3 \\ \hline
All                                & 1.6 & 2.3 \\
\enddata
\tablecomments{Peak value refers to 
the histogram in Fig.~\ref{f_rv_internal_error_histo}.
}
\end{deluxetable}

\begin{deluxetable}{lrrr}
\tablecaption{Datasets used to check the radial velocity accuracy 
\label{t:rvaccuracy}}
\tablewidth{0pt}
\tablecolumns{4}
\tablehead{
\colhead{Reference Dataset} & $N$ & $<\!\!\Delta RV\!\!>$ & $\sigma(\Delta RV)$ \\
                            &     & (\kms) & (\kms)
}
\startdata
Geneva Copenhagen Survey & 144 &  0.34  & 1.83 \\
Sophie observations      &  33 &  0.63  & 1.18 \\
Asiago observations      &  21 &--0.71  & 1.09 \\
Elodie observations      &  15 &  0.07  & 1.32 \\ \hline
All                      & 213 &  0.26  & 1.68 \\
Last 3 datasets          &  69 &  0.10  & 1.30 \\
\enddata
\tablecomments{$\Delta RV$ 
is the difference between radial velocities derived by RAVE 
and those from the reference dataset.}
\end{deluxetable}

\subsection{Accuracy of stellar parameters}

\label{accuracystellarparameters}
For the vast majority of the stars in this data release, there is no prior spectroscopic
information available. Some photometric information
is available (see Section \ref{photometry}) but after a detailed investigation 
we concluded that this external information is not of sufficient quality 
to be used as a prior on any of the stellar parameters. Unknown extinction 
presents a further problem. This situation is expected to continue until 
high quality multi-epoch photometry becomes available for the Southern sky 
from the SkyMapper project \citep{skymapper}. RAVE is 
therefore the first large spectroscopic survey to use only spectroscopic data 
to derive the values of stellar parameters. So it is appropriate to make a 
detailed check of the results with external datasets coming both from 
the literature and our own custom observations. 

\subsubsection{External datasets}

\label{externaldatasets}
RAVE stars are generally too faint to have data available in the literature, 
so we obtained a separate set of RAVE observations of stars from three 
reference sets in the literature. In addition we obtained custom 
observations of regular RAVE targets with two Northern hemisphere telescopes, at Observatory 
in Asiago and at Apache Point Observatory. In the coming months we plan to expand 
the comparison using the observing time allocated at UCLES at Siding Spring 
and at ESO. Here we describe the presently available datasets which 
contain altogether 331 stars. In all cases the corresponding RAVE observations 
were obtained and processed in the same way as for the other stars in this 
data release. 

The ARC echelle spectrograph at Apache Point Observatory (APO) 3.5-m telescope was 
used to observe 45 RAVE stars. These spectra cover the entire optical wavelength
range (3,500 -- 10,000~\AA) in 107 orders with an effective resolving power  
of about 35,000. The spectra were extracted using standard IRAF routines 
incorporating bias and scattered light removal plus flat-field corrections. The 
wavelength calibration was obtained using ThAr hollow cathode lamps. Temperature and 
gravity were derived by a least squares fit to the same library of stellar 
spectra \citep{munari2005} as used for RAVE catalog, but using a  
procedure \citep{munari2005a} independent from the RAVE method described in 
Sec.~\ref{estimationofstellarparameters}. 
The results are consistent with those
resulting from the analysis using the excitation temperature and equivalent widths 
of Fe~I and Fe~II lines to derive iron abundance, temperature and gravity 
(Fulbright et al. 2006, 2007). The metallicity was derived 
by both the least square fit of the whole spectrum and by the method based on 
equivalent widths of the Fe lines. The latter yields an iron abundance, but 
the metallicity can be calculated assuming that the $\alphaenh$ ratio, 
which influences the even-Z elements between O and Ti, increases linearly 
from zero for stars with [Fe/H] = 0 to $+0.3$ for stars with [Fe/H] $= -1$,
and stays constant outside these ranges. 
In Table~\ref{t:B1} we list the temperature and gravity as derived by the least 
square fit method, and metallicity from the Fe line method. 

The echelle spectrograph at the 1.8-m telescope, operated by INAF 
Osservatorio Astronomico di Padova on top of Mt. Ekar in Asiago was
used to observe 24 RAVE stars. These spectra cover the range from 3,300 to 
7,300~\AA, but the analysis was limited to the three echelle orders
around 5,200~\AA\ with the highest signal. The resolving power was 
around 20,000. The spectra were carefully treated for scattered light,
bias and flat-field and reduced using standard IRAF routines. 
They were analyzed with the same least square 
procedure as the APO data. The results are given in Table~\ref{t:B2}.

The RAVE spectrograph was used to observe three additional sets of stars with 
parameters known from the literature. We observed 60 stars from the
\citet{soubiran2005} catalog and obtained 49 spectra useful to check the 
metallicity and the temperature values. The reported gravity values were  
not used for checks as the catalog does not estimate their accuracy. 
\citet{soubiran2005} do not report 
metallicity, so its value was derived from a weighted sum of quoted 
element abundances of Fe, O, Na, Mg, Al, Si, Ca, Ti, and Ni, 
assuming Solar abundance ratios from \citet{andersgrevesse89}, in accordance 
with classical Kurucz models. The choice of a reference Solar abundance model 
is not critical. Newer Solar abundance scales introduce only a small 
shift in the mean metallicity of the \citet{soubiran2005} stars if  
compared to typical errors of RAVE observations: 
$\Delta \MH = -0.002$ for Solar abundances given by \citet{grevessesauval98} and 
$\Delta \MH = -0.005$ for \citet{asplund06} Solar abundances. 
The standard deviation of metallicities, derived from new compared 
to classical abundances, is 0.005 for \citet{grevessesauval98} and 
0.012 for \citet{asplund06}. The parameter values as derived from the 
literature and from RAVE spectra are listed in Table~\ref{t:B3}. 

We also observed 12 members of the 
M~67 cluster (Table~\ref{t:B4}) for which we adopted the metallicity 
of $+0.01$. This value of metallicity is a weighted sum of its modern 
metallicity determinations (\citet{randich06} and references therein). 
Finally Table~\ref{t:B5} reports on the comparison of temperatures for 
201 stars from the Geneva Copenhagen Survey \citep{nordstrom2004}.
This catalog does not include metallicities but only iron abundances. The 
two values are not identical, so a comparison on a star by star basis could 
not be made (but see below for a general comparison of the two values).

\begin{figure}[hbtp]
\centering
\includegraphics[width=10.7cm,angle=270]{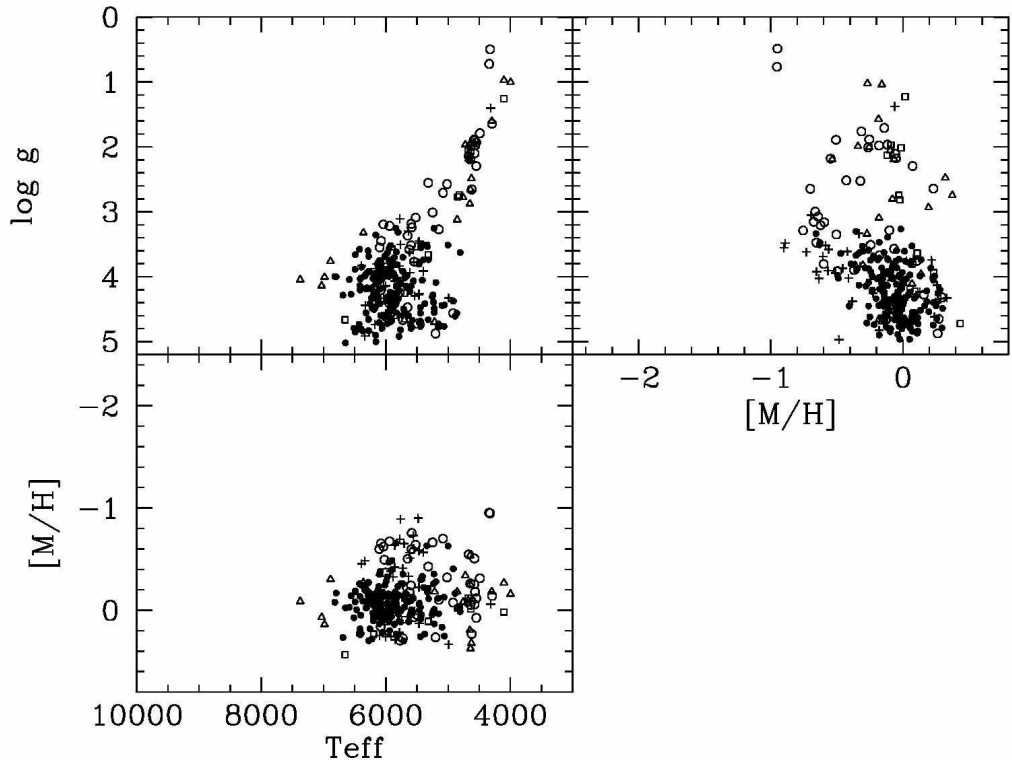}
\caption{Objects from external datasets on the temperature--gravity--metallicity
wedge using the values determined from RAVE spectra. Symbols code individual datasets
which were used to check the values of stellar parameters:
GCS ($\bullet$), Apache Point ($\circ$), Soubiran ($+$), M67 ($\Box$), and Asiago 
($\triangle$).
}
\label{f:params_comparison_datasets}
\end{figure}

Table \ref{t:externalparametersets} summarizes the properties of individual 
datasets. $N$ is the number of stars in a given dataset and the $\surd$ sign  
marks parameters that could be checked. Temperatures, gravities and metallicities 
of stars in these datasets are plotted in Fig.~\ref{f:params_comparison_datasets}.
The values are those determined from RAVE spectra, as some parameter values 
are not known for the datasets from the literature. The distributions of external 
dataset objects in the temperature--gravity--metallicity space can be compared to 
the ones of the whole data release (Fig.~\ref{paramscalib}).

\begin{deluxetable}{lrccc}
\tablecaption{Datasets used to check stellar parameters 
\label{t:externalparametersets}}
\tablewidth{0pt}
\tablecolumns{5}
\tablehead{
\colhead{Reference Dataset} & $N$ & $\teff$    & $\logg$ & [M/H] \\
}
\startdata
Apache Point echelle         &  45    & $\surd$ &$\surd$ &$\surd$ \\
Asiago echelle               &  24    & $\surd$ &$\surd$ &$\surd$ \\
Soubiran \&\ Girard catalog  &  49    & $\surd$ &        &$\surd$ \\
M67 members                  &  12    &         &        &$\surd$ \\
Geneva Copenhagen Survey     & 211    & $\surd$ &        &        \\
\enddata
\end{deluxetable}

\subsubsection{Comparison of external and RAVE parameter values}

\label{zeropointoffsets}
The first property to check is the consistency of values derived 
by the RAVE pipeline with those from the reference datasets. 
Table~\ref{t:zeropointparameteroffsets} lists mean offsets 
and dispersions around the mean for individual stellar parameters. 

\begin{deluxetable}{lrr}
\tablecaption{Zero point offsets and dispersions of the differences 
between RAVE and reference stellar parameter values 
\label{t:zeropointparameteroffsets}}
\tablewidth{0pt}
\tablecolumns{3}
\tablehead{
\colhead{Parameter} & Zero Point & Dispersion \\
}
\startdata
Temperature (without GCS dataset) & $-7 \pm 18$~K& 188~K \\
Temperature (with GCS dataset included) & $53 \pm 14$~K& 238~K \\
Gravity                  & $-0.06 \pm 0.04$ & 0.38 \\
Uncalibrated metallicity ($\mh$) & $-0.26 \pm 0.03$ & 0.37 \\
Calibrated metallicity   ($\MH$) & $  0.0  \pm 0.02$ & 0.18 \\
\enddata
\tablecomments{
In the case of metallicity the reference values are those obtained by measurement of
equivalent widths of absorption lines in the APO observations, as derived from 
the \citet{soubiran2005} catalog and from the adopted metallicity of M67.
}
\end{deluxetable}

The temperature shows no offsets when the reference sets of echelle 
observations at APO and Asiago are used, together with our 
observations of \citet{soubiran2005} stars. However if the GCS 
dataset is included the RAVE temperatures appear too hot on average,
and also the dispersion is increased. We believe this is a consequence 
of somewhat larger errors introduced by the photometric determination of 
the temperature in the GCS and not a consequence of errors 
of the RAVE pipeline. 
Gravity shows a negligible offset and a dispersion of 0.4~dex. 
However the metallicity as derived by the RAVE pipeline (in 
Table~\ref{t:zeropointparameteroffsets} we refer to it as 'uncalibrated')
appears to have a significant offset. The values derived by the 
RAVE pipeline are generally more metal poor than those obtained by 
measurement of equivalent widths of absorption lines in APO observations, 
as derived from the \citet{soubiran2005} catalog, and also compared to the 
metallicity of M67. So it seems worthwhile to explore the possibility of a calibration 
that would make metallicities derived by RAVE consistent with the 
values in these reference datasets. 

\subsubsection{Calibrating metallicity}

\label{calibratingmetallicity}
\begin{figure}[hbtp]
\centering
\includegraphics[width=10.7cm,angle=270]{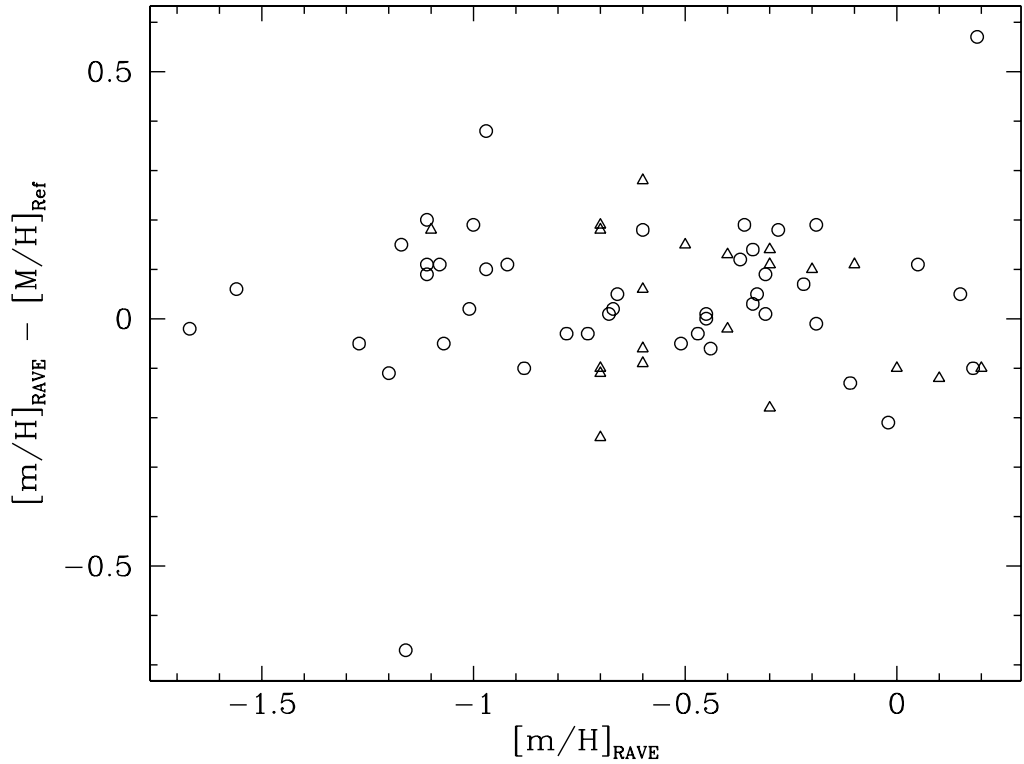}
\caption{Comparison of metallicities derived by the RAVE pipeline with those from 
an alternative $\chi^2$ analysis \citep{munari2005a} used as a reference. 
The circles are the results for the APO stars, while the triangles 
are the ones for the Asiago stars (Table~\ref{t:B2}). Metallicities 
derived by the two methods match very well. The mean values of the metallicity 
difference do not exceed 0.1~dex in the studied range of $-1.7 < \MH < 0.2$. 
}
\label{munarianalysisAsidata}
\end{figure}

The RAVE pipeline derives metallicity as any other parameter, i.e.\ by 
a penalized $\chi^2$ technique finding an optimal match between the 
observed spectrum and the one constructed from a library of pre-computed 
synthetic spectra. The results match even for the metallicity 
if a similar analysis method is used. This is demonstrated by 
Figure~\ref{munarianalysisAsidata}. The results of the analysis using an independent 
$\chi^2$ procedure \citep{munari2005a} yield metallicities which are
entirely consistent with the RAVE pipeline results (mean offset of 
$0.04 \pm 0.02$ dex and a standard deviation of $0.17$ dex). 
RAVE metallicities as derived from the RAVE pipeline are part of a 
self-consistent native RAVE system of stellar parameters which is 
tied to a $\chi^2$ analysis using a library of Kurucz template spectra.
The system is unlikely to change in the future. So metallicities, as derived 
by the pipeline, are also a part of the final data release. 

\begin{figure}[hbtp]
\centering
\includegraphics[angle=270,width=16.5cm]{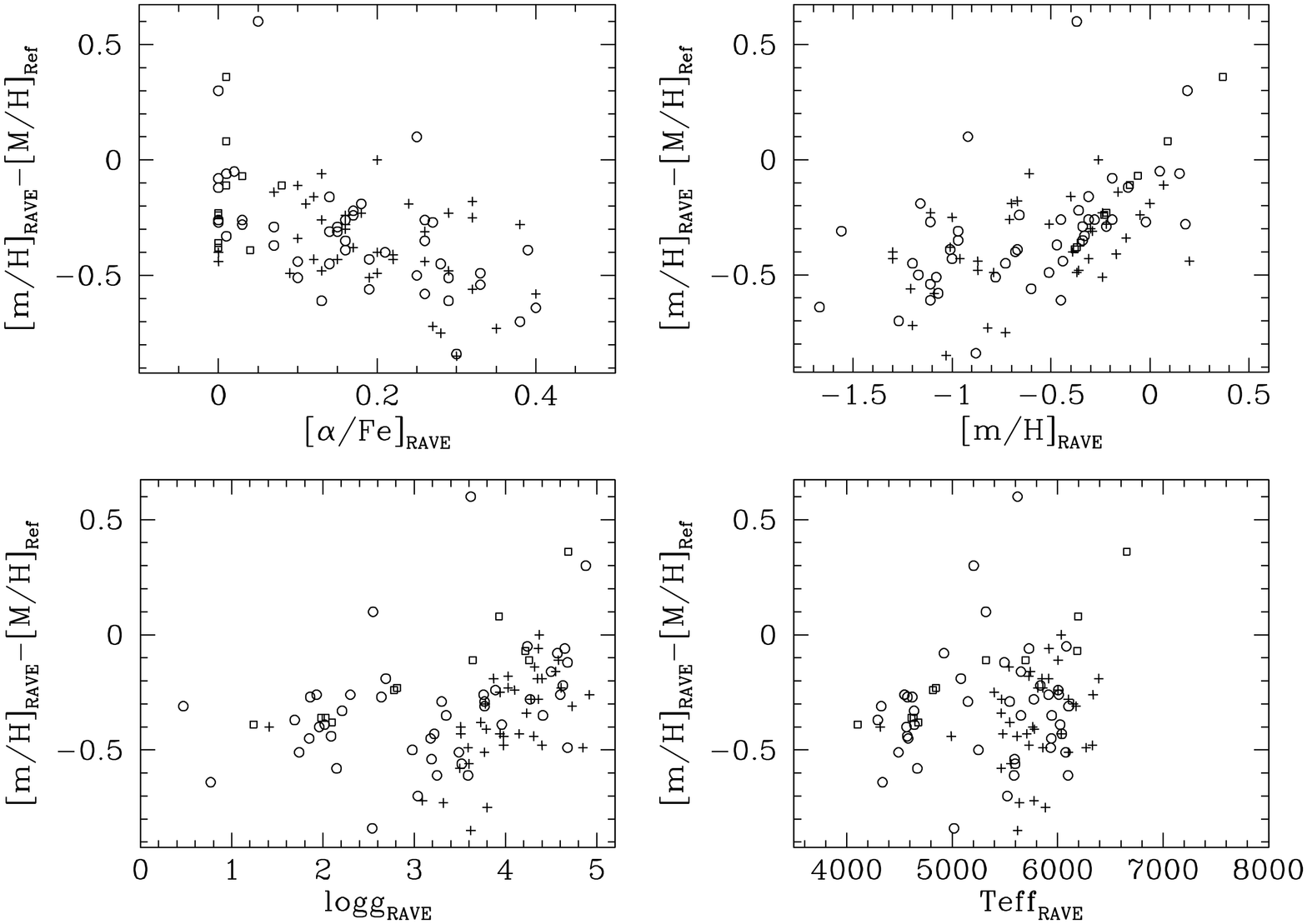}
\caption{Comparison of uncalibrated metallicities 
derived by RAVE to the reference values obtained by the 
measurement of the equivalent widths of absorption lines or 
from the literature. Symbol types distinguish 
between the reference datasets and are the same as in 
Fig.~\ref{f:params_comparison_datasets}.
}
\label{origmetcomparison}
\end{figure}

However other spectral methods, which derive metallicities from the 
strengths of individual spectral lines and not from a $\chi^2$ match of 
synthetic and observed spectra, do not yield results so consistent with 
those of the RAVE pipeline. Figure~\ref{origmetcomparison}
shows some obvious trends:
\begin{itemize}
\item[(i)] the difference between the RAVE and the reference metallicity increases with 
an increased $\alpha$ enhancement, in the sense that RAVE values become too metal 
poor;
\item[(ii)] the difference is also larger at lower metallicities; 
\item[(iii)] the difference is larger for giants than for main sequence stars, though
the variation is much weaker than for $\alpha$ enhancement or metallicity; 
\item[(iv)] the difference does not seem to depend on temperature. 
\end{itemize}
The aim of this section is to provide a calibration relation that transforms the 
uncalibrated metallicities, derived by the $\chi^2$ method, to the calibrated ones, 
which are in line with the metallicity system of the above mentioned datasets. 
The trends can be represented with a linear relationship, there is no indication 
of quadratic terms. So we assume that the calibrated metallicity 
$\MH$ is given by the relation
\begin{equation}
\MH = c_0 \ \mh + c_1 \ \alphaenh + 
c_2 \ \logg + c_3 \ \teff + c_4
\label{formmetallicityrelation}
\end{equation}
where all parameters on the right refer to the values derived by the RAVE 
pipeline (Section \ref{estimationofstellarparameters}) and $c_i$ are constants. 
Figure \ref{origmetcomparison} contains a few outliers, so there 
is a danger that the fit is driven by these points and not by general trends. 
The fit is therefore performed twice and after the first fit we reject 
5\%\ of the most deviating points. Such a clipping does not decrease 
the number of calibration points significantly, still it effectively avoids 
outliers.

It is not obvious whether all parameters in equation \ref{formmetallicityrelation} 
need to be used. So we tested a range of solutions, using between 1 and 5 free 
parameters. It turns out that the main parameters are metallicity, 
$\alpha$ enhancement, and gravity, while for the temperature parameter 
($c_3$) improvement of the goodness of fit is not significant. Also, the 
calibrating datasets cover a limited range in temperature, so this 
parameter is not sampled over its whole physical span.
So we decided not to use temperature for the calibration of the metallicity. 
The final form of the calibration relation is 
\begin{equation}
\MH = 0.938 \ \mh + 0.767 \ \alphaenh 
-0.064 \ \logg + 0.404
\label{calibrationrelation}
\end{equation}
where $\MH$ and $\mh$ denote the calibrated and the uncalibrated metallicities, 
respectively. This convention shall be used throughout the paper. Calibration 
nicely removes the trends mentioned before. Note that the gravity term 
nearly cancels the constant offset for main sequence stars. Its inclusion 
in the relation \ref{calibrationrelation} is further justified by the fact 
that larger discrepancies in metallicity are constrained to lower gravities. 

Inclusion of $\alpha$ enhancement ($\alphaenh$) in the calibration relation 
may seem a bit problematic. Its value is not known a priori, and we 
said in Sec.~\ref{methodvalidation} that it cannot be accurately recovered 
by the RAVE pipeline (see the upper right panel of Fig.~\ref{f:reconserror}). 
A typical recovery error of up to 0.15~dex makes $\alphaenh$ values 
derived by RAVE hardly useful to decide if a certain star has an enhanced 
abundance of elements produced by capture of $\alpha$ particles or not. 
The reason is that the whole range of this parameter amounts to only 0.4~dex, 
i.e.\ not much larger than the recovery error. On the other hand the 
$\alphaenh$ values derived by RAVE are not random, so they statistically 
improve the accuracy of derived metallicity. A factor of 0.767 implies that 
they increase it by up to 0.3~dex in extremely $\alpha$ enhanced stars. So, 
even though an accurate value of $\alphaenh$ cannot be derived by RAVE, 
we know that its value is changing from star to star. In fact the 
enhancement of $\alpha$ elements is the first improvement on the abundance 
modelling of stars which reaches past the uniform scaling of 
Solar abundances. RAVE stars are expected to show much of a variation in 
this parameter, as we are covering a wide range of stars from local 
dwarfs to the rather distant 
supergiants well above the Galactic plane. This is also the reason we 
included variation of $\alpha$ enhancement in the method to determine 
stellar parameters. If the value of $\alphaenh$ were held fixed, or if 
it were calculated by some arbitrary relation, the resulting metallicity 
would be biased, with values shifted by up to 0.3~dex. We try to avoid 
such biases, so $\alphaenh$ is part of the spectral processing, even 
though it cannot be accurately recovered. 

The need for a metallicity calibration can be partly also due to our 
choice of the wavelength range. The largest contributors of strong 
absorption lines in RAVE spectra (for stars dominating
the observed stellar population) are Ca~II, Si~I, Mg~I, Ti~I, and Fe~I.
All but the last one are produced by the capture of $\alpha$ particles.
For the spectral type K0~III
we have 54 prominent spectral lines of 3 $\alpha$-elements (Si~I, Mg~I, 
and Ti~I) and 60 Fe~I lines of similar strength. So $\alpha$-elements 
produce a similar number of lines as iron, not counting 
the {\it very} strong lines of $\alpha$-element Ca~II which actually 
dominate any $\chi^2$ fit. So, when the RAVE pipeline tries to match the 
metallic content, the fits pointing to an enhanced $\alpha$ abundance or
an increased metallicity are similar. As a result the pipeline may 
split the effect of metallicity in two parts, in the sense that it 
partly modifies the metallicity and partly adjusts the $\alpha$ enhancement. 
This may explain the large correlation between the $\alpha$ enhancement and 
metallicity, reflected in a large value of the coefficient $c_1$ in the 
calibration relation (eq.~\ref{calibrationrelation}). The ambiguity 
could be broken only by a higher S/N spectra covering a wider spectral 
range. This is also the reason why analysis methods involving 
equivalent widths  of individual lines, could not be used on a vast 
majority of RAVE spectra. A $\chi^2$ method described in 
Sec.~\ref{estimationofstellarparameters} was chosen 
because it uses the whole spectrum and so makes the best use of the 
available information. 

\begin{figure}[hbtp]
\centering
\includegraphics[angle=270,width=16.5cm]{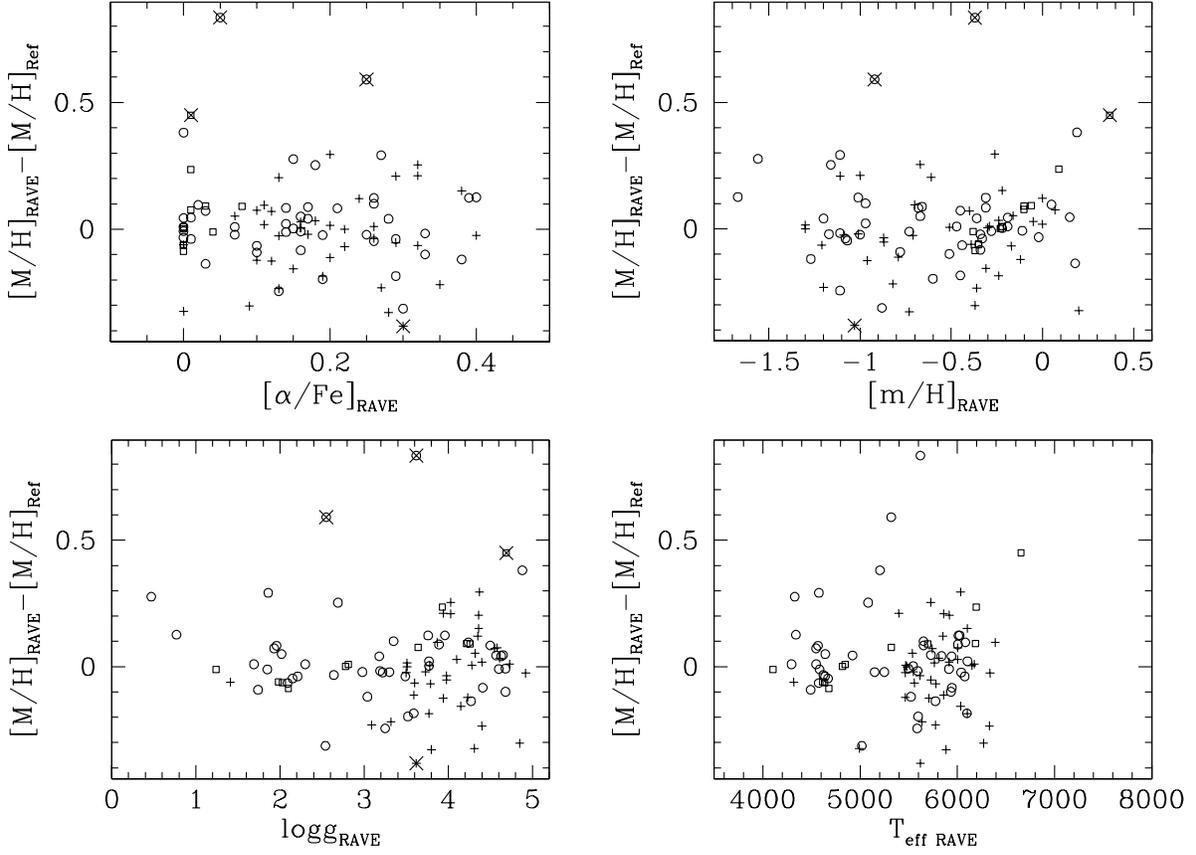}
\caption{Comparison of calibrated metallicities 
derived by RAVE to the reference values. Symbol types are as in 
Figure \ref{origmetcomparison}. Points rejected during iterative calculation 
of the metallicity calibration are crossed out. 
}
\label{corrmetcomparison}
\end{figure}

Figure~\ref{corrmetcomparison} 
shows the situation after application of the calibration 
relation (\ref{calibrationrelation}). All trends and offsets in the 
metallicity values have disappeared and the scatter between the derived 
and the reference metallicity is reduced from 0.37 to 0.18 dex 
(Table \ref{t:zeropointparameteroffsets}).

\begin{figure}[hbtp]
\centering
\includegraphics[angle=0,width=14.0cm]{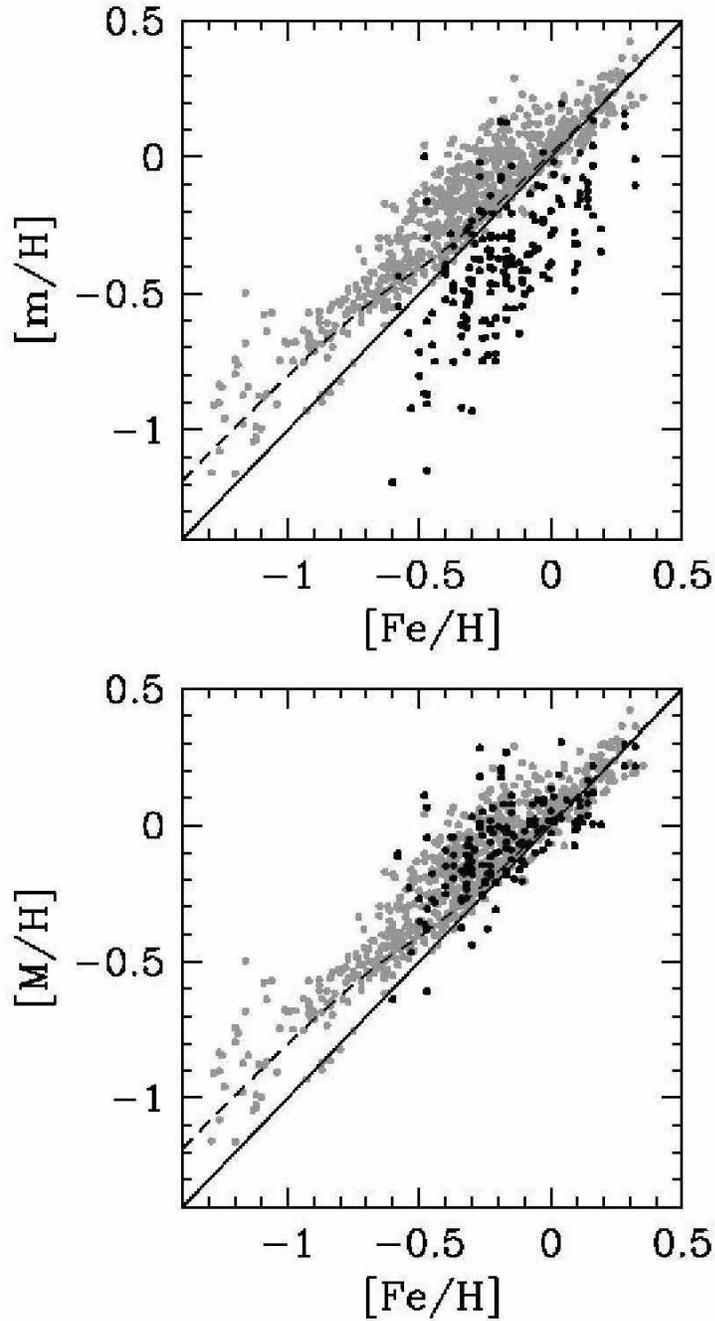}
\caption{Relation between iron abundance and metallicity. The grey points 
mark the positions of all stars in the \citet{soubiran2005} catalog. The 
black ones are RAVE observations of stars from the GCS survey, with 
the uncalibrated values of metallicity in the top graph and the calibrated 
ones in the bottom one. The solid line traces the 1:1 relation, while 
the dashed one is the mean relation between the iron abundance and the metallicity
derived from the APO observations (eq.~\ref{eq:MHfromFeH}). 
}
\label{genevasoubiran}
\end{figure}

We used Soubiran stars, APO observations and M67 members to derive 
the calibration relation. The GCS stars can be used to 
check what we obtained. The GCS does not report metallicity ($\MH$) but only 
iron abundance ([Fe/H]). As mentioned before the two are not identical. A 
substantial scatter in the metallicity vs. iron abundance relation (as demonstrated 
in Fig.~\ref{genevasoubiran} for the Soubiran stars) prevents us from deriving 
a unique iron abundance to metallicity relation in the absence of additional 
information, as is the case with the GC survey. In Figure~\ref{genevasoubiran}
we therefore plot RAVE metallicity vs.\ iron abundance from the GCS catalog. 
The uncalibrated RAVE metallicities (top panel) make the Soubiran and Geneva
Copenhagen surveys 
occupy different regions of the metallicity/iron abundance diagram. But 
the calibrated RAVE metallicities (bottom panel) provide an 
excellent match. As said before the GC survey stars were not used in 
derivation of the calibration relation. The match is therefore a further 
evidence that the relation (\ref{calibrationrelation}) can be trusted.

\begin{figure}[hbtp]
\centering
\includegraphics[width=12.7cm,angle=0]{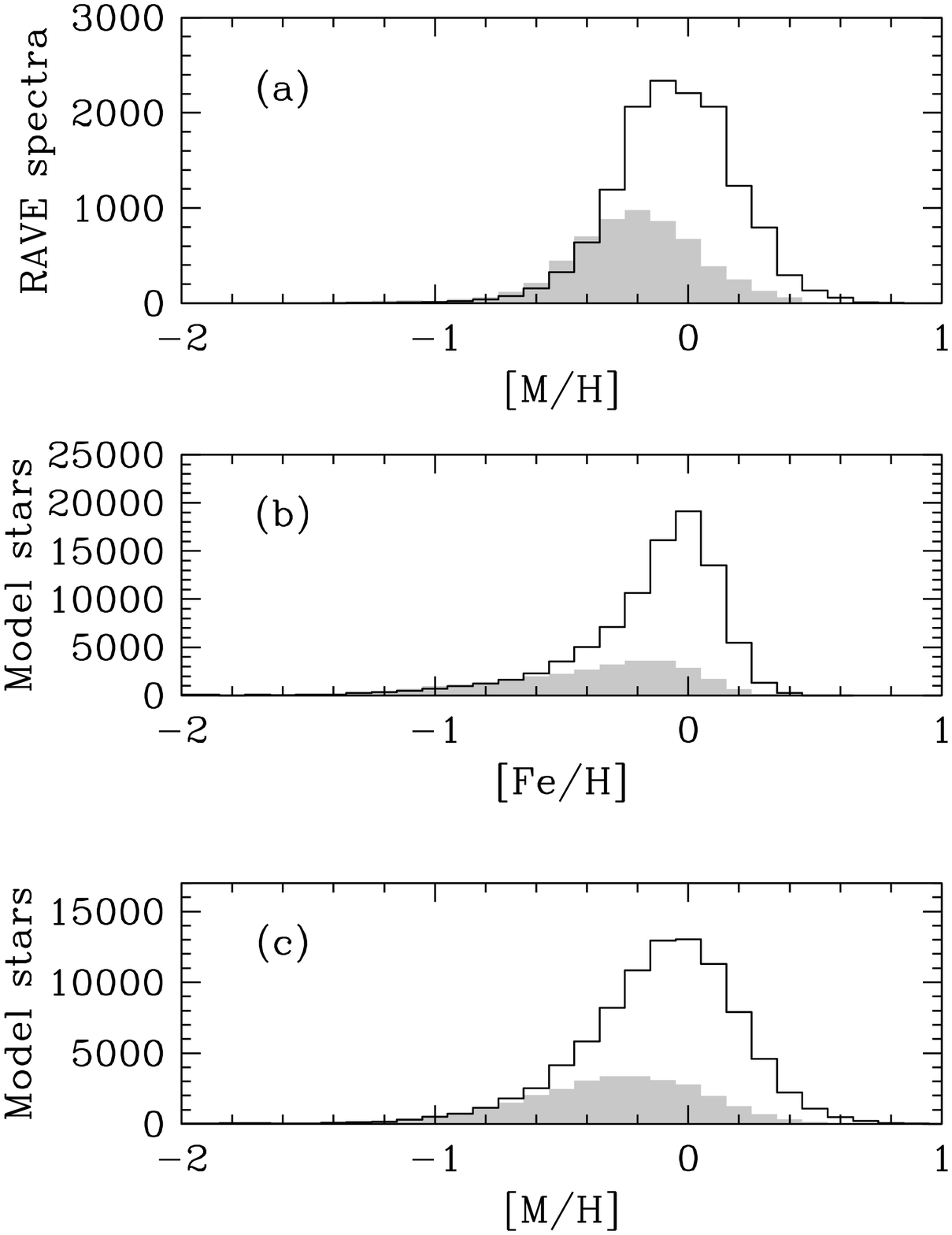}
\caption{
Comparison of the metallicity of the observed targets to the iron abundances 
from the Besan\c{c}on model. The clear and the shaded histograms mark high gravity 
($\logg > 3.0$) and low gravity ($\logg \le 3.0$) objects. Panel (a) plots 
all RAVE spectra with $|b| \ge 25^\mathrm{o}$. Panel (b) is a distribution 
of stars drawn at random from the Besan\c{c}on model. The stars are more than 
$25^\mathrm{o}$ from the plane 
and have the same distribution of $I$ apparent magnitudes as $I_\mathrm{DENIS}$
magnitudes in RAVE.  Panel (c) is a histogram 
from the same Besan\c{c}on model, using the iron abundance to metallicity relation 
from eq.~\ref{eq:MHfromFeH} and convolved with typical RAVE observational errors. 
}
\label{besanconcomparison}
\end{figure}

The calibrated metallicity can be checked also against predictions of 
semi-empirical models. Figure \ref{besanconcomparison}.a plots the distribution 
of the calibrated metallicity determined from RAVE spectra, while 
\ref{besanconcomparison}.b is an empirical prediction of the distribution 
of iron abundance. The latter was calculated using the 
Besan\c{c}on Galactic model \citep{besancon03} with the apparent 
$I_\mathrm{DENIS}$ magnitude distribution of RAVE stars and a random sample 
of objects more than $25^\mathrm{o}$ from the Galactic plane, except for 
the inaccessible region $60^\mathrm{o} < l < 210^\mathrm{o}$. 
The observed distribution in metallicity is more symmetric than 
its theoretical iron--abundance counterpart. The reason lies in the differences 
of the two quantities. Figure~\ref{genevasoubiran} shows that the 
metallicity is usually higher than iron abundance due to an enhanced presence 
of $\alpha$ elements. APO observations of RAVE stars (Table~\ref{t:B1}) yield
both iron abundance and metallicity, so they allow us to fit a statistical 
relation between metallicity and iron abundance
\begin{equation}
\MH = \feh + 0.11 [1 \pm (1 - e^{-3.6 |\feh+0.55|})]
\label{eq:MHfromFeH}
\end{equation}
where the plus sign applies for $\feh < -0.55$ and the minus sign otherwise. 
The relation is plotted with a dashed line in Fig.~\ref{genevasoubiran}.
It makes the metallicity 0.22~dex larger than the iron abundance for very metal poor 
stars with $\alphaenh = 0.3$, while the difference vanishes when approaching 
the Solar metallicity. The relation is very similar to the one of
\citet{salaris93}. If this relation, together with metallicity errors 
typical for the RAVE observations (equation~\ref{eq:sigma} and figure~\ref{parerrors}), 
is used, the resulting histograms (Fig.~\ref{besanconcomparison}.c) are 
very similar to the observed ones (Fig.~\ref{besanconcomparison}.a). Peaks of 
the histograms match to within 0.06~dex, while the width is $\sim 25$~\%\ 
larger in the model compared to the observations. A somewhat 
larger width of the model histograms suggests that the error estimates 
for the RAVE metallicity are conservative. Note 
however that the Besan\c{c}on model predicts a smaller fraction of low gravity 
stars ($\logg \le 3.0$) than observed. 

The description of stellar chemical composition by metallicity 
and $\alpha$-enhancement values is a simplification. Generally, the individual 
stellar elemental abundances (including those of the alpha elements) do not scale linearly 
or in a constant ratio with those of the Sun, and spectral lines of some elements are 
not present in the RAVE wavelength range. Individual element abundances frequently 
scatter by 0.2 or 0.3~dex if compared to the iron abundance \citep{soubiran2005}. This fact 
of nature is also the cause of a large scatter of metallicity vs.\ iron abundance in the 
Soubiran sample, depicted by grey points in Figure~\ref{genevasoubiran}. The metallicity 
change of 0.2--0.3~dex, as introduced by the calibration relation, is therefore comparable 
to the intrinsic scatter of individual element abundances in stars. So it would be very 
difficult to provide a detailed physical explanation for the calibration relation between 
the metallicities derived by equivalent width or photometry methods and those obtained by 
a $\chi^2$ analysis. Equation~\ref{eq:MHfromFeH} therefore reflects only approximate general 
trends. Nevertheless it allows us to check that the distribution of the calibrated 
metallicities derived by RAVE is consistent with the predictions of the Besan\c{c}on 
Galactic model.

\subsubsection{Method for stellar parameter error estimation}

Errors associated with a given stellar parameter depend on the S/N ratio 
of the spectrum and on the spectral properties of the star. We discuss 
them in turn. 
The calibration data have very different values of S/N, in general higher 
than typical RAVE survey data. The average S/N ratio for the survey stars 
for which we publish values of stellar parameters is 41. So we choose 
S/N~=~40 as the reference value of S/N. The error estimate 
$\sigma_{40}$ below therefore refers to a star with S/N~=~40. Extensive 
Monte Carlo simulations show that the error $\sigma$ for a stellar 
parameter has the following scaling with the S/N of the observed spectrum:
\begin{equation}
\sigma =  r^k \, \sigma_{40} 
\label{eq:sigma}
\end{equation}
where 
\begin{equation}
r = \left\{ \begin{array}{rl}
(S/N) / 40,  & \mathrm{if}\,\,\, S/N < 80;\\
80 / 40,     & \mathrm{otherwise,}
\end{array} \right.
\label{requation}
\end{equation}
and the coefficient $k$ has the value of $-0.848$ for temperature, 
$-0.733$ for gravity and $-0.703$ for metallicity. The 
simulations used 63 high S/N spectra observed by RAVE for which also 
high resolution echelle spectroscopy has been obtained in Asiago or 
at the APO. We assumed that the analysis of echelle spectroscopy yields 
the true values of the 
parameters for these stars and studied how the values derived by 
the RAVE pipeline would worsen if additional Gaussian noise was added 
to the RAVE spectra. We found that the offsets in mean values of stellar 
parameters appear only at $S/N < 6$ (an offset  
in temperature at S/N=6 is 100~K) and disappear at higher S/N ratios. 
Gaussian noise is not the only source of the problems with weak signal 
spectra. Systematic effects due to scattered light, fiber crosstalk, 
and incomplete removal of flatfield interference patterns are 
preventing a reliable parameter determination in a large fraction of 
spectra with $S/N < 13$.  So we decided to publish radial velocities 
down to $S/N = 6$, while stellar parameter values are published only 
for spectra with $S/N > 13$. The latter decision influences 
$\sim 13$\%\ of RAVE spectra which have $6 < S/N < 13$. 

Simulations also show that the errors on the 
parameters do not continue to improve for stars 
with $S/N>80$, because systematic errors tend to dominate over statistical 
noise in such low--noise cases. So we flatten out the error decrease for 
$S/N > 80$ in equation (\ref{requation}). 

The choice of the reference signal to noise ratio of 40 means that the errors
discussed below should be about twice larger for the noisiest 
spectra with published parameters, and about twice smaller for 
spectra with the largest ratio of signal to noise.

The calibration datasets (Table~\ref{t:externalparametersets})
contain only stars hotter than 4000~K and cooler than 7500~K. 
The majority of these stars are on or close to the main sequence with 
a metallicity similar to the Solar value. Many of the RAVE program stars 
are of this type, but not all. For example one cannot 
judge the errors of hot stars or very metal poor stars from these datasets. 
So we need to use simulations to estimate the value of $\sigma_{40}$ in 
Eq.~\ref{eq:sigma}, i.e.\ how the error depends on the type of star that is 
observed. Relative errors are estimated from a theoretical grid of Kurucz 
models, but the observed calibration datasets are used for the scaling 
of the relative to absolute error values and for verifying the results.

We start with a theoretical normalized spectrum from the pre-computed 
Kurucz grid and investigate the increase of the root mean square difference (RMS) 
when we compare it with grid-point spectra 
in its vicinity in the 5-dimensional space of $\teff$, $\logg$, 
$\MH$, $\alphaenh$, and $V_\mathrm{rot}$. If we denote the values of five 
parameters for the initial spectrum as $P_i$ ($i=1,..,5$), and 
if $p_i$ ($i=1,..,5$) denote their values at a grid point in its vicinity, 
the estimate $\sigma_j$ of an error of parameter $j$ for the 
initial spectrum can be obtained from the minimum of  
$\mathrm{RMS}_j = min\{\mathrm{RMS} (p_1,p_2,..,p_5), p_j \neq P_j \}$. We 
assume that an increase of RMS has a similar effect on the parameter 
estimation as an increase of a noise level. So $\mathrm{RMS}_j$ is 
inversely proportional to the S/N of the normalized spectrum, 
but the dependence of the error $\sigma_j$ of the parameter $j$ on the 
S/N ratio is given by the value of the coefficient $k$ in the equation 
\ref{eq:sigma}.  The only remaining factor is the proportionality 
constant. It is derived by the assumption that 68.2\%\ of all calibration 
spectra should have the value of the parameter $j$ determined by the RAVE 
pipeline within $\pm \sigma_j$ of the reference value. 

The scheme allows us to estimate errors in all corners of the parameter space 
covered by Kurucz models, i.e.\ even in parts where we lack any calibration spectra. 
Calibration spectra are used exclusively for scaling of the $\sigma_{40}$ value 
of a given stellar parameter in eq.~\ref{eq:sigma}. This scaling was done 
assuming that $\sim 2/3$ of RAVE calibration objects should have a given parameter 
within one standard deviation of the true value obtained from high resolution observations. 
So we can check if the relative number of calibration objects within e.g.\ 0.5 or 2 
standard deviations conforms to the normal distribution. A positive answer would 
support the results. Next we discuss the accuracy of each stellar parameter in turn.

\subsubsection{Temperature accuracy}

\begin{figure}[hbtp]
\centering
\includegraphics[angle=00,width=14.0cm]{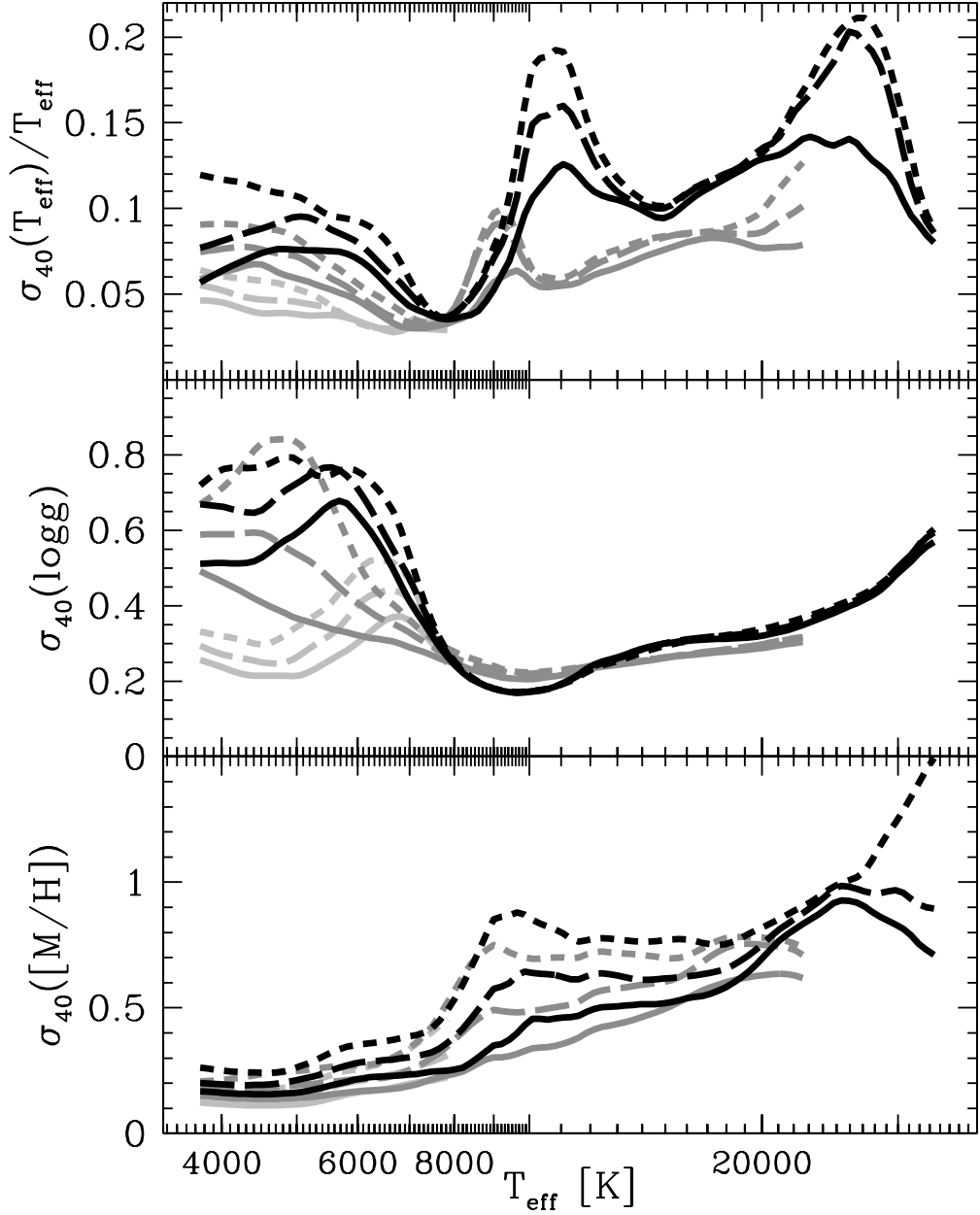}
\caption{Errors for temperature, gravity and metallicity as a function 
of temperature. The nine curves are errors 
for three values of gravity and three values of metallicity: 
black curves are errors for main sequence stars ($\logg = 4.5$),
grey for subgiants ($\logg = 3.0$), and light-grey 
curves for supergiants ($\logg = 1.0$). Solid lines 
are for Solar metallicity, long dashed ones are for $\MH = -0.5$ and 
short dashed  ones for $\MH = -1.0$. All errors apply for a star 
with $S/N = 40$, the ones for other noise levels follow from Eq.~\ref{eq:sigma}. 
}
\label{parerrors}
\end{figure}

The top panel of Figure \ref{parerrors} plots the standard deviation of temperature 
as a function of temperature for stars with S/N = 40. The value of the standard 
deviation is divided by temperature. So an ordinate value of 0.05 at 6000~K 
denotes a standard deviation of 300~K. The nine curves are errors 
for three values of gravity and three values of metallicity. Light grey 
curves are for supergiants ($\logg = 1.0$), grey ones for subgiants 
($\logg = 3.0$), and black ones for MS stars ($\logg = 4.5$). Solid lines 
are for Solar metallicity, while long dashed ones are for $\MH = -0.5$ and 
short dashed ones for $\MH = -1.0$. 

Typical errors for stars cooler than 9000~K are around 400~K. The errors are 
the smallest for supergiants. Their atmospheres are the most transparent ones,
so that a wealth of spectral lines arising at different optical depths can 
improve the temperature accuracy. Understandably the errors for metal poor 
stars are larger than for their Solar counterparts. The errors get 
considerably worse for hot stars ($T>9000$~K) where most metal lines are 
missing and the spectrum is largely dominated by hydrogen lines.
All these trends can be seen from Figure~\ref{f06} where wavelength 
intervals affected by temperature change are marked by red lines.

These error estimates are rather conservative because we assumed that 
any discrepancy arises only because of RAVE errors, i.e.\ that the calibration 
datasets are error free. As mentioned already in the discussion on zero point
offsets (Sec.~\ref{zeropointoffsets}) this is not always the case. In particular, 
the errors in temperature would be 20\%\ smaller if we did not use 
the GCS stars in error estimation. 

\begin{figure}[hbtp]
\centering
\includegraphics[angle=270,width=9.0cm]{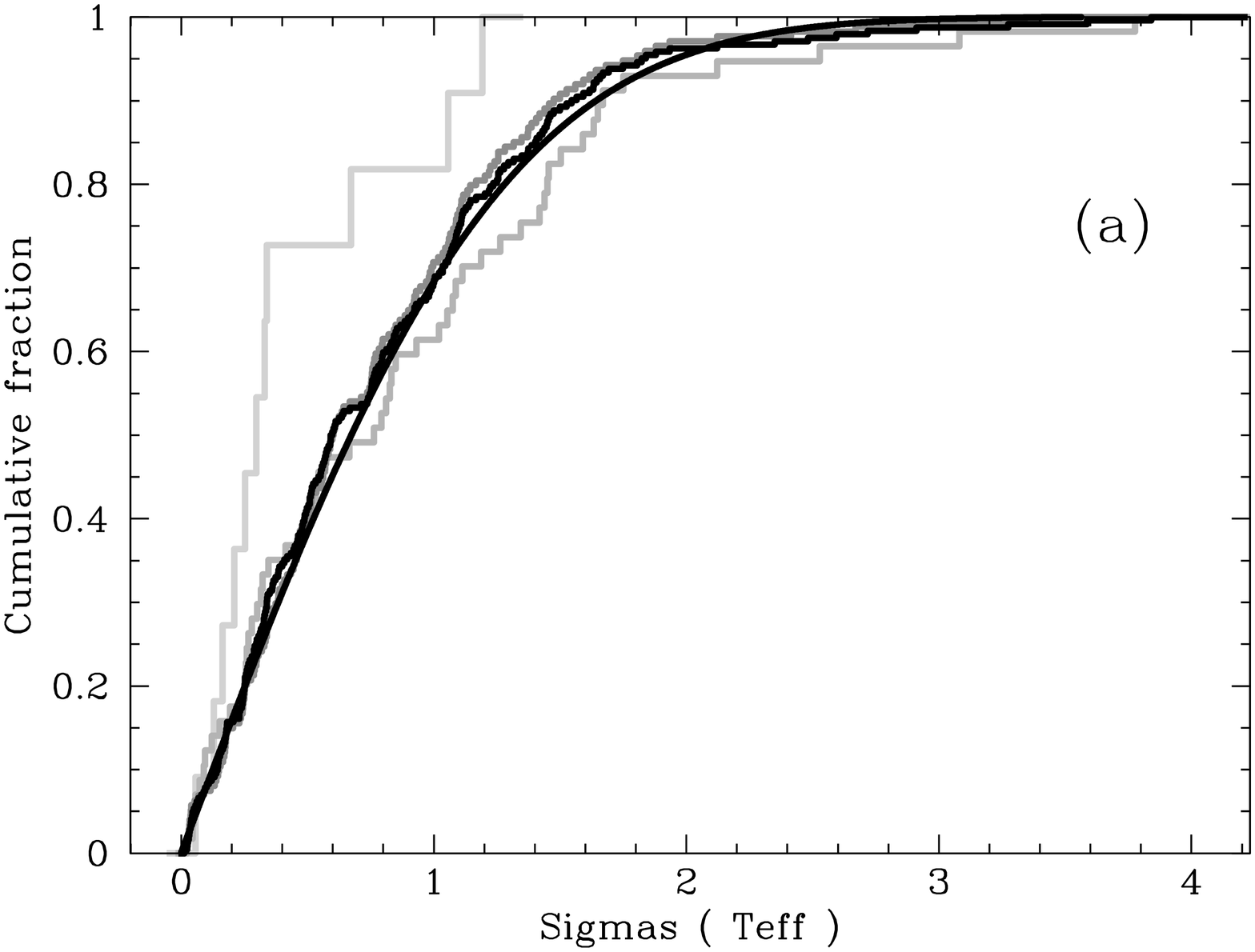}\\
\includegraphics[angle=270,width=9.0cm]{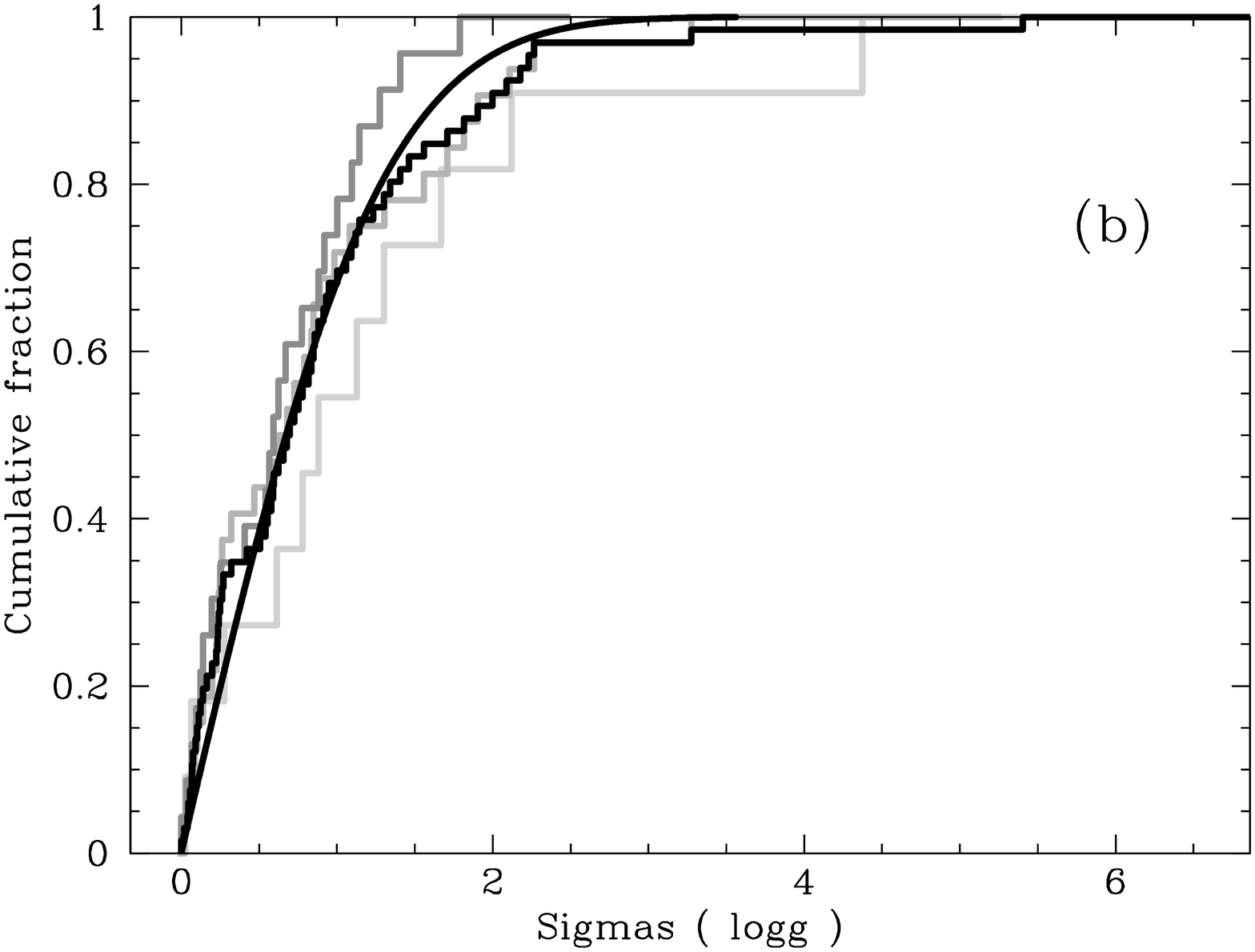}\\
\includegraphics[angle=270,width=9.0cm]{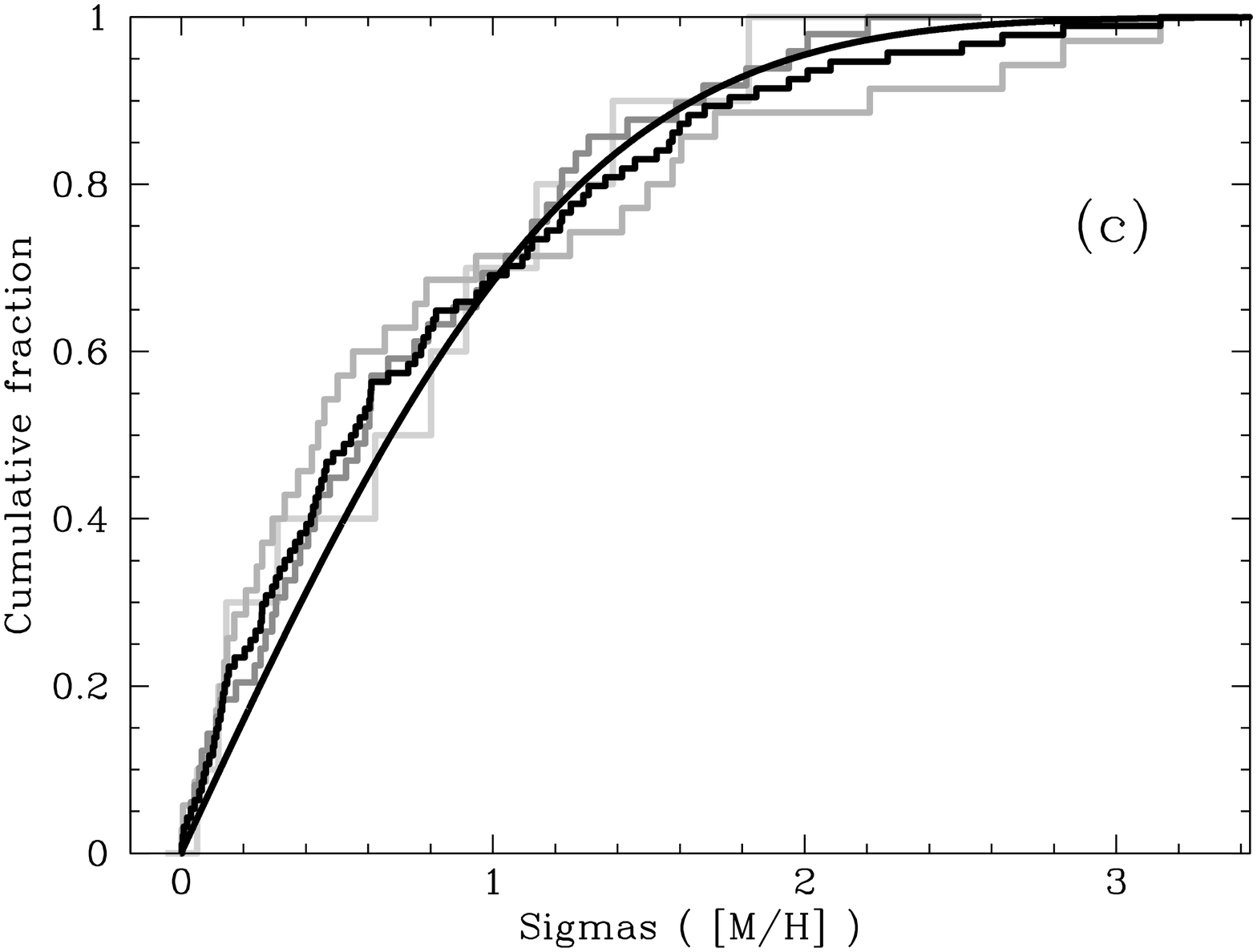}\\
\caption{Cumulative fraction of reference spectra with the difference between the 
RAVE and the reference value within the given number of standard deviations:
(a) for temperature, (b) for gravity, (c) for metallicity. 
The black histogram is for all calibration stars, while the 
grey ones denote different gravity ranges: dark grey marks $\logg \ge 3.75$,
middle grey is for $2.0 \le \logg < 3.75$, and light grey corresponds to  
$\logg < 2.0$. The smooth curve plots an ideal normal distribution. 
}
\label{errorscumul}
\end{figure}

Figure \ref{errorscumul}a plots the cumulative distribution of
errors for the calibration stars used to derive the temperature 
errors. Line types and greyscale tones are the same as in Figure \ref{parerrors}.
We see that 68\%\ of our stars have their error within one 
sigma, a condition we used for scaling. But also the distribution of 
stars along the error curve closely follows the normal distribution. This 
supports the error estimates given in Fig.~\ref{parerrors}.

\subsubsection{Gravity accuracy}

The middle panel of Figure \ref{parerrors} plots errors in gravity as a function of temperature. 
Strong wings of Hydrogen lines which are sensitive to gravity allow 
small gravity errors in hot stars (see blue marks in Figure~\ref{f06}
which mark gravity--sensitive regions). On the other hand rather narrow 
metallic lines in the RAVE wavelength range, including the ones of Ca~II, do not allow an accurate 
determination of gravity in cool stars. The gravity error in cool stars has a strong 
gravity dependence: in dwarfs it is large, but the rather transparent atmospheres 
of giant stars still allow for a reasonably accurate gravity determination. In any 
case the errors in gravity do not exceed 0.8~dex, which still allows 
determination of a luminosity class. 

Figure \ref{errorscumul}b is similar to Figure \ref{errorscumul}a
in the sense that it plots the errors of calibration stars. Again we have 
68\%\ of the stars with errors smaller than the standard deviation, a condition 
used to calibrate the errors in Figure \ref{parerrors}. Departures 
from the normal distribution of errors can be explained by a rather small 
number of spectra used to determine the gravity errors.

\subsubsection{Metallicity accuracy}

The bottom panel of Figure \ref{parerrors} plots standard deviations of 
the calibrated metallicity
($\MH$). The typical error for stars cooler than 7000~K is 0.2~dex. The 
error for hotter stars is understandably much larger, as these stars lack 
most of the metallic lines in their spectra (lack of green marks in 
hot spectra in Figure~\ref{f06}). Figure~\ref{errorscumul}c 
shows that the distribution of errors is very close to the normal one.

\subsubsection{Errors on other parameters}

\label{errorsotherparams}
The rotational velocity will be a topic of a separate paper which will discuss 
fast rotating stars, so we do not estimate its error here. 
The $\alpha$ enhancement value is part of this data release, but given the 
fact that the Kurucz grid covers only two values (0.0 and 0.4) it is very 
hard to estimate its error. We note that our metallicity has a typical error of
0.2 dex, so it seems likely that the statistical error on $\alpha$ enhancement 
is larger. Note that this is comparable to the value of $\alphaenh = 0.22$ 
reached in typical metal poor stars. So although the $\alphaenh$ parameter 
is useful to improve the accuracy of derived metallicities 
(eq.~\ref{calibrationrelation}) its value is not accurate enough to be 
trusted for individual stars.

\subsection{Detection of peculiar and problematic spectra}
 
Errors on temperature, gravity and metallicity have been presented for 
a range of normal stars. We estimate that these errors are statistically 
accurate to $\sim 30$\%. Errors for other normal stars could be derived 
by linear interpolation. But not all stars have normal spectra. RAVE observed 
a number of binaries, emission type objects and other peculiar stars, 
while occasionally a spectrum of a normal star is jeopardized by 
systematic errors. So it is vital to identify such objects. 

Radial velocity information is present 
in all spectral lines. Still, a very noisy spectrum, too uncertain a wavelength 
solution or other systematic errors could lead to unreliable results. The
simulations showed that the radial velocity is not systematically affected by 
noise if the S/N ratio is larger than 6. At lower S/N ratios the best 
template identified by our matching method would be systematically 
offset (for 100~K or more in temperature) therefore affecting RV 
accuracy. This effect is not present at higher S/N ratios. So we calculated 
the S/N ratio for each spectrum and visually checked if the calcium 
lines (and at higher S/N ratios also others) show a mutually consistent radial 
velocity. 698 spectra were rejected, mostly because their $S/N<6$, 
and are not part of this data release. 

The measurement of stellar parameters requires a higher S/N ratio. We adopted 
a limit of $S/N = 13$ as the minimum value. Note that this limit is 
still quite conservative as it corresponds to $\sim 8$\%\ error in the 
flux of each pixel. So the published values of stellar parameters are 
statistically correct, but parameters for individual stars with 
$S/N < 20$ should be considered as preliminary. This data release
contains 3411 such relatively noisy spectra.

All spectra new to this release were visually checked. 
The goal is to avoid systematic errors, as well as to identify 
types of objects which are not properly covered by our grid of 
theoretical models. In the latter case large and arbitrary errors
in values of stellar parameters could result. Such objects include double 
lined spectroscopic binary stars, emission type objects and other peculiar 
stars. We do not publish values of stellar parameters for such objects, 
but only the values of their radial velocity which is calculated in the 
same way as for normal stars. So we are consistent with 
the first data release. We also avoid arbitrary decisions in cases of 
undetected or marginally detected binaries. Their published 
radial velocity is somewhere between the instantaneous velocities of 
the two components and does not correspond necessarily to the barycentric one. 
The physical analysis of detected double lined spectroscopic binaries will 
be presented in a separate paper. But \citet{seabroke2008} showed that they 
do not affect statistical kinematic Galactic studies significantly. 

The first data release contained 26,079 spectra for which we published 
radial velocities but no stellar parameters. Also in the data new 
to this data release there are 3,343 spectra without published stellar 
parameters. From these there are 140 emission type spectra, 135 double 
lined binary spectra and 86 spectra of peculiar stars. Other spectra 
without published parameters have the $S/N < 13$ or are affected by 
systematic problems. Table \ref{t:classes} summarizes the results. 
The last column quotes the number of different objects with a given 
classification. Some stars occasionally show normal spectra and we publish 
the values of their stellar parameters, but in other occasions they show some 
kind of peculiarity or systematic problem, so that their parameter values 
are not published. So the first number in the last column is not an exact 
sum of the two numbers below it. 

\begin{deluxetable}{lcccrrrr}
\tablecaption{Number of entries with a given stellar type in this data 
release 
\label{t:classes}}
\tablewidth{0pt}
\tablecolumns{5}
\tablehead{
\colhead{Type} & SpectraFLAG & \multicolumn{2}{c}{Published}&\multicolumn{2}{c}{Number of}
 \\
               &     & RVs & Parameters &Spectra & Stars   \\
}
\startdata
All entries with RVs                         &    &$\surd$&       & 51,829 & 49,327 \\
Entries with normal type spectra             &    &$\surd$&$\surd$& 22,407 & 21,121 \\
Any type without parameters                  &    &$\surd$&       & 29,422 & 28,747 \\ \hline
Emission line                                &e   &$\surd$&       &    140 &    136 \\
Double--lined binaries                       &p   &$\surd$&       &    135 &    132 \\
Peculiar                                     &x   &$\surd$&       &     86 &     75 \\ 
\enddata
\end{deluxetable}

\subsection{Repeated observations}

\label{repeatedobservations}
Most stars are observed by RAVE only once, but some observations 
are repeated for calibration purposes. 1,893 objects in the 
present data release have more than one spectrum. Table \ref{t:classes}
explains that the present release contains 51,829 spectra of 49,327 
different stars. Note that the latter number is smaller than the 
number of stars of individual types. This is a consequence of the 
fact that a spectrum of a star may appear as a double lined 
binary star in one spectrum, and as an entirely normal single star
in another one (taken close to conjunction). So the star would 
be counted as a member of two types. A definite classification of all stars 
in this data release is beyond the scope of this paper. We plan to pursue 
follow-up studies for particular types of objects, like spectroscopic 
binaries, and present them in separate papers.

Repeated observations allow a comparison of the measured properties
of these stars. If we assume that values for a given star do not 
change with time, the scatter can be used to estimate errors on 
radial velocity and the values of the stellar parameters. This 
assumption may not always be true, for example in the case of 
binaries or intrinsically variable stars.  
So we assumed that the sigma of a parameter is the value which comprises
68.2\%\ of the differences between the measured values of a parameter and 
its average value for a given star. This way we minimize the effect of large 
deviations of (rare) variable objects and measure an effective standard 
deviation of a given parameter. 

The data release contains 1,893 objects with 2 or more measurements of 
radial velocity. The dispersion of measurements for a particular object 
is smaller than 1.80~\kms\ in 68.2\%\ of the cases, and smaller than
7.9~\kms\ in 95\%\ of the cases.

For 822 objects we have also 2 or more spectra with published 
stellar parameters. In this case the dispersion of velocities 
is within 1.66~\kms\ (68.2\%\ of objects) and 6.1~\kms\ (95\%\ 
of cases). The corresponding scatter in the  
temperature is 135~K (68.2\%) and 393~K (95\%), for 
gravity 0.2~dex (68.2\%) and 0.5~dex (95\%), and for the 
calibrated metallicity 0.1~dex (68.2\%) and 0.2~dex (95\%).

Spectra of repeated objects share the same distribution of S/N ratio 
as all RAVE stars. Their typical S/N ratio of 40 is smaller than for 
the reference datasets (see Fig.~\ref{f10}), still the above quoted value 
for the dispersion of radial velocities is similar to the errors of the 
reference datasets (Table~\ref{t:rvaccuracy}). Also the  
dispersions of stellar parameter values as derived from the repeated 
observations are smaller than the dispersions for the reference 
datasets (Table~\ref{t:zeropointparameteroffsets}). One expects a higher 
internal consistency of the repeated observations, as these are free from 
zero point errors. But the zero point errors are very small for both 
radial velocity and stellar parameters (Tables \ref{t:rvaccuracy} and 
\ref{t:zeropointparameteroffsets}). Note that our error estimates of 
radial velocity and stellar parameter values are derived assuming that 
the reference values from the external datasets are error free, and this 
may not always be the case.  We conclude that the error estimates on 
radial velocity (Sec.\ \ref{radialvelocityaccuracy}) and stellar 
parameters (Sec.\ \ref{accuracystellarparameters}) are quite conservative.

\section{Second data release}

\label{secondrelease}

\subsection{Global properties}

The second public data release of the RAVE data (RAVE DR2) is accessible 
online. It can be queried or retrieved from the Vizier database at the CDS, 
as well as from the RAVE collaboration website (www.rave--survey.org). 
Table~\ref{t:A1} describes its column entries. The tools to query and extract 
information are described in Paper~I. 

The result of the RAVE survey are radial velocities and values of 
stellar parameters (temperature, gravity and metallicity). Metallicity 
is given twice: as coming from the data reduction pipeline ([m/H])
and after application of calibration equation \ref{calibrationrelation} 
([M/H], see Section \ref{calibratingmetallicity} for details). The latter 
includes also the value of $\alpha$ enhancement. So the catalog includes 
also the estimated values of $\alphaenh$. As explained in 
Section \ref{errorsotherparams} this is provided mainly for calibration 
purposes and is not intended to infer properties of individual objects.

\begin{figure}[hbtp]
\centering
\includegraphics[width=14.7cm,angle=0]{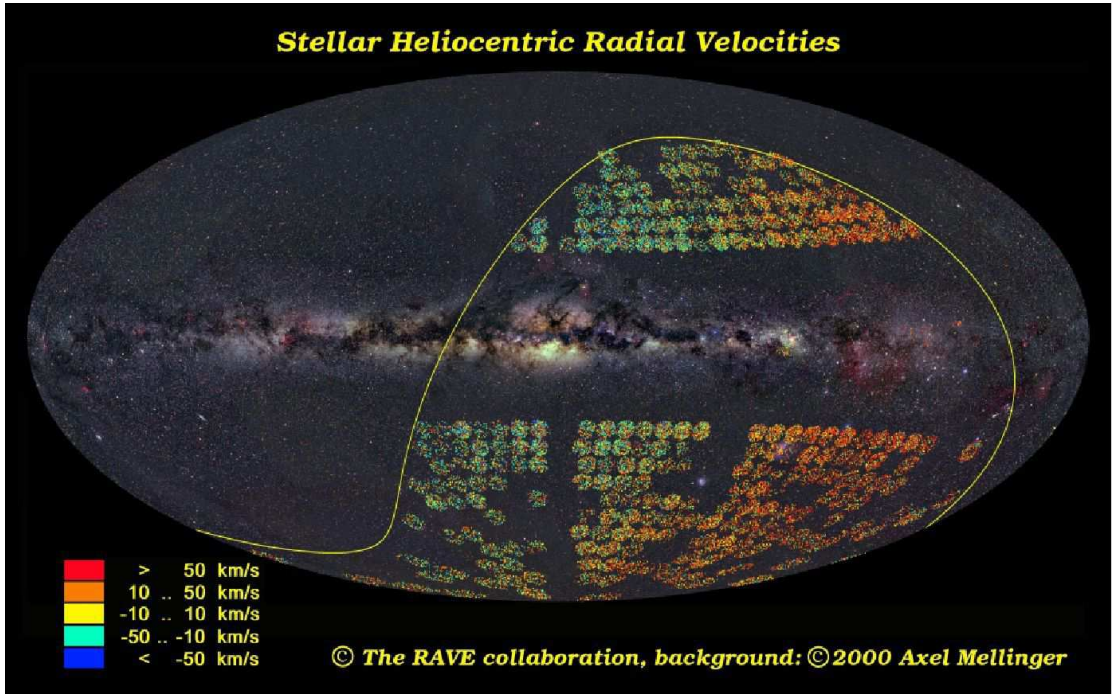}
\caption{Aitoff projection in Galactic coordinates of RAVE 2nd data 
release fields. The yellow line represents the celestial equator and the 
background is from Axel Mellinger's all-sky panorama.  
}
\label{aitoffRVs}
\end{figure}

Figure~\ref{aitoffRVs} plots the general pattern of 
(heliocentric) radial velocities. The dipole distribution is  
due to Solar motion with respect to the Local Standard of Rest. 
Spatial coverage away from the Galactic plane is rather good, with the 
exception of stars at small Galactic longitudes. These areas have already 
been observed and will be part of the next data release. 

\begin{figure}[hbtp]
\centering
\includegraphics[width=6.6cm,angle=270]{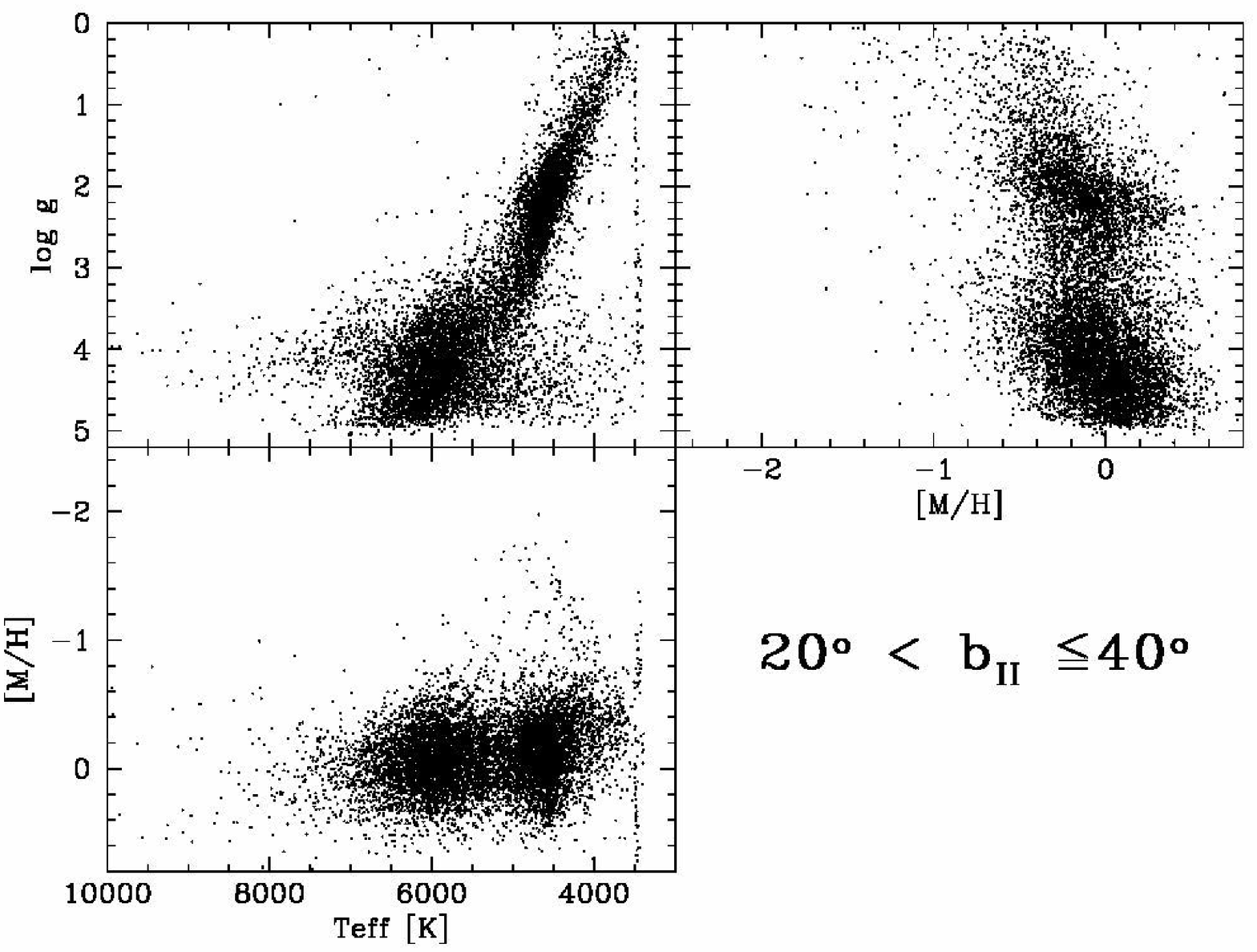}\\
\includegraphics[width=6.6cm,angle=270]{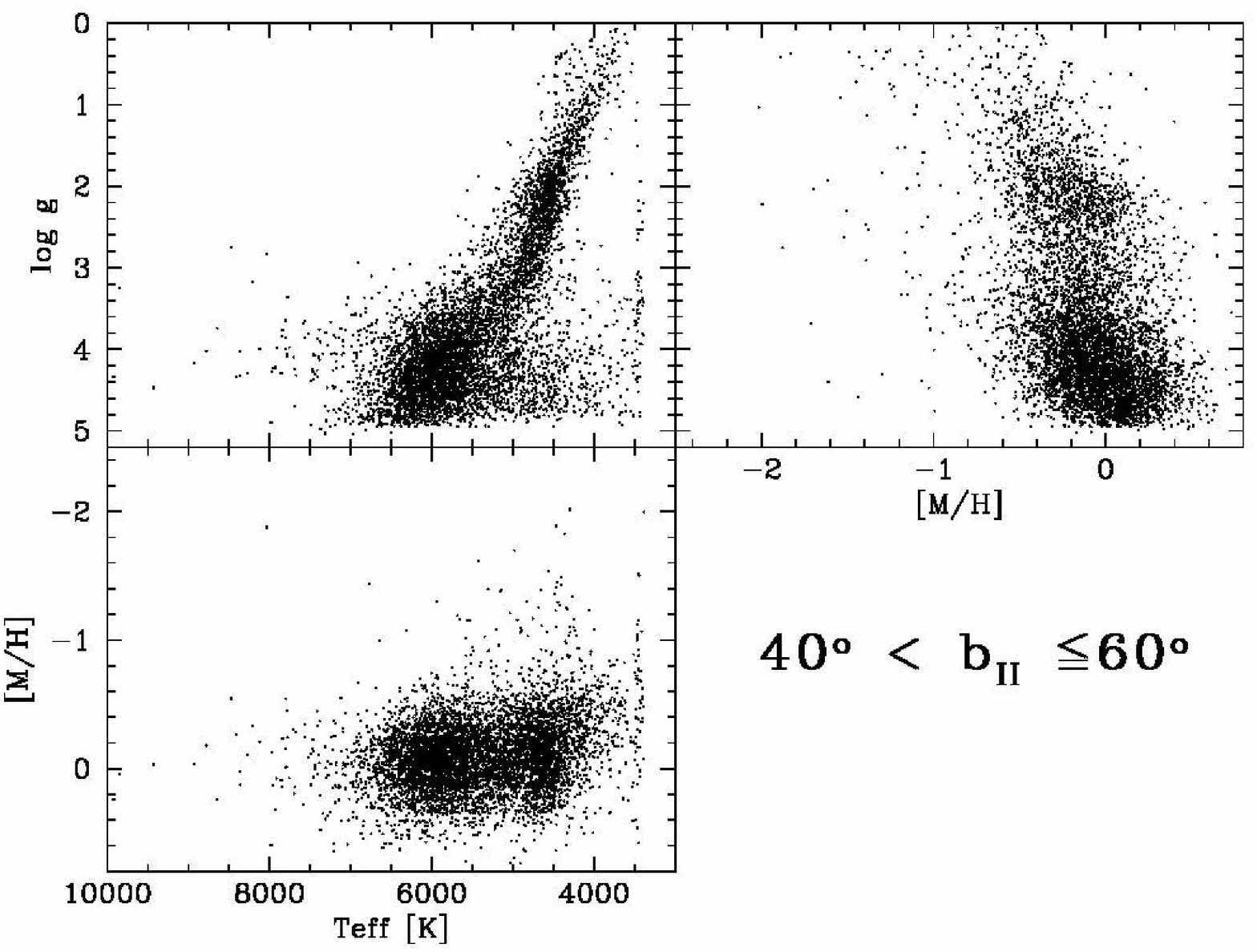}\\
\includegraphics[width=6.6cm,angle=270]{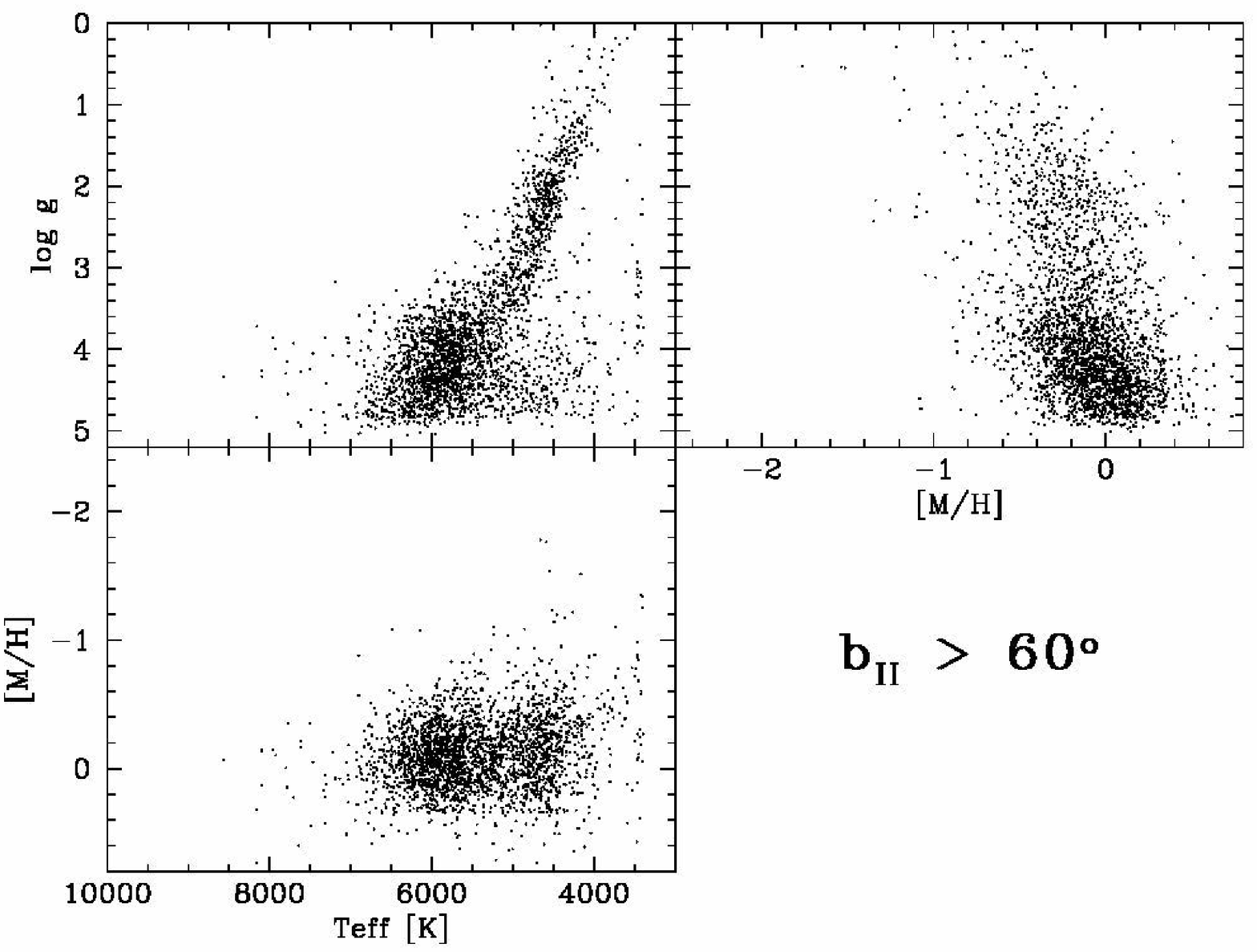}\\
\caption{
Temperature--gravity--metallicity wedge for 3 bands in Galactic 
latitude. Spectra with 
$\teff \sim 3500$~K are at the edge of the grid of spectral templates: 
so their temperatures should be used with caution, usually as an upper 
limit to the real value. 
}
\label{paramscalib}
\end{figure}

\begin{figure}[hbtp]
\centering
\includegraphics[width=14.7cm,angle=0]{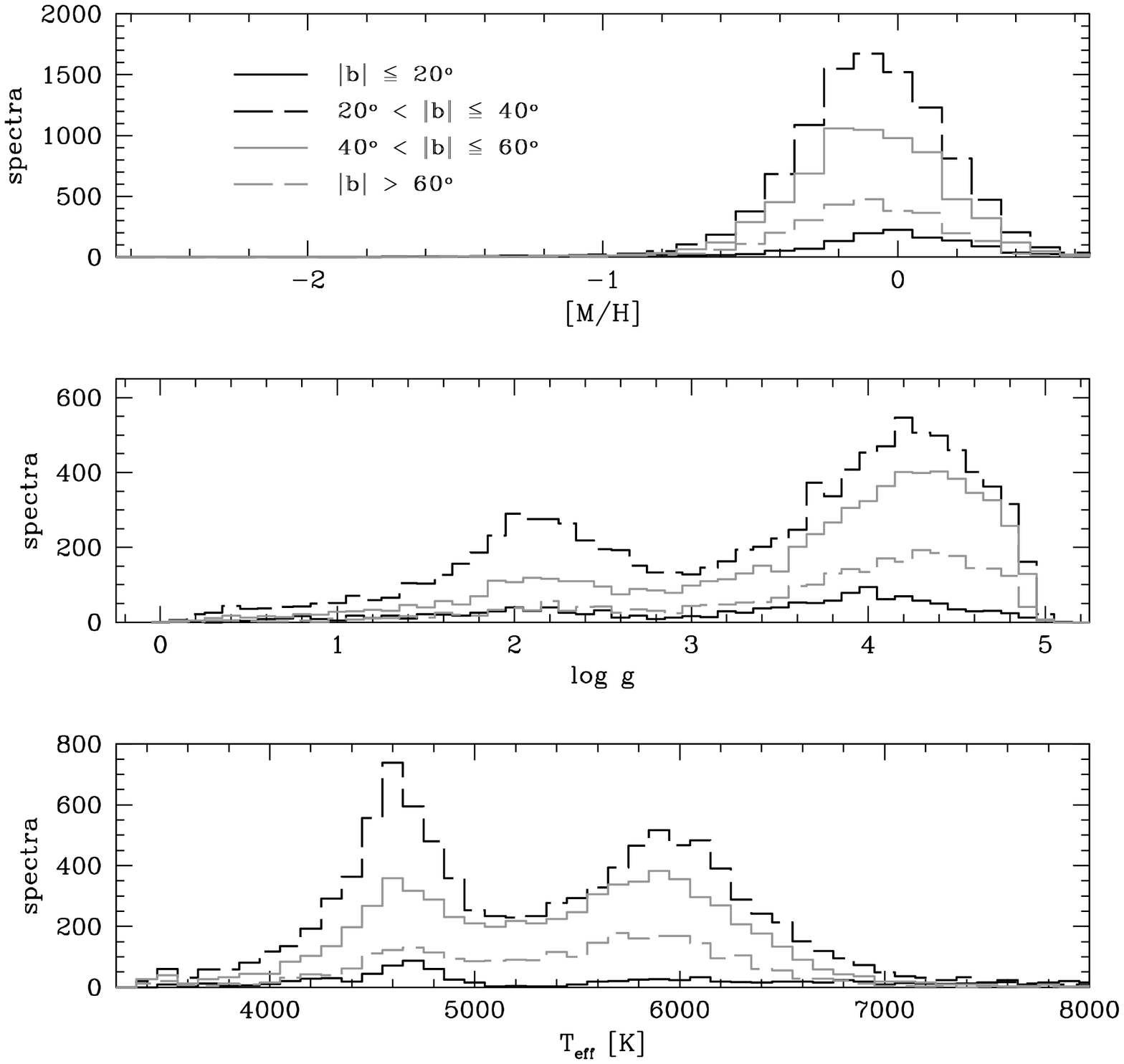}
\caption{Temperature, gravity, and metallicity histograms for spectra 
with published stellar parameters. Histograms for individual Galactic 
latitude bands are plotted separately with the key given in the 
top panel. Spectra with $|b| \le 20^o$ are calibration fields. Note 
the increasing fraction of main sequence stars at high Galactic latitudes. 
}
\label{parameterhisto}
\end{figure}

The investigation of properties of the stellar parameters
and their links to Galactic dynamics and formation history are beyond 
the scope of this paper. To illustrate the situation we outline just 
two plots. Figure \ref{paramscalib} shows the location of all spectra on the  
temperature--gravity--metallicity wedge. Note the main sequence and 
giant groups, their relative frequency and metallicity distribution 
for three bands in Galactic latitude. 
Figure~\ref{parameterhisto} plots histograms of the parameters, 
again for different bands in Galactic latitude.  
The fraction of main sequence stars increases with the distance from 
the Galactic plane. This can be understood by the fact that the RAVE 
targets have rather similar apparent magnitudes (Figure~\ref{f01}). 
Giants therefore trace a more distant population, and those at high 
latitudes would be already members of the (scarcely populated) Galactic 
halo. 

\begin{deluxetable}{ccccccc}
\tablecaption{Number and fraction of RAVE database entries with a counterpart 
in the photometric catalogs
\label{t:photometricquality}}
\tablewidth{0pt}
\tablecolumns{7}
\tablehead{
\colhead{Catalog name} & Number of & \%\ of Entries &\multicolumn{4}{c}{\%\ with quality flag}\\
                       & Entries   & with Counterpart& A & B & C & D \\
}
\startdata
2MASS & 51,813 & 99.97\% & 99.6\% &    0\% &   0\% & 0.4\% \\
DENIS & 40,106 & 77.4 \% & 73.7\% & 23.5\% & 2.3\% & 0.5\% \\
USNO-B& 51,466 & 99.3 \% & 99.2\% & 0.5\%  &   0\% & 0.3\% \\ \hline
\enddata
\end{deluxetable}

\begin{deluxetable}{rlrrrr}
\tablecaption{Summary of proper motion sources and their 
average and 90\%\ errors
\label{t:PMquality}}
\tablewidth{0pt}
\tablecolumns{6}
\tablehead{
\colhead{SPM} & Catalog & Number of & Fraction  & Average & 90\% \\
         Flag & Name    & Entries   & of entries& PM error & PM error\\
              &         &           &           & [mas~yr$^{-1}]$&[mas~yr$^{-1}$]
}
\startdata
0 & No proper motion &    74 & 0.1\% &     &      \\
1 & Tycho-2          &   879 & 1.7\% & 2.9 &  4.0 \\
2 & SSS              & 3,427 & 6.6\% &23.7 & 31.7 \\
3 & STARNET 2.0      &31,739 &61.2\% & 3.3 &  4.6 \\
4 & 2MASS$+$GSC 1.2  &    62 & 0.1\% &18.7 & 26.1 \\
5 & UCAC2            &15,047 &29.0\% & 6.7 & 11.1 \\ \hline
1-5 & all with proper motion   
                     &51,154 &98.7\% & 5.7 & 10.6 \\ 
\enddata
\end{deluxetable}

\subsection{Photometry}

\label{photometry}

The data release includes 
cross-identification with optical and near-IR catalogs (USNO-B, 
DENIS, 2MASS) where the nearest neighbor criterion was used 
for matching. Similar to the first data release we provide the 
distance to the nearest neighbor and a quality flag on the reliability 
of the match. Note that this is important as RAVE uses optical 
fibers with a projected diameter of 6.7~arc~sec on the sky. 
Table \ref{t:photometricquality} shows that nearly all stars were 
successfully matched for the 2MASS and USNO-B catalogs, while only 
about 3/4 of the stars lie in the sky area covered by the DENIS catalog.
For the matched stars we include USNO-B B1, R1, B2, R2 and $I$ magnitudes,
DENIS I, J, and K magnitudes, and 2MASS J, H, and K magnitudes. 
As mentioned our wavelength range is best represented by the $I$ 
filter. With the publication of the second release of the DENIS catalog 
we decided to use the DENIS $I$ magnitude as our reference in planning of 
future observations.

We note here that the DENIS $I$ magnitudes appear to be affected
by saturation for stars with $I < 10$.  Following a comment from 
a member of the DENIS team, we compared the DENIS and 2MASS magnitude
scales. 2MASS does not provide an $I-$magnitude. However the transformation
\begin{equation}
I_\mathrm{2MASS} = J_\mathrm{2MASS} + 1.103\,\,(J-K)_\mathrm{2MASS} + 0.07 
\end{equation}
gives an approximate $I$ magnitude
on the DENIS system from the 2MASS $JK$ photometry for giants and dwarfs
with $(J-K)<0.65$.  First we confirmed that the $(J-K)_\mathrm{2MASS}$ colors 
are consistent with the temperature derived by RAVE for all objects.  
We then compared the DENIS and 2MASS $I-$magnitudes for
all stars in the current data release having errors $< 0.05$ in both 
of these $I-$ magnitudes. For most stars with $I_\mathrm{DENIS} > 10$, the 
magnitudes agree within the expected errors.  However we note that 
(1) the relation between the two magnitudes becomes non-linear for the 
$\sim 16$\% of the brightest stars with $I_\mathrm{DENIS} < 10$, and (2) about 8\% 
of the fainter stars with apparently well-determined magnitudes from 
both catalogs have differences $|(I_\mathrm{DENIS} - I_\mathrm{2MASS})| > 0.2$. Some 
stars have differences greater than $\pm 3$ magnitudes. We therefore 
propose to avoid to use $I_\mathrm{DENIS}$ magnitudes when the 
condition 
$-0.2 < (I_\mathrm{DENIS} - J_\mathrm{2MASS}) - 
        (J_\mathrm{2MASS} - K_\mathrm{2MASS}) < 0.6$ is not met.
Figure~\ref{f:countsIdenis} follows this advice and avoids the scatter 
due to some problematic $I_\mathrm{DENIS}$ magnitude values.

\subsection{Proper motions}

\label{propermotion}

Similarly to the first data release the proper motions are 
taken from Starnet 2.0, and Tycho-2 catalogs (see Paper~I for a 
complete discussion). These values are however not available 
for $\sim 30\%$ of the spectra and in Paper~I we bridged the 
gap with proper motions from the SSS catalog. 
The SSS catalog suffers from substantial uncertainties, 
so we now attempted a cross-identification with the UCAC2 catalog 
\citep{Zacharias04}. RAVE coordinates were used to search for the 
nearest two neighbors in the UCAC2 catalogue. It turned 
out that it suffices to use the data for the first next 
neighbor, as there were no cases where the matching distance 
to the first neighbor was less than 3 arcsec while that to the 
second one was less than 6 arcsec. The UCAC2 counterpart within
3~arcsec search radius was identified for 94\%\ of the spectra, 
many of the remaining objects have large errors in reported 
proper motion.  
Note that UCAC2 values are systematically offset from 
the Starnet~2 measurements. The difference is $\sim 2$~mas~yr$^{-1}$ in 
right ascension and $\sim 1$~mas~yr$^{-1}$ in declination (with the UCAC2 
values being smaller than the Starnet~2 ones). The final catalog 
therefore includes the UCAC2 proper motion if the Starnet~2 or 
Tycho-2 values are not available ($\sim 23$\%\ of cases). The 
source of proper motions is flagged, so the systematic differences 
could be taken into account. Table~\ref{t:PMquality} gives details 
on the use of proper motion catalogs in the present data release 
and their reported average and 90 percentile errors. In all cases 
this data release includes proper motion from the source with the 
best value of reported accuracy.

\section{Conclusions}

This second data release reports radial velocities of 51,829 spectra of 
49,327 different stars, randomly selected in the magnitude range of 
$9 \simlt I \simlt 12$ and located more than $25^\mathrm{o}$ away from 
the Galactic plane (except for a few test observations). It covers
an area of $\sim 7,200$ square degrees. These numbers approximately 
double the sample reported in Paper~I. Moreover, this data release
is the first to include values of stellar parameters as determined 
from stellar spectra. We report temperature, gravity and metallicity 
for 21,121 normal stars, all observed after the first data release. Stars 
with a high rotational velocity or peculiar type (e.g.\ binary stars and 
emission stars) will be discussed separately. 

Radial velocities for stars new in this data release are more accurate 
than before, with typical errors between 1.3 and 1.7~\kms. These values 
are confirmed both by repeated observations and by external datasets and 
have only a weak dependence on the S/N ratio. We used five separate 
external datasets to check values of stellar parameters derived from 
RAVE spectra. These included observations with different instruments 
at different resolving powers and in different wavelength regimes, as 
well as data from the literature. The uncertainty of stellar parameter values 
strongly depends on stellar type. Despite considerable effort our calibration 
observations do not cover (yet) the entire parameter space. We plan to 
improve on this using dedicated calibration observations with at least 
4 telescopes. For this data release we had to resort to extensive 
simulations which are however tuned by calibration observations. 
A typical RAVE star has an uncertainty of 
400~K in temperature, 0.5~dex in gravity, and 0.2~dex in metallicity.
The error depends on the signal to noise ratio and can be  
$\simgt 2$ times better/worse for stars at extremes of the noise range. 
Repeated observations show that these error estimates are  
rather conservative, possibly due to intrinsic variability of 
the observed stars and/or non-negligible errors of reference 
values from the calibration datasets.

Future data releases will follow on an approximately 
yearly basis. They will benefit from our considerable and 
ongoing effort to obtain calibration datasets using other telescopes 
and similar or complimentary observing techniques. Notably we expect 
that SkyMapper \citep{skymapper}, an all-southern-sky survey just starting 
at the Siding Spring Observatory, will provide accurate photometry and 
temporal variability information for all RAVE stars. 

RAVE is planned to observe up to a million spectra of stars away from 
the Galactic plane. It represents an unprecedented sample of 
stellar kinematics and physical properties in the range of magnitudes 
probing scales between the very local surveys 
(Geneva Copenhagen Survey and \citet{famaey05}) and more distant
ones (SDSSII/SEGUE), complementing the planned AAOmega efforts closer to 
the Galactic plane. So it helps to complete our picture of the Milky 
Way, paving the way for the next decade endeavors, like Gaia.

\acknowledgments
{\bf Acknowledgments}

We are most grateful to our referee, prof.\ David W.\ Latham, for his 
detailed and very relevant comments which improved the quality of 
the presentation of the paper. 

Funding for RAVE has been provided by the Anglo-Australian Observatory, 
the Astrophysical Institute Potsdam, the Australian Research Council, 
the German Research foundation, the National Institute for Astrophysics 
at Padova, The Johns Hopkins University, the Netherlands Research School 
for Astronomy, the Natural Sciences and Engineering Research Council of 
Canada, the Slovenian Research Agency, the Swiss National Science 
Foundation, the National Science Foundation of the USA (AST-0508996), 
the Netherlands Organisation for Scientific Research, the Particle 
Physics and Astronomy Research Council of the UK, Opticon, Strasbourg 
Observatory, and by the Universities of Basel, Cambridge, and Groningen. 
The RAVE web site is at www.rave-survey.org.

KCF, QAP, BG, RC, WR and ECW acknowledge support from Australian Research 
Council grants DP0451045 and DP0772283.
A.~Siviero and EKG are supported by the Swiss National Science Foundation under the
grants 200020-105260 and 200020-113697.
JPF acknowledges support from the Keck Foundation, through a grant to JHU. 
RFGW acknowledges seed money from the School of Arts and Sciences at JHU,
plus NSF grant AST-0508996.
GMS was funded by a Particle Physics and Astronomy Research Council PhD
Studentship. 
OB acknowledges financial support from the CNRS/INSU/PNG.
PRF is supported by the European Marie Curie RTN ELSA, contract 
MRTN-CT-2006-033481.

This  research  has made  use  of the  VizieR  catalogue  access tool,  CDS,
Strasbourg, France. This publication makes use of data products from the Two 
Micron All Sky Survey, which is a joint project of the University of Massachusetts 
and the IPAC/Caltech, funded by NASA and NSF.
The results are 
based partly on observations obtained at the Asiago 1.82-m telescope (Italy) and 
at the Observatoire de Haute Provence (OHP) (France) which is operated by the 
French CNRS. 
The cross-identification of the RAVE data release 
with the UCAC2 catalogue was done using the electronic version 
of the UCAC2 kindly provided by Norbert Zacharias.

\newpage

\section*{Appendix A}

Table \ref{t:A1} describes the contents of individual columns of the second data 
release catalog. The catalog is accessible online at www.rave--survey.org and
via the Strasbourg astronomical Data Center (CDS) services.

\begin{deluxetable}{cclcll}
\tabletypesize{\footnotesize}
\tablecaption{Catalog description \label{t:A1}}
\tablehead{
Column & Character & Format & Units & Symbol & Description\\ 
number & range     &        &      &        &     }
\startdata
 1&  1--16 &A16   &---   & Name            &Target designation\\
 2& 18--29 &F12.8 &deg   & RAdeg           &Right ascension (J2000.0)\\
 3& 31--42 &F12.8 &deg   & DEdeg           &Declination (J2000.0)\\
 4& 44--52 &F9.5  &deg   & GLON            &Galactic longitude\\
 5& 54--61 &F9.5  &deg   & GLAT            &Galactic latitude\\
 6& 64--70 &F7.1  &\kms  & HRV             &Heliocentric radial velocity\\
 7& 72--77 &F6.1  &\kms  & eHRV            &HRV error\\
 8& 79--84 &F6.1  &mas yr$^{-1}$& pmRA            &proper motion RA\\
 9& 86--91 &F6.1  &mas yr$^{-1}$& epmRA           &error proper motion RA\\
10& 93--98 &F6.1  &mas yr$^{-1}$& pmDE            &proper motion DE\\
11&100--105&F6.1  &mas yr$^{-1}$& epmDE           &error proper motion DE\\
12&107--107&I1    &---   & Spm             &source of proper motion (1)\\
13&109--113&F5.2  &mag   & Imag            &Input catalog I magnitude\\
14&115--122&A8    &---   & Obsdate         &Date of observation yyyymmdd\\
15&124--133&A10   &---   & FieldName       &Name of RAVE field\\
16&135--135&I1    &---   & PlateNumber     &Plate number used\\
17&137--139&I3    &---   & FiberNumber     &Fiber number [1,150]\\
18&141--144&I5    &K     & Teff            &Effective Temperature\\
19&146--150&F4.2  &dex   & logg            &Gravity\\
20&152--156&F5.2  &dex   & Met             &Uncalibrated [M/H]\\
21&158--161&F4.2  &dex   & alpha           &[Alpha/Fe]\\
22&163--167&F5.2  &dex   & cMet            &Calibrated [M/H]\\
23&169--176&F8.1  &---   & CHISQ           &chi square\\
24&178--182&F5.1  &---   & S2N             &Corrected Signal to noise S2N\\
25&184--188&F5.1  &---   & CorrelationCoeff&Tonry-Davis $R$ correlation coefficient\\
26&190--193&F4.2  &---   & PeakHeight      &Height of correlation peak\\
27&195--200&F6.1  &\kms  & PeakWidth       &Width of correlation peak\\
28&202--207&F6.1  &\kms  & CorrectionRV    &Zero point correction applied\\
29&209--214&F6.1  &\kms  & SkyRV           &Measured HRV of sky\\
30&216--221&F6.1  &\kms  & SkyeRV          &error HRV of sky\\
31&223--227&F5.1  &---   & SkyCorrelation  &Sky Tonry-Davis correl. coefficient\\
32&229--233&F5.1  &---   & SNRatio         &Spectra signal to noise ratio\\
33&235--240&F6.3  &mag   & BT              &Tycho-2 BT magnitude\\
34&242--247&F6.3  &mag   & eBT             &error BT\\
35&249--254&F6.3  &mag   & VT              &Tycho-2 VT magnitude\\
36&256--261&F6.3  &mag   & eVT             &error VT\\
37&263--276&A12   &---   & USNOID          &USNO-B designation\\
38&278--283&F6.3  &mas   & DisUSNO         &Distance to USNO-B source\\
39&285--289&F5.2  &mag   & B1              &USNO-B B1 magnitude\\
40&291--295&F5.2  &mag   & R1              &USNO-B R1 magnitude\\
41&297--301&F5.2  &mag   & B2              &USNO-B B2 magnitude\\
42&303--307&F5.2  &mag   & R2              &USNO-B R2 magnitude\\
43&309--313&F5.2  &mag   & IUSNO           &USNO-B I magnitude\\
44&315--315&A1    &---   & XidQualityUSNO  &Cross-identification flag (2)\\
45&317--332&A16   &---   & DENISID         &DENIS designation\\
46&334--339&F6.3  &mas   & DisDENIS        &Distance to DENIS source\\
47&341--346&F6.3  &mag   & IDENIS          &DENIS I magnitude\\
48&348--351&F4.2  &mag   & eIDENIS         &error DENIS I magnitude\\
49&353--359&F6.3  &mag   & JDENIS          &DENIS J magnitude\\
50&360--363&F4.2  &mag   & eJDENIS         &error DENIS J magnitude\\
51&365--370&F6.3  &mag   & KDENIS          &DENIS K magnitude\\
52&372--375&F4.2  &mag   & eKDENIS         &error DENIS K magnitude\\
53&377--377&A1    &---   & XidQualityDENIS &Cross-identification flag (2)\\
54&379--394&A16   &---   & TWOMASSID       &2MASS designation\\
55&396--401&F6.3  &mas   & Dis2MASS        &Distance to 2MASS source\\
56&403--408&F6.3  &mag   & J2MASS          &2MASS J magnitude\\
57&410--413&F4.2  &mag   & eJ2MASS         &error 2MASS J magnitude\\
58&415--420&F6.3  &mag   & H2MASS          &2MASS H magnitude\\
59&422--425&F4.2  &mag   & eH2MASS         &error 2MASS H magnitude\\
60&427--432&F6.3  &mag   & K2MASS          &2MASS K magnitude\\
61&434--437&F4.2  &mag   & eK2MASS         &error 2MASS K magnitude\\
62&439--441&A3    &---   & TWOMASSphotFLAG &2MASS photometric flag\\
63&443--443&A1    &---   & XidQuality2MASS &Cross-identification flag (2)\\
64&445--447&A3    &---   & ZeroPointFLAG   &Zero point correction flag (3)\\
65&449--456&A8    &---   & SpectraFLAG     &Spectra quality flag (4)\\
\enddata
\tablecomments{
(1): Flag value between 0 and 4:
    0- no proper motion,
    1- Tycho-2 proper motion,
    2- Supercosmos Sky Survey proper motion,
    3- STARNET2.0 proper motion,
    4- GSC1.2 x 2MASS proper motion,
    5- UCAC-2 proper motions. 
\\
(2): Flag value is A,B,C,D or X:
    A- good association,
    B- 2 solutions within 1 arcsec,
    C- more than two solutions within 1 arcsec,
    D- nearest neighbor more than 2 arcsec away,
    X- no possible counterpart found. 
\\
(3): Flag value of the form FGH, F being for the entire plate, G for the 
    50 fibers group to which the fiber belongs. If H is set to * the fiber is
    close to a 15 fiber gap. 
    For F and G the values can be A, B, C, D or E:
    A- dispersion around correction lower than 1 \kms,
    B- dispersion between 1 and 2 \kms,
    C- dispersion between 2 and 3 \kms,
    D- dispersion larger than 3 \kms,
    E- less than 15 fibers available for the fit. 
\\    
(4): Flag identifying possible problem in the spectra (values can be 
    combined):
    a- asymmetric Ca lines,
    c- cosmic ray pollution,
    e- emission line spectra,
    n- noise dominated spectra,
    l- no lines visible,
    w- weak lines,
    g- strong ghost,
    t- bad template fit,
    s- strong residual sky emission,
    cc- bad continuum,
    r- red part of the spectra shows problem,
    b- blue part of the spectra shows problem,
    p- possible binary/doubled lined,
    x- peculiar object.
}
\end{deluxetable}

\newpage

\section*{Appendix B: External data}

Tables~\ref{t:B1}--\ref{t:B5} compare the results of RAVE observations with 
those from the external datasets. The latter are discussed in  
Sec.~\ref{externaldatasets}.

\begin{deluxetable}{lllrrlrrrrrrrr}
\tablecaption{Results of the re-observation of 45 RAVE stars at Apache Point 
Observatory \label{t:B1}}
\tablewidth{0pt}
\tabletypesize{\footnotesize}
\rotate
\tablehead{
  \multicolumn{1}{c}{ }&\multicolumn{4}{c}{Echelle results}&\multicolumn{9}{|c}{RAVE values}\\
\cline{2-5} \cline{6-14}
  Name &          $\teff$&$\logg$&[Fe/H]&[M/H]&Obsdate  & FieldName &FibNum&$\teff$&$\logg$&Met&alpha&CHISQ&S2N
}
\startdata
  T4671\_00811\_1 &  5943&3.55&-0.53&-0.42&20040629 & 0030m06   &    43&  6006&  +3.89&-0.66& +0.17&   935.21  & 44\\
  T4701\_00802\_1 &  4808&2.59&-0.62&-0.49&20041202 & 0238m05   &    78&  4670&  +2.15&-1.07& +0.26&   7332.54&  94\\
  T4926\_00806\_1 &  4633&2.89& 0.25& 0.25&20050322 & 1119m04   &   110&  4622&  +2.64&-0.02& +0.00&   3253.97&  60\\
  T4927\_01523\_1 &  5500&4.23&-0.19&-0.15&20050321 & 1058m07   &    97&  5652&  +4.50&-0.31& +0.14&   3163.09&  71\\
  T4931\_00266\_1 &  5235&2.44&-1.24&-1.02&20040629 & 1146m01   &    45&  5320&  +2.55&-0.92& +0.25&   955.08 &  37\\
  T5231\_00846\_1 &  5801&3.70&-0.17&-0.14&20040629 & 2212m04   &   134&  5838&  +4.63&-0.36& +0.17&   2101.47&  74\\
  T5279\_00819\_1 &  4627&1.98&-0.35&-0.28&20041022 & 0136m15   &    76&  4642&  +2.02&-0.67& +0.16&   5406.29&  74\\
  T5279\_00819\_1 &  4627&1.98&-0.35&-0.28&20041023 & 0136m15   &    76&  4568&  +1.96&-0.68& +0.21&   970.44 &  63\\
  T5279\_01652\_1 &  5565&2.80&-0.05&-0.04&20041022 & 0136m15   &   108&  5019&  +2.54&-0.88& +0.30&   2348.39&  65\\
  T5279\_01652\_1 &  5565&2.80&-0.05&-0.04&20041023 & 0136m15   &   108&  5598&  +3.52&-0.60& +0.19&   4490.77&  75\\
  T5310\_00259\_1 &  4370&0.63&-1.47&-1.25&20041202 & 0352m13   &    28&  4328&  +0.47&-1.56& +0.15&   388.73 &  34\\
  T5310\_00788\_1 &  4627&2.30&-0.24&-0.19&20041202 & 0352m13   &    95&  4548&  +1.93&-0.45& +0.03&   319.58 &  32\\
  T5491\_01056\_1 &  5986&3.67& 0.16& 0.16&20040510 & 1025m08   &   141&  6100&  +3.59&-0.45& +0.29&   3596.59&  69\\
  T5496\_00127\_1 &  4594&2.12& 0.00& 0.00&20040501 & 1014m13   &    88&  4575&  +2.09&-0.44& +0.10&   1099.67&  47\\
  T5496\_00127\_1 &  4594&2.12& 0.00& 0.00&20040502 & 1014m13   &    88&  4640&  +2.21&-0.33& +0.01&   1748.35&  62\\
  T5499\_00076\_1 &  5944&3.90&-0.79&-0.62&20040531 & 1058m07   &    38&  6025&  +3.96&-1.01& +0.39&   1574.16&  53\\
  T5507\_01406\_1 &  6075&3.69&-0.84&-0.66&20040530 & 1101m15   &    89&  6105&  +3.77&-0.97& +0.14&   1742.15&  81\\
  T5543\_00567\_1 &  5497&4.27& 0.21& 0.21&20050330 & 1309m11   &   146&  5729&  +4.65&+0.15& +0.01&   6685.34&  94\\
  T5562\_00279\_1 &  5090&3.24&-0.06&-0.05&20040607 & 1418m11   &    77&  5150&  +3.30&-0.34& +0.07&   4356.54&  76\\
  T5762\_00685\_1 &  5191&3.08&-0.86&-0.67&20040629 & 2034m12   &    40&  5250&  +2.98&-1.17& +0.25&   2550.69&  70\\
  T5789\_00559\_1 &  4501&1.41&-1.06&-0.84&20040627 & 2159m08   &    47&  4577&  +1.86&-1.11& +0.27&   627.04 &  50\\
  T5803\_01091\_1 &  5905&3.72&-0.06&-0.05&20040627 & 2159m08   &   119&  6011&  +3.76&-0.31& +0.26&   2908.89&  70\\
  T5806\_01423\_1 &  5784&4.37& 0.01& 0.01&20040628 & 2216m13   &    56&  5945&  +4.41&-0.34& +0.16&   1188.81&  48\\
  T5852\_00128\_1 &  4792&4.26&-0.14&-0.11&20041022 & 0136m15   &    18&  4922&  +4.57&-0.19& +0.00&   4162.91&  80\\
  T5852\_00128\_1 &  4792&4.26&-0.14&-0.11&20041023 & 0136m15   &    18&  5204&  +4.88&+0.19& +0.00&   443.61 &  35\\
  T5852\_00673\_1 &  5412&2.95&-0.72&-0.57&20041022 & 0136m15   &     2&  5523&  +3.04&-1.27& +0.38&   2772.21&  65\\
  T5852\_00673\_1 &  5412&2.95&-0.72&-0.57&20041023 & 0136m15   &     2&  5592&  +3.19&-1.11& +0.33&   4107.41&  67\\
  T5852\_01716\_1 &  5574&3.59&-1.19&-0.97&20041022 & 0136m15   &    34&  5084&  +2.69&-1.16& +0.18&   1567.50&  51\\
  T5852\_01716\_1 &  5574&3.59&-1.19&-0.97&20041023 & 0136m15   &    34&  5621&  +3.62&-0.37& +0.05&   498.88 &  26 \\
  T5866\_00288\_1 &  5535&3.34&-0.64&-0.50&20040826 & 0243m17   &   114&  5587&  +3.25&-1.11& +0.13&   8243.82& 103 \\
  T5875\_00738\_1 &  4366&1.00&-1.25&-1.03&20041022 & 0313m20   &    79&  4341&  +0.77&-1.67& +0.40&   4006.72& 103\\
  T6077\_00047\_1 &  5707&4.27& 0.46& 0.46&20050301 & 1101m15   &   117&  5775&  +4.27&+0.18& +0.03&   3442.99&  65\\
  T6092\_00615\_1 &  5560&4.59& 0.01& 0.01&20050228 & 1144m20   &    22&  5496&  +4.68&-0.11& +0.00&   2096.18&  48\\
  T6109\_01354\_1 &  6042&3.95& 0.10& 0.10&20050228 & 1232m22   &    72&  6083&  +4.24&+0.05& +0.02&   2425.02&  61\\
  T6135\_00087\_1 &  4250&1.68&-0.12&-0.10&20050301 & 1345m21   &    92&  4295&  +1.69&-0.47& +0.07&   2188.12&  70\\
  T6412\_00004\_1 &  5971&2.92&-0.76&-0.57&20040923 & 0014m21   &    17&  6041&  +3.22&-1.00& +0.19&   571.02 &  57\\
  T6412\_00004\_1 &  5971&2.92&-0.76&-0.57&20041024 & 0014m21   &    19&  6077&  +3.49&-1.08& +0.29&   2600.99&  93\\
  T6459\_00058\_1 &  4526&1.92&-0.34&-0.27&20041230 & 0414m29   &    66&  4491&  +1.74&-0.78& +0.10&   4411.66&  74\\
  T6473\_00818\_1 &  4594&2.42& 0.07& 0.07&20041023 & 0504m26   &    84&  4549&  +2.30&-0.19& +0.00&   7898.21&  84\\
  T6478\_00245\_1 &  5503&3.54& 0.07& 0.07&20041023 & 0504m26   &   106&  5546&  +3.77&-0.22& +0.15&   1435.45&  51\\
  T6484\_00022\_1 &  4599&2.01&-0.36&-0.28&20050128 & 0535m29   &    73&  4583&  +1.85&-0.73& +0.14&   4421.43&  82\\
  T6705\_00713\_1 &  5813&4.57&-0.02&-0.02&20040528 & 1252m28   &    84&  5914&  +4.60&-0.28& +0.16&   1167.70&  66\\
  T6705\_00713\_1 &  5813&4.57&-0.02&-0.02&20040529 & 1252m28   &    84&  5935&  +4.68&-0.51& +0.33&   3947.94&  86\\
  T6904\_00180\_1 &  5889&3.21&-0.96&-0.75&20040607 & 2008m28   &    42&  5941&  +3.18&-1.20& +0.28&   6137.75&  95\\
  T6975\_00058\_1 &  5294&2.96&-0.79&-0.62&20041024 & 2315m25   &    16&  5652&  +3.35&-0.97& +0.26&   648.78 &  41 \\ 
\enddata
\tablecomments{
All stars passed visual inspection. Column names follow symbol
names in Table~\ref{t:A1}, so {\it Met} marks the original (uncalibrated) value of 
the metallicity.
} 
\end{deluxetable}

\begin{deluxetable}{lrrrrrrrrrrrl}
\tabletypesize{\footnotesize}
\tablecaption{Results of the re-observation of 24 RAVE stars with the Asiago 
Observatory echelle spectrograph \label{t:B2}}
\tablewidth{0pt}
\small
\rotate
\tablehead{
\multicolumn{1}{c}{ }&\multicolumn{3}{c}{Echelle results}&\multicolumn{9}{|c}{RAVE values}\\
\cline{2-4} \cline{5-13}
Name           &  $\teff$&$\logg$ &[M/H] & Obsdate & FieldName & FibNum & $\teff$ & $\logg$&  Met & alpha& S2N & SpectraFLAG
}
\startdata
T4678\_00087\_1&  3938& 3.00& -1.03&  20040629  &  0030m06 & 38 &3818&  3.2&  -0.9&   0.0&    76&  e \\
T4679\_00388\_1&  6291& 4.00& -0.88&  20040629  &  0030m06 &117 &6102&  3.8&  -0.7&   0.2&    73&  \\
T4701\_00802\_1&  4865& 2.19& -1.28&  20041202  &  0238m05 & 78 &4652&  2.2&  -1.1&   0.3&    94&  \\
T4702\_00944\_1&  5851& 4.10& -0.38&  20041202  &  0238m05 & 79 &5704&  4.7&  -0.4&   0.2&    80&  \\
T4704\_00341\_1&  5757& 4.20& -0.45&  20041202  &  0238m05 & 50 &5667&  4.0&  -0.6&   0.2&    76&  p \\
T4749\_00016\_1&  5178& 3.06& -0.65&  20041202  &  0500m08 & 64 &4676&  2.2&  -0.5&   0.2&    90&  \\
T4749\_00143\_1&  6928& 3.97& -0.30&  20041202  &  0500m08 & 59 &7034&  4.1&  -0.2&   0.1&    73&  \\
T4763\_01210\_1&  4451& 2.01& -0.51&  20041202  &  0500m08 & 99 &4301&  1.6&  -0.6&   0.1&    72&  \\
T5178\_01006\_1&  4498& 2.23& -0.53&  20040626  &  2054m02 & 65 &4682&  2.1&  -0.4&   0.0&    68&  \\
T5186\_01028\_1&  4777& 2.88&  0.22&  20040626  &  2054m02 & 25 &4628&  2.5&   0.1&   0.0&    62&  \\
T5198\_00021\_1&  4898& 3.15&  0.10&  20040629  &  2119m03 & 71 &4653&  2.9&  -0.0&   0.0&    77&  \\
T5198\_00784\_1&  7426& 3.78& -0.44&  20040629  &  2119m03 & 98 &7375&  4.0&  -0.3&   0.0&    83&  \\
T5199\_00143\_1&  7102& 4.24& -0.21&  20040629  &  2119m03 & 97 &6986&  4.0&  -0.1&   0.1&    99&  \\
T5201\_01410\_1&  4901& 2.98&  0.30&  20040629  &  2119m03 & 23 &4641&  2.7&   0.2&   0.0&    68&  \\
T5207\_00294\_1&  4105& 1.33& -0.59&  20040628  &  2133m08 & 61 &3999&  1.0&  -0.7&   0.2&    80&  \\
T5225\_01299\_1&  4241& 1.97& -0.60&  20040629  &  2212m04 & 86 &4104&  1.0&  -0.7&   0.1&    80&  \\
T5227\_00846\_1&  5239& 3.34& -0.88&  20040629  &  2212m04 & 23 &4856&  3.1&  -0.6&   0.2&    60&  \\
T5228\_01074\_1&  5098& 4.10& -0.41&  20040629  &  2212m04 & 98 &5219&  4.7&  -0.3&   0.0&    72&  \\
T5231\_00546\_1&  7104& 3.52& -0.46&  20040629  &  2212m04 &129 &6891&  3.8&  -0.7&   0.3&    85&  \\
T5232\_00783\_1&  4906& 3.01& -0.12&  20040629  &  2212m04 &127 &4760&  2.8&  -0.3&   0.0&    72&  \\
T5242\_00324\_1&  3915& 3.37& -0.91&  20040626  &  2313m03 & 67 &3530&  4.0&  -0.7&   0.0&    73&  e \\
T5244\_00102\_1&  6523& 3.28& -0.54&  20040626  &  2313m03 & 96 &6363&  3.3&  -0.6&   0.1&    64&  \\
T5246\_00361\_1&  4890& 2.44& -0.89&  20040626  &  2313m03 & 81 &4724&  2.0&  -0.7&   0.1&    78&  \\
T5323\_01037\_1&  4880& 2.48& -0.66&  20041202  &  0500m08 &139 &4646&  2.0&  -0.6&   0.1&    83&  
\enddata
\tablecomments{
Column names follow symbol names in 
Table~\ref{t:A1}, so {\it Met} marks the original (uncalibrated) values of the  
metallicity. Only stars with an empty SpectraFLAG 
were retained for further analysis.
} 
\end{deluxetable}

\begin{deluxetable}{lrrrrrrrrrrrr}
\tablecaption{Results of the observation of 49 stars from the 
\citet{soubiran2005} catalog \label{t:B3}}
\tablewidth{0pt}
\small
\rotate
\tablehead{
\multicolumn{1}{c}{ }&\multicolumn{4}{c}{Echelle results}&\multicolumn{8}{|c}{RAVE values}\\
\cline{2-4} \cline{5-13}
Name        &  $\teff$ &$\logg$&[Fe/H]& [M/H] & Obsdate  &  FieldName&FibNum  & $\teff$ &  $\logg$&  Met  & alpha & S2N \\ \hline
}
\startdata
  BD-213420 &  5946& 4.41& -1.04& -0.90 & 20070422 &  1155m22  &  114   &  5765 &  +3.51& -1.30 & +0.20 &   90 \\
  HD136351  &  6341& 4.04&  0.01&  0.00 & 20070422 &  1522m47  &  114   &  6151 &  +3.78& -0.30 & +0.16 &  235 \\ 
  HD157467  &  6016& 3.72&  0.11&  0.27 & 20070422 &  1726m03  &   33   &  6103 &  +3.77& -0.24 & +0.19 &  212 \\
  HD156635  &  6136& 4.28& -0.10&  0.12 & 20070422 &  1726m03  &   41   &  6330 &  +4.40& -0.36 & +0.13 &  243 \\
  HD157347  &  5687& 4.38&  0.00& -0.01 & 20070422 &  1726m03  &   67   &  5813 &  +4.61& -0.24 & +0.18 &  206 \\
  HD158809  &  5464& 3.80& -0.77& -0.49 & 20070422 &  1726m03  &  109   &  5727 &  +4.03& -0.67 & +0.32 &  147 \\
  HD159307  &  6227& 3.94& -0.71& -0.51 & 20070422 &  1726m03  &  114   &  6391 &  +3.87& -0.70 & +0.11 &  206 \\
  HD126681  &  5540& 4.49& -1.17& -0.87 & 20070423 &  1425m18  &  120   &  5481 &  +3.51& -1.30 & +0.22 &  144 \\
  HD149612  &  5680& 4.53& -0.48& -0.43 & 20070423 &  1650m57  &   18   &  5615 &  +3.98& -0.87 & +0.26 &  208 \\
  HD153075  &  5770& 4.17& -0.57& -0.39 & 20070423 &  1650m57  &  114   &  5728 &  +3.98& -0.87 & +0.29 &  214 \\
  HD131117  &  6001& 4.09&  0.13&  0.06 & 20070425 &  1450m30  &  120   &  6102 &  +4.36& -0.22 & +0.38 &  165 \\ 
  HD172051  &  5552& 4.49& -0.29& -0.24 & 20070425 &  1835m21  &  120   &  5742 &  +4.55& -0.40 & +0.12 &  257 \\
  HD112164  &  5953& 4.00&  0.24&  0.24 & 20070506 &  1254m44  &  116   &  5781 &  +3.79& -0.17 & +0.22 &  202 \\ 
  HD119173  &  5905& 4.48& -0.63& -0.53 & 20070506 &  1340m03  &  144   &  5709 &  +3.94& -0.96 & +0.12 &  171 \\
  HD144585  &  5856& 4.12&  0.28&  0.18 & 20070506 &  1607m14  &  120   &  6005 &  +4.58& +0.07 & +0.10 &  238 \\
  HD153240  &  6135& 4.31& -0.09&  0.12 & 20070506 &  1655m04  &  116   &  6271 &  +4.85& -0.37 & +0.09 &  172 \\
  HD160691  &  5800& 4.30&  0.32&  0.19 & 20070506 &  1744m51  &  120   &  5916 &  +4.40& -0.00 & +0.11 &  246 \\
  HD113679  &  5632& 4.01& -0.67& -0.51 & 20070507 &  1305m38  &  120   &  5466 &  +3.50& -1.09 & +0.40 &  131 \\
  HD121004  &  5635& 4.39& -0.73& -0.55 & 20070507 &  1353m46  &  116   &  5918 &  +4.36& -0.61 & +0.13 &   94 \\
  HD156365  &  5820& 3.91&  0.23&  0.19 & 20070507 &  1718m24  &  120   &  6004 &  +4.10& -0.05 & +0.16 &  226 \\
  HD161098  &  5617& 4.30& -0.27& -0.23 & 20070507 &  1743m03  &  116   &  5468 &  +4.28& -0.51 & +0.16 &  196 \\
  HD108510  &  5929& 4.31& -0.06&  0.02 & 20070508 &  1227m08  &  120   &  6176 &  +4.73& -0.29 & +0.26 &  196 \\
  HD125184  &  5629& 4.11&  0.22&  0.19 & 20070508 &  1418m07  &  116   &  5852 &  +4.35& -0.00 & +0.24 &  147 \\
  HD150177  &  6200& 3.98& -0.56& -0.48 & 20070508 &  1638m09  &  120   &  5775 &  +3.09& -1.20 & +0.27 &  263 \\
  HD103891  &  5978& 3.75& -0.25& -0.09 & 20070509 &  1159m09  &   70   &  5638 &  +3.32& -0.82 & +0.35 &  117 \\
  HD104304  &  5361& 4.47&  0.14&  0.22 & 20070509 &  1159m09  &  116   &  5466 &  +4.23& -0.12 & +0.10 &  164 \\
  HD163799  &  5764& 4.02& -0.92& -0.65 & 20070509 &  1758m22  &  116   &  5559 &  +3.60& -1.21 & +0.32 &  160 \\
  HD091345  &  5663& 4.43& -1.09& -0.88 & 20070505 &  1020m71  &  120   &  5860 &  +4.03& -1.11 & +0.29 &  150 \\
  HD102365  &  5558& 4.55& -0.34& -0.26 & 20070505 &  1145m40  &  117   &  6034 &  +4.37& -0.26 & +0.20 &   37 \\
  HD120559  &  5390& 4.48& -0.94& -0.75 & 20070505 &  1350m57  &  127   &  5399 &  +3.94& -1.00 & +0.32 &   61 \\
  HD134088  &  5625& 4.37& -0.87& -0.63 & 20070505 &  1508m08  &  121   &  5547 &  +3.73& -1.01 & +0.17 &  144 \\
  HD152449  &  6096& 4.18& -0.05&  0.12 & 20070505 &  1647m02  &   96   &  6034 &  +4.15& -0.31 & +0.15 &  239 \\
  HD152986  &  6074& 4.25& -0.17&  0.02 & 20070505 &  1647m02  &  108   &  5885 &  +3.80& -0.73 & +0.28 &  160 \\
  HD162396  &  6079& 4.15& -0.37& -0.30 & 20070505 &  1752m42  &  120   &  5859 &  +3.59& -0.79 & +0.20 &  319 \\
  HD177565  &  5625& 4.21&  0.03& -0.02 & 20070505 &  1906m37  &  116   &  5539 &  +4.32& -0.16 & +0.07 &  155 \\
  HD106516  &  6208& 4.39& -0.71& -0.45 & 20070521 &  1210m10  &  120   &  6337 &  +4.92& -0.71 & +0.13 &   44 \\
  HD125072  &  4671& 4.62&  0.49&  0.64 & 20070521 &  1418m59  &  116   &  4992 &  +4.31& +0.20 & +0.00 &   32 \\
  HD145937  &  5813& 4.07& -0.60& -0.18 & 20070522 &  1610m06  &  120   &  5621 &  +3.62& -1.03 & +0.30 &  196 \\ 
\enddata
\tablecomments{
All stars passed visual inspection. 
Column names follow symbol names in Table~\ref{t:A1}, so 
{\it Met} marks the original (uncalibrated) value of the metallicity.
} 
\end{deluxetable}

\begin{deluxetable}{lrrrrrrrrr}
\tablecaption{Results of the observation of 12 members of the open 
cluster M67 \label{t:B4}}
\tablewidth{0pt}
\small
\rotate
\tablehead{
\multicolumn{1}{c}{ }&Adopted&\multicolumn{8}{|c}{RAVE values}\\
\cline{2-2} \cline{3-10}
Name        & [M/H]  & Obsdate  &  FieldName&FibNum   &  $\teff$ & $\logg$&  Met  & alpha & S2N 
}
\startdata
  M67-6469  &  0.01  & 20070409 &   0851p11 &     42  &  4318 &  +1.41& -0.39 & +0.00 &  138 \\
  M67-0084  &  0.01  & 20070409 &   0851p11 &     72  &  4614 &  +1.98& -0.35 & +0.00 &   86 \\
  M67-6495  &  0.01  & 20070409 &   0851p11 &     76  &  4105 &  +1.24& -0.38 & +0.04 &  117 \\
  M67-0223  &  0.01  & 20070409 &   0851p11 &     91  &  4634 &  +2.03& -0.35 & +0.00 &   73 \\
  M67-0218  &  0.01  & 20070409 &   0851p11 &    105  &  4820 &  +2.78& -0.23 & +0.00 &   75 \\
  M67-0286  &  0.01  & 20070409 &   0851p11 &    120  &  4678 &  +2.10& -0.37 & +0.00 &  102 \\
  M67-0135  &  0.01  & 20070409 &   0851p11 &    146  &  4847 &  +2.81& -0.22 & +0.00 &   64 \\
  M67-0115  &  0.01  & 20070601 &   0851p11 &     13  &  6189 &  +4.22& -0.06 & +0.03 &   25 \\
  M67-0046  &  0.01  & 20070601 &   0851p11 &     58  &  5697 &  +4.26& -0.10 & +0.08 &   22 \\
  M67-7859  &  0.01  & 20070601 &   0851p11 &     72  &  6656 &  +4.69& +0.37 & +0.01 &   18 \\
  M67-0192  &  0.01  & 20070601 &   0851p11 &    101  &  6194 &  +3.93& +0.09 & +0.01 &   24 \\
  M67-0227  &  0.01  & 20070601 &   0851p11 &    143  &  5320 &  +3.64& -0.10 & +0.01 &   24 
\enddata
\tablecomments{
[M/H] marks the adopted value of metallicity from the literature, 
while other values, including the uncalibrated metallicity ({\it Met})
were obtained from RAVE observations. Column names follow symbol names in Table~\ref{t:A1}.
The M67 numbering system is summarized in 
http://www.univie.ac.at/webda//cgi-bin/ocl\_page.cgi?cluster=m67. 
Stars with numbers between 0001 and 0295 were numbered by
\citet{Fagerholm1906}. Stars with 64xx numbers are from 
\citet{Montgomery1993} and star 7859 is from \citet{Fan1996}.
} 
\end{deluxetable}

\begin{deluxetable}{llrrrrlrrrrrrrrl}
\tabletypesize{\scriptsize}
\tablecaption{Results of observations of Geneva Copenhagen Survey stars \label{t:B5}}
\tablewidth{0pt}
\rotate
\tablehead{
\multicolumn{1}{c}{ }&\multicolumn{4}{c}{GC survey}&\multicolumn{11}{|c}{RAVE values}\\
\cline{2-5} \cline{6-16}
Name     &log($\teff$)&[Fe/H]&HRV & eHRV  & Obsdate  &FieldName&FibNum &       HRV &   eHRV & $\teff$ &$\logg$& Met & alpha& S2N & SpectraFLAG 
}
\startdata
HD 13386  & 3.708&  0.28 &   31.9 & 0.2 & 20050827 &0220m29b &  021  &   36.758  &  1.440 & 5445& +4.6& +0.2& +0.0 &    38& \\
HD 14294  & 3.796& -0.30 &  -11.0 & 0.2 & 20050827 &0220m29b &  066  &   -8.309  &  1.087 & 6213& +4.1& -0.7& +0.2 &    35& t \\ 
HD 14555  & 3.720& -0.27 &    0.9 & 0.4 & 20050827 &0220m29b &  079  &    1.942  &  1.483 & 5463& +4.7& -0.2& +0.1 &    37& \\
HD 15337  & 3.707&  0.14 &   -4.6 & 0.3 & 20050827 &0220m29b &  093  &   -5.358  &  1.033 & 5244& +4.3& -0.1& +0.0 &    42& \\
HD 14868  & 3.760& -0.17 &   28.8 & 0.3 & 20050827 &0220m29b &  117  &   29.347  &  0.684 & 5892& +4.5& -0.4& +0.3 &    85& \\
HD 14680  & 3.699& -0.03 &   51.5 & 0.2 & 20050827 &0220m29b &  149  &   56.242  &  2.313 & 5582& +4.6& +0.1& +0.0 &    18& cc \\ 
HD 13386  & 3.708&  0.28 &   31.9 & 0.2 & 20050827 &0220m29  &  021  &   32.172  &  1.009 & 5376& +4.4& +0.1& +0.0 &   111& \\
HD 14294  & 3.796& -0.30 &  -11.0 & 0.2 & 20050827 &0220m29  &  066  &  -10.571  &  1.028 & 5952& +3.6& -0.9& +0.3 &    97& \\
HD 14555  & 3.720& -0.27 &    0.9 & 0.4 & 20050827 &0220m29  &  079  &    1.308  &  1.259 & 5608& +4.7& -0.1& +0.0 &    95& \\
HD 15337  & 3.707&  0.14 &   -4.6 & 0.3 & 20050827 &0220m29  &  093  &   -4.636  &  0.960 & 5255& +4.3& -0.1& +0.0 &    98& \\
HD 14868  & 3.760& -0.17 &   28.8 & 0.3 & 20050827 &0220m29  &  117  &   29.436  &  0.665 & 5972& +4.6& -0.3& +0.2 &   159& \\ 
HD 14680  & 3.699& -0.03 &   51.5 & 0.2 & 20050827 &0220m29  &  149  &   51.758  &  0.759 & 5066& +4.8& -0.2& +0.0 &    89& \\
HD 21216  & 3.797& -0.11 &   12.3 & 1.1 & 20050827 &0328m06b &  043  &   18.183  &  1.412 & 6320& +4.6& -0.5& +0.2 &    57& \\ 
HD 21977  & 3.764&  0.11 &   26.6 & 0.2 & 20050827 &0328m06b &  085  &   28.603  &  1.371 & 5977& +4.4& -0.1& +0.0 &    32& \\
HD 21543  & 3.749& -0.60 &   63.7 & 0.1 & 20050827 &0328m06b &  119  &   62.293  &  0.896 & 5004& +3.5& -1.2& +0.4 &    38& \\
HD 21995  & 3.767& -0.21 &  -16.2 & 0.2 & 20050827 &0328m06b &  143  &  -17.914  &  1.011 & 5807& +4.2& -0.7& +0.4 &    56& \\
HD 21216  & 3.797& -0.11 &   12.3 & 1.1 & 20050827 &0328m06  &  043  &   16.003  &  1.220 & 6494& +4.9& -0.4& +0.2 &   146& \\
HD 21977  & 3.764&  0.11 &   26.6 & 0.2 & 20050827 &0328m06  &  085  &   27.628  &  1.082 & 6062& +4.7& -0.2& +0.2 &    75& \\
HD 21543  & 3.749& -0.60 &   63.7 & 0.1 & 20050827 &0328m06  &  119  &   64.017  &  0.977 & 5344& +3.5& -1.2& +0.4 &    71& \\
HD 21995  & 3.767& -0.21 &  -16.2 & 0.2 & 20050827 &0328m06  &  143  &  -15.020  &  0.761 & 5676& +4.1& -0.7& +0.4 &   119& \\
HD 150875 & 3.814& -0.32 &  -16.5 & 0.3 & 20050827 &1652m03b &  047  &  -13.025  &  0.993 & 6491& +3.7& -0.5& +0.2 &    79& \\
HD 151258 & 3.781& -0.65 &  -16.6 & 0.3 & 20050827 &1652m03b &  064  &  -13.372  &  1.011 & 5972& +4.1& -0.9& +0.2 &    50& cc \\ 
HD 152986 & 3.783& -0.23 &   -8.0 & 0.3 & 20050827 &1652m03b &  093  &   -5.371  &  1.214 & 5953& +4.0& -0.4& +0.1 &    78& \\
HD 153479 & 3.797& -0.25 &   24.5 & 0.4 & 20050827 &1652m03b &  124  &   25.621  &  0.909 & 6124& +4.3& -0.6& +0.3 &    38& \\
HD 153240 & 3.788& -0.15 &  -22.9 & 0.2 & 20050827 &1652m03b &  132  &  -29.001  &  0.929 & 6472& +4.8& -0.3& +0.2 &    69& \\
HD 150875 & 3.814& -0.32 &  -16.5 & 0.3 & 20050827 &1652m03  &  047  &  -14.034  &  0.744 & 6423& +3.6& -0.6& +0.2 &   150& \\
HD 151258 & 3.781& -0.65 &  -16.6 & 0.3 & 20050827 &1652m03  &  064  &  -14.952  &  0.711 & 5884& +3.6& -0.8& +0.0 &   107& t \\
HD 152986 & 3.783& -0.23 &   -8.0 & 0.3 & 20050827 &1652m03  &  093  &   -6.667  &  1.054 & 6168& +4.1& -0.3& +0.1 &   130& \\ 
HD 153479 & 3.797& -0.25 &   24.5 & 0.4 & 20050827 &1652m03  &  124  &   25.676  &  0.795 & 6150& +4.2& -0.7& +0.4 &   106& \\
HD 153240 & 3.788& -0.15 &  -22.9 & 0.2 & 20050827 &1652m03  &  132  &  -29.131  &  0.758 & 6374& +4.8& -0.4& +0.2 &   143& \\
HD 181177 & 3.754& -0.23 &   48.3 & 0.2 & 20050827 &1930m62b &  004  &   51.749  &  1.155 & 5890& +4.5& -0.3& +0.2 &    63& \\
HD 181249 & 3.697&  0.12 &  -42.9 & 0.3 & 20050827 &1930m62b &  013  &  -42.012  &  2.068 & 5098& +4.2& +0.0& +0.0 &    17& cc \\ 
HD 180120 & 3.815& -0.23 &   -0.7 & 0.8 & 20050827 &1930m62b &  022  &    0.740  &  0.983 & 6339& +4.0& -0.6& +0.3 &    47& \\
HD 177104 & 3.811& -0.32 &  -17.0 & 0.7 & 20050827 &1930m62b &  035  &  -16.017  &  0.687 & 6329& +4.2& -0.7& +0.4 &    67& \\
HD 185260 & 3.734&  0.04 &  -34.7 & 0.2 & 20050827 &1930m62b &  091  &  -30.514  &  1.883 & 5717& +4.5& +0.2& +0.0 &    17& \\
HD 186784 & 3.794& -0.01 &   -0.5 & 0.2 & 20050827 &1930m62b &  104  &    2.611  &  1.150 & 6133& +3.8& -0.4& +0.3 &    45& \\
HD 185579 & 3.806& -0.03 &   -2.0 & 0.6 & 20050827 &1930m62b &  110  &    0.087  &  1.688 & 6515& +4.8& -0.1& +0.1 &    32& \\
HD 185142 & 3.810& -0.17 &   17.5 & 0.4 & 20050827 &1930m62b &  120  &   21.527  &  0.947 & 6095& +3.9& -0.5& +0.3 &    66& \\
HD 171278 & 3.818& -0.26 &  -16.8 & 0.4 & 20050827 &1930m62b &  149  &  569.046  & 28.089 & 1629& +0.1& +0.1& +0.1 &   000& n \\
HD 181177 & 3.754& -0.23 &   48.3 & 0.2 & 20050827 &1930m62  &  004  &   48.909  &  0.785 & 5924& +4.3& -0.4& +0.3 &   133& \\
HD 181249 & 3.697&  0.12 &  -42.9 & 0.3 & 20050827 &1930m62  &  013  &  -42.520  &  1.161 & 5156& +4.7& -0.1& +0.0 &    50& \\
HD 180120 & 3.815& -0.23 &   -0.7 & 0.8 & 20050827 &1930m62  &  022  &    2.021  &  0.701 & 6419& +4.1& -0.5& +0.2 &   121& \\
HD 177104 & 3.811& -0.32 &  -17.0 & 0.7 & 20050827 &1930m62  &  035  &  -15.111  &  0.661 & 6559& +4.3& -0.5& +0.3 &   136& \\
HD 185260 & 3.734&  0.04 &  -34.7 & 0.2 & 20050827 &1930m62  &  091  &  -33.909  &  1.314 & 5617& +4.2& -0.2& +0.1 &    65& \\
HD 186784 & 3.794& -0.01 &   -0.5 & 0.2 & 20050827 &1930m62  &  104  &   -0.487  &  0.643 & 6127& +3.9& -0.4& +0.2 &   100& \\
HD 185579 & 3.806& -0.03 &   -2.0 & 0.6 & 20050827 &1930m62  &  110  &   -2.211  &  0.866 & 6303& +4.2& -0.4& +0.3 &   106& \\
HD 185142 & 3.810& -0.17 &   17.5 & 0.4 & 20050827 &1930m62  &  120  &   19.737  &  0.900 & 6101& +3.7& -0.6& +0.3 &   157& \\
HD 171278 & 3.818& -0.26 &  -16.8 & 0.4 & 20050827 &1930m62  &  149  &  464.747  &  9.245 & 2787& +0.5& +0.4& +0.3 &   000& n \\
HD 192628 & 3.763& -0.30 &  -53.7 & 0.2 & 20050827 &2017m15b &  007  &   48.773  &  1.593 & 5697& +4.0& -0.3& +0.0 &    26& \\
HD 192266 & 3.784& -0.37 &    8.8 & 0.2 & 20050827 &2017m15b &  026  &   12.689  &  0.852 & 6092& +4.4& -0.5& +0.2 &    45& \\
HD 190613 & 3.764&  0.19 &  -15.8 & 0.4 & 20050827 &2017m15b &  043  &  -10.239  &  0.771 & 5800& +4.3& -0.3& +0.2 &    55& \\
HD 192428 & 3.772& -0.53 &   15.8 & 0.2 & 20050827 &2017m15b &  059  &   22.577  &  1.481 & 5794& +3.7& -0.9& +0.3 &    51& cc \\
HD 194601 & 3.728& -0.21 &   -8.6 & 0.2 & 20050827 &2017m15b &  108  &   -8.925  &  0.952 & 5334& +3.5& -0.5& +0.1 &    71& \\
HD 194581 & 3.708& -0.37 &  -59.0 & 0.1 & 20050827 &2017m15b &  121  &  -57.864  &  0.592 & 5315& +3.7& -0.5& +0.2 &    77& \\
HD 207467 & 3.733& -0.04 &  -16.2 & 0.4 & 20050827 &2017m15b &  135  & -466.692  & 25.600 &    0& +0.0& +0.0& +0.0 &    00& n \\
HD 192772 & 3.802& -0.34 &   12.7 & 0.3 & 20050827 &2017m15b &  148  &   23.171  &  1.067 & 5913& +3.8& -0.9& +0.4 &    55& \\
HD 192628 & 3.763& -0.30 &  -53.7 & 0.2 & 20050827 &2017m15  &  007  &   51.996  &  0.741 & 5931& +4.2& -0.5& +0.2 &    90& \\
HD 192266 & 3.784& -0.37 &    8.8 & 0.2 & 20050827 &2017m15  &  026  &    8.453  &  0.570 & 6196& +4.3& -0.5& +0.3 &   109& \\
HD 190613 & 3.764&  0.19 &  -15.8 & 0.4 & 20050827 &2017m15  &  043  &  -16.719  &  0.775 & 5878& +4.3& -0.3& +0.2 &   117& \\
HD 192428 & 3.772& -0.53 &   15.8 & 0.2 & 20050827 &2017m15  &  059  &   16.577  &  1.363 & 5906& +4.0& -0.9& +0.3 &    92& \\
HD 192117 & 3.731& -0.07 &   30.9 & 0.3 & 20050827 &2017m15  &  075  &  755.172  &  9.722 &    0& +0.0& +0.0& +0.0 &     0& n \\
HD 194601 & 3.728& -0.21 &   -8.6 & 0.2 & 20050827 &2017m15  &  108  &   -8.509  &  0.692 & 5229& +3.3& -0.7& +0.2 &   120& \\
HD 194581 & 3.708& -0.37 &  -59.0 & 0.1 & 20050827 &2017m15  &  121  &  -58.443  &  0.595 & 5464& +3.8& -0.3& +0.1 &   148& \\
HD 207467 & 3.733& -0.04 &  -16.2 & 0.4 & 20050827 &2017m15  &  135  &  -45.205  &  4.302 &    0& +0.0& +0.0& +0.0 &     0& n \\ 
HD 192772 & 3.802& -0.34 &   12.7&  0.3 & 20050827 &2017m15  &  148  &   13.494 &   0.655 & 6114& +4.0& -0.7& +0.3  &  132& \\
HD 216531 & 3.764& -0.26 &   -2.8&  0.2 & 20050827 &2255m44b &  002  &   12.698 &   2.510 & 6880& +4.8& +0.2& +0.0  &   18& cc \\
HD 215468 & 3.832& -0.31 &  -16.5&  1.0 & 20050827 &2255m44b &  016  &  -13.316 &   1.895 & 6801& +4.0& -0.5& +0.2  &   35& \\
HD 215877 & 3.753& -0.20 &   37.8&  0.3 & 20050827 &2255m44b &  028  &   37.065 &   0.991 & 5843& +4.4& -0.4& +0.2  &   48& \\
HD 216568 & 3.799& -0.47 &   27.0&  0.3 & 20050827 &2255m44b &  075  &   20.612 &   1.460 & 6171& +3.8& -0.6& +0.0  &   48& cc \\
HD 217025 & 3.714&  0.01 &   -7.2&  0.3 & 20050827 &2255m44b &  082  &  -14.490 &   1.079 & 5428& +4.7& -0.1& +0.0  &   54& \\
HD 217844 & 3.792& -0.11 &   44.0&  0.3 & 20050827 &2255m44b &  139  &   50.159 &   0.695 & 6104& +3.8& -0.4& +0.2  &   69& \\
HD 216531 & 3.764& -0.26 &   -2.8&  0.2 & 20050827 &2255m44  &  002  &   -2.133 &   0.955 & 5988& +4.6& -0.6& +0.4  &   57& \\
HD 215468 & 3.832& -0.31 &  -16.5&  1.0 & 20050827 &2255m44  &  016  &  -14.253 &   0.795 & 6821& +4.0& -0.4& +0.2  &  141& \\
HD 215877 & 3.753& -0.20 &   37.8&  0.3 & 20050827 &2255m44  &  028  &   38.380 &   1.060 & 5922& +4.4& -0.3& +0.2  &  114& \\
HD 216568 & 3.799& -0.47 &   27.0&  0.3 & 20050827 &2255m44  &  075  &   28.317 &   1.005 & 6105& +3.9& -0.9& +0.4  &  127& \\
HD 217025 & 3.714&  0.01 &   -7.2&  0.3 & 20050827 &2255m44  &  082  &   -7.593 &   1.083 & 5447& +4.7& -0.0& +0.0  &   92& \\
HD 217844 & 3.792& -0.11 &   44.0&  0.3 & 20050827 &2255m44  &  139  &   45.229 &   0.485 & 5950& +3.7& -0.5& +0.3  &  129& \\
HD 223121 & 3.697&  0.10 &  -17.6&  0.3 & 20050827 &2348m33b &  072  &  -13.219 &   1.617 & 4808& +3.6& -0.2& +0.0  &   23& \\
HD 223723 & 3.763& -0.19 &    4.3&  0.2 & 20050827 &2348m33b &  080  &    7.586 &   1.360 & 5940& +4.5& -0.1& +0.2  &   46& \\
HD 223691 & 3.733& -0.17 &    1.6&  0.2 & 20050827 &2348m33b &  113  &    0.946 &   0.951 & 5558& +4.1& -0.4& +0.2  &   76& \\
HD 223641 & 3.728& -0.26 &   14.3&  0.2 & 20050827 &2348m33b &  147  &   16.481 &   1.132 & 5440& +3.6& -0.7& +0.3  &   63& \\
HD 223121 & 3.697&  0.10 &  -17.6&  0.3 & 20050827 &2348m33  &  072  &  -16.241 &   1.198 & 5036& +4.4& -0.1& +0.1  &   62& cc \\
HD 223723 & 3.763& -0.19 &    4.3&  0.2 & 20050827 &2348m33  &  080  &    4.062 &   1.316 & 6432& +4.7& +0.1& +0.0  &  102& \\
HD 223691 & 3.733& -0.17 &    1.6&  0.2 & 20050827 &2348m33  &  113  &    1.667 &   0.531 & 5468& +3.8& -0.4& +0.2  &  130& \\
HD 223641 & 3.728& -0.26 &   14.3&  0.2 & 20050827 &2348m33  &  147  &   14.033 &   0.801 & 5394& +3.7& -0.7& +0.3  &  101& \\
HD 4989   & 3.704& -0.24 &    1.6&  0.3 & 20050828 &0103m42b &  027  &    6.112 &   1.275 & 4919& +4.4& -0.7& +0.2  &   46& \\
HD 5510   & 3.791&  0.09 &   -4.2&  0.3 & 20050828 &0103m42b &  048  &    0.977 &   1.239 & 6092& +4.2& -0.3& +0.2  &   62& \\
HD 7052   & 3.717& -0.08 &   10.1&  0.3 & 20050828 &0103m42b &  089  &    8.378 &   1.568 & 5395& +4.7& +0.0& +0.0  &   28& t \\
HD 6444   & 3.837& -0.09 &   -7.2&  0.7 & 20050828 &0103m42b &  121  &  -10.387 &   1.126 & 6497& +3.9& -0.4& +0.2  &   74& \\
HD 6768   & 3.818& -0.12 &   -4.5&  2.0 & 20050828 &0103m42b &  142  &    1.099 &   1.330 & 6252& +3.6& -0.5& +0.4  &   56& t \\ 
HD 4989   & 3.704& -0.24 &    1.6&  0.3 & 20050828 &0103m42  &  027  &    3.216 &   1.126 & 5320& +4.7& -0.3& +0.0  &  103& \\
HD 5510   & 3.791&  0.09 &   -4.2&  0.3 & 20050828 &0103m42  &  048  &   -3.262 &   0.998 & 5831& +3.8& -0.5& +0.3  &  123& \\
HD 7052   & 3.717& -0.08 &   10.1&  0.3 & 20050828 &0103m42  &  089  &   10.239 &   0.923 & 5448& +4.8& -0.1& +0.0  &   70& \\
HD 6444   & 3.837& -0.09 &   -7.2&  0.7 & 20050828 &0103m42  &  121  &   -8.370 &   0.966 & 6586& +4.0& -0.4& +0.2  &  150& \\
HD 6768   & 3.818& -0.12 &   -4.5&  2.0 & 20050828 &0103m42  &  142  &   -2.720 &   0.581 & 6163& +3.4& -0.6& +0.4  &  110& \\
HD 10166  & 3.717& -0.40 &   -1.8&  0.3 & 20050828 &0145m25b &  012  &   -4.625 &   2.978 & 5341& +4.4& +0.1& +0.1  &   16& cc \\
HD 10037  & 3.774& -0.50 &  -20.5&  0.3 & 20050828 &0145m25b &  028  &  -22.128 &   1.262 & 5729& +3.7& -0.8& +0.3  &   56& \\
HD 9769   & 3.786& -0.33 &   31.5&  0.4 & 20050828 &0145m25b &  054  &   29.559 &   0.955 & 5958& +3.9& -0.6& +0.3  &   52& \\
HD 11523  & 3.760& -0.23 &   20.6&  0.2 & 20050828 &0145m25b &  138  &   18.261 &   1.035 & 5834& +4.4& -0.5& +0.3  &   58& \\
HD 11020  & 3.721& -0.06 &   22.2&  0.3 & 20050828 &0145m25b &  144  &   20.625 &   1.125 & 5701& +4.6& -0.2& +0.1  &   45& \\
HD 10166  & 3.717& -0.40 &   -1.8&  0.3 & 20050828 &0145m25  &  012  &   -1.670 &   0.892 & 5236& +4.4& -0.4& +0.0  &   55& \\
HD 10037  & 3.774& -0.50 &  -20.5&  0.3 & 20050828 &0145m25  &  028  &  -22.784 &   0.633 & 5995& +3.9& -0.7& +0.3  &  140& \\
HD 9769   & 3.786& -0.33 &   31.5&  0.4 & 20050828 &0145m25  &  054  &   30.835 &   0.748 & 6162& +4.1& -0.5& +0.2  &  121& \\
HD 11523  & 3.760& -0.23 &   20.6&  0.2 & 20050828 &0145m25  &  138  &   19.082 &   0.838 & 5848& +4.4& -0.5& +0.3  &  118& \\
HD 11020  & 3.721& -0.06 &   22.2&  0.3 & 20050828 &0145m25  &  144  &   22.150 &   0.965 & 5574& +4.7& -0.3& +0.1  &  108& \\
HD 14680  & 3.699& -0.03 &   51.5&  0.2 & 20050828 &0230m30  &  026  &   50.609 &   1.377 & 5159& +4.8& -0.2& +0.0  &   68& \\ 
HD 14868  & 3.760& -0.17 &   28.8&  0.3 & 20050828 &0230m30  &  058  &   32.101 &   0.653 & 5885& +4.4& -0.4& +0.3  &  115& \\
HD 15337  & 3.707&  0.14 &   -4.6&  0.3 & 20050828 &0230m30  &  075  &   -1.753 &   1.200 & 5223& +4.5& -0.2& +0.2  &  112& \\
HD 16297  & 3.726& -0.03 &   -1.7&  0.2 & 20050828 &0230m30  &  090  &   -0.629 &   1.428 & 5765& +4.8& -0.0& +0.0  &   79& \\
HD 16784  & 3.772& -0.54 &   40.3&  0.3 & 20050828 &0230m30  &  114  &   40.749 &   1.089 & 6280& +4.7& -0.6& +0.3  &   47& \\
HD 169812 & 3.771& -0.15 &  -59.5&  0.2 & 20050828 &1830m40b &  013  &  -59.071 &   2.344 & 6091& +3.9& -0.2& +0.0  &   19& cc \\
HD 169499 & 3.708& -0.47 &  -15.0&  0.2 & 20050828 &1830m40b &  016  &  -12.065 &   2.377 & 5591& +4.1& -0.2& +0.0  &   33& \\
HD 170865 & 3.755& -0.14 &   52.2&  0.2 & 20050828 &1830m40b &  079  &   55.214 &   1.208 & 6112& +4.7& -0.4& +0.3  &   68& \\
HD 170869 & 3.757& -0.30 &  -63.8&  0.3 & 20050828 &1830m40b &  098  &  -63.249 &   0.892 & 6103& +4.1& -0.2& +0.1  &   49& \\
HD 172283 & 3.795& -0.37 &  -22.5&  0.3 & 20050828 &1830m40b &  130  &  -20.023 &   2.155 & 6190& +3.8& -0.6& +0.2  &   39& cc \\
HD 169812 & 3.771& -0.15 &  -59.5&  0.2 & 20050828 &1830m40  &  013  &  -58.058 &   0.894 & 5847& +4.0& -0.5& +0.3  &   71& \\
HD 169499 & 3.708& -0.47 &  -15.0&  0.2 & 20050828 &1830m40  &  016  &  -12.742 &   1.772 & 5420& +3.5& -0.6& +0.1  &   99& \\
HD 170865 & 3.755& -0.14 &   52.2&  0.2 & 20050828 &1830m40  &  079  &   53.598 &   0.636 & 5821& +4.2& -0.6& +0.3  &  150& \\
HD 170869 & 3.757& -0.30 &  -63.8&  0.3 & 20050828 &1830m40  &  098  &  -63.634 &   0.523 & 5876& +3.9& -0.5& +0.2  &  115& \\
HD 172283 & 3.795& -0.37 &  -22.5&  0.3 & 20050828 &1830m40  &  130  &  -21.514 &   1.899 & 6390& +3.9& -0.7& +0.4  &  103& \\ 
HD 175114 & 3.801& -0.47 &   30.9&  0.4 & 20050828 &1900m30b &  007  &   29.620 &   1.040 & 5844& +3.3& -1.1& +0.3  &   43& \\
HD 173858 & 3.762& -0.58 &   66.0&  0.2 & 20050828 &1900m30b &  044  &   67.152 &   1.229 & 6169& +4.5& -0.4& +0.2  &   50& \\
HD 175568 & 3.776& -0.32 &  -20.3&  0.3 & 20050828 &1900m30b &  072  &  -22.102 &   0.985 & 6187& +4.3& -0.3& +0.1  &   49& \\
HD 175979 & 3.774& -0.40 &   -1.3&  0.3 & 20050828 &1900m30b &  075  &   -3.173 &   0.830 & 6054& +4.0& -0.4& +0.1  &   68& \\
HD 176367 & 3.778&  0.11 &   -5.8&  0.5 & 20050828 &1900m30b &  078  &   -7.187 &   2.102 & 6252& +4.7& -0.2& +0.1  &   70& cc \\
HD 176612 & 3.774& -0.29 &  -16.7&  0.3 & 20050828 &1900m30b &  082  &  -18.859 &   0.870 & 6231& +4.3& -0.4& +0.2  &   58& \\
HD 177033 & 3.679& -0.23 &  -46.9&  0.3 & 20050828 &1900m30b &  085  &  -48.394 &   0.990 & 4861& +4.6& -0.1& +0.0  &   43& \\
HD 178673 & 3.772& -0.27 &   27.6&  0.3 & 20050828 &1900m30b &  112  &   26.270 &   1.073 & 6691& +4.3& -0.0& +0.2  &   66& \\
HD 177122 & 3.768& -0.31 &  -31.6&  0.2 & 20050828 &1900m30b &  139  &  -33.593 &   0.528 & 5953& +4.4& -0.5& +0.3  &   86& \\
HD 175114 & 3.801& -0.47 &   30.9&  0.4 & 20050828 &1900m30  &  007  &   29.692 &   0.767 & 6091& +3.8& -0.9& +0.4  &  101& \\
HD 173858 & 3.762& -0.58 &   66.0&  0.2 & 20050828& 1900m30   & 044  &   65.839 &   0.831&  6125& +4.4& -0.5& +0.3 &   108 &\\
HD 175568 & 3.776& -0.32 &  -20.3&  0.3 & 20050828& 1900m30   & 072  &  -21.259 &   0.658&  5760& +4.1& -0.6& +0.2 &   108 &\\
HD 175979 & 3.774& -0.40 &   -1.3&  0.3 & 20050828& 1900m30   & 075  &   -2.194 &   0.495&  6016& +3.7& -0.5& +0.1 &   137 &\\
HD 176367 & 3.778&  0.11 &   -5.8&  0.5 & 20050828& 1900m30   & 078  &   -6.836 &   2.214&  6439& +4.8& -0.1& +0.1 &   119 &cc \\
HD 176612 & 3.774& -0.29 &  -16.7&  0.3 & 20050828& 1900m30   & 082  &  -17.171 &   0.592&  6224& +4.5& -0.3& +0.1 &   121 &\\
HD 177033 & 3.679& -0.23 &  -46.9&  0.3 & 20050828& 1900m30   & 085  &  -48.422 &   0.562&  4889& +4.6& -0.3& +0.0 &   112 &\\
HD 178673 & 3.772& -0.27 &   27.6&  0.3 & 20050828& 1900m30   & 112  &   26.467 &   0.928&  6416& +4.2& -0.2& +0.3 &   120 &\\
HD 177122 & 3.768& -0.31 &  -31.6&  0.2 & 20050828& 1900m30   & 139  &  -33.000 &   0.503&  6048& +4.6& -0.5& +0.3 &   165 &\\
HD 189389 & 3.761&  0.03 &   14.3&  0.3 & 20050828& 2005m43   & 044  &   13.769 &   1.223&  5837& +4.1& -0.2& +0.3 &    74 &\\
HD 188903 & 3.782&  0.01 &   -3.3&  0.3 & 20050828& 2005m43   & 058  &   -2.607 &   0.737&  6061& +4.0& -0.3& +0.2 &   109 &\\
HD 190649 & 3.745& -0.47 &  -55.7&  0.2 & 20050828& 2005m43   & 079  &  -55.229 &   1.591&  6194& +4.5& -0.3& +0.3 &   102 &\\
HD 192071 & 3.752& -0.45 &  -16.4&  0.3 & 20050828& 2005m43   & 129  &  -16.474 &   0.663&  5730& +3.9& -0.7& +0.3 &   138 &\\
HD 190269 & 3.815& -0.06 &    1.9&  0.2 & 20050828& 2005m43   & 149  &   -0.640 &   0.647&  6322& +4.0& -0.3& +0.2 &   111 &\\
HD 205187 & 3.769& -0.18 &   24.2&  0.2 & 20050828& 2142m41b  & 062  &   22.655 &   0.924&  5922& +4.3& -0.4& +0.3 &    63 &\\
HD 207790 & 3.778&  0.14 &   28.3&  1.0 & 20050828& 2142m41b  & 104  &   24.974 &   1.908&  6257& +4.4& -0.2& +0.3 &    64 &\\
HD 206303 & 3.790& -0.19 &  -14.3&  0.6 & 20050828& 2142m41b  & 121  &  -14.624 &   0.987&  6412& +4.0& -0.1& +0.2 &    65 &\\
HD 206667 & 3.767& -0.26 &   19.3&  0.2 & 20050828& 2142m41b  & 135  &   19.705 &   0.650&  6162& +5.0& -0.2& +0.2 &    81 &\\
HD 206682 & 3.792& -0.15 &    9.2&  0.3 & 20050828& 2142m41b  & 148  &   13.098 &   1.327&  6403& +4.8& -0.3& +0.3 &    57 &\\
HD 205187 & 3.769& -0.18 &   24.2&  0.2 & 20050828& 2142m41   & 062  &   23.812 &   0.578&  5971& +4.3& -0.3& +0.3 &   146 &\\
HD 207790 & 3.778&  0.14 &   28.3&  1.0 & 20050828& 2142m41   & 104  &   27.269 &   1.118&  6102& +4.0& -0.2& +0.2 &   137 &\\
HD 206303 & 3.790& -0.19 &  -14.3&  0.6 & 20050828& 2142m41   & 121  &  -14.083 &   0.786&  6401& +4.0& -0.1& +0.2 &   129 &\\
HD 206667 & 3.767& -0.26 &   19.3&  0.2 & 20050828& 2142m41   & 135  &   18.444 &   0.709&  6165& +4.9& -0.3& +0.2 &   155 &\\
HD 206682 & 3.792& -0.15 &    9.2&  0.3 & 20050828& 2142m41   & 148  &    9.145 &   0.950&  6202& +4.3& -0.3& +0.3 &   118 &\\
HD 2404   & 3.724& -0.44 &  -43.1&  0.3 & 20050829& 0030m31   & 005  &  -49.767 &   1.178&  5416& +4.1& -0.7& +0.3 &    49 &\\
HD 2348   & 3.787& -0.40 &   32.5&  0.3 & 20050829& 0030m31   & 021  &   31.169 &   1.363&  6186& +4.3& -0.4& +0.2 &    47 &\\
HD 1557   & 3.798& -0.45 &   25.1&  0.3 & 20050829& 0030m31   & 043  &   23.863 &   1.498&  6289& +3.9& -0.6& +0.3 &    37 &\\
HD 1674   & 3.784& -0.17 &   -5.7&  0.7 & 20050829& 0030m31   & 057  &   -4.178 &   0.814&  5850& +4.0& -0.6& +0.3 &    62 &\\
HD 3810   & 3.750& -0.31 &   37.4&  0.2 & 20050829& 0030m31   & 101  &   36.477 &   1.159&  5963& +4.6& -0.4& +0.2 &    49 &\\
HD 3560   & 3.788& -0.31 &  -10.6&  0.4 & 20050829& 0030m31   & 127  &  -12.069 &   1.317&  6167& +3.9& -0.3& +0.1 &    30 &\\
HD 157884 & 3.804& -0.38 &    9.1&  0.5 & 20050829& 1730m30   & 003  &   12.029 &   0.593&  6651& +5.0& -0.3& +0.2 &   144 &\\
HD 156423 & 3.719& -0.34 &  -30.2&  0.3 & 20050829& 1730m30   & 043  &  -31.890 &   0.843&  5199& +4.1& -0.6& +0.2 &   106 &\\
HD 160573 & 3.799& -0.10 &    3.7&  1.4 & 20050829& 1730m30   & 118  &    8.937 &   1.111&  6287& +3.6& -0.5& +0.2 &   132 &\\
HD 159784 & 3.781&  0.32 &   -8.8&  0.3 & 20050829& 1730m30   & 127  &   -8.392 &   1.008&  6226& +4.2& +0.0& +0.2 &   114 &\\
HD 159882 & 3.729& -0.15 &   12.3&  0.3 & 20050829& 1730m30   & 136  &   12.660 &   0.897&  5597& +4.8& -0.2& +0.0 &   111 &\\
HD 158884 & 3.789& -0.48 &   53.7&  0.5 & 20050829& 1730m30   & 144  &   54.156 &   0.509&  5979& +3.6& -0.9& +0.4 &   118 &\\
HD 157884 & 3.804& -0.38 &    9.1&  0.5 & 20050829& 1930m35b  & 002  &   18.933 &   3.780&  6090& +4.5& +0.1& +0.1 &    12 &cc \\
HD 156423 & 3.719& -0.34 &  -30.2&  0.3 & 20050829& 1930m35b  & 012  &  -32.357 &   1.672&  6278& +3.7& -0.6& +0.3 &    54 &\\
HD 160573 & 3.799& -0.10 &    3.7&  1.4 & 20050829& 1930m35b  & 026  &   23.692 &   1.639&  5874& +4.4& -0.4& +0.3 &    52 &\\
HD 159784 & 3.781&  0.32 &   -8.8&  0.3 & 20050829& 1930m35b  & 106  &   17.940 &   1.340&  6050& +3.9& -0.1& +0.2 &    45 &\\
HD 159882 & 3.729& -0.15 &   12.3&  0.3 & 20050829& 1930m35b  & 121  &   -6.329 &   0.824&  5964& +4.6& -0.0& +0.0 &    38 &\\
HD 158884 & 3.789& -0.48 &   53.7&  0.5 & 20050829& 1930m35b  & 140  &    5.617 &   1.068&  5635& +4.6& -0.0& +0.0 &    37 &\\
HD 183198 & 3.725& -0.09 &   -7.6&  0.3 & 20050829& 1930m35   & 002  &   -9.661 &   2.264&  5459& +4.8& -0.1& +0.1 &    37 &t \\ 
HD 181893 & 3.810& -0.21 &  -42.1&  0.6 & 20050829& 1930m35   & 012  &  -42.955 &   0.908&  6270& +3.9& -0.6& +0.4 &   133 &\\
HD 181452 & 3.749& -0.27 &   25.8&  0.2 & 20050829& 1930m35   & 026  &   23.519 &   1.576&  6267& +4.6& +0.0& +0.1 &   110 &cc \\
HD 184514 & 3.769&  0.00 &   10.4&  0.3 & 20050829& 1930m35   & 106  &   10.729 &   1.162&  5866& +3.7& -0.2& +0.2 &   100 &\\
HD 185679 & 3.745&  0.06 &  -10.8&  0.2 & 20050829& 1930m35   & 121  &  -10.358 &   0.967&  5949& +4.6& -0.1& +0.1 &    88 &\\
HD 184374 & 3.737&  0.11 &   15.8&  0.3 & 20050829& 1930m35   & 140  &   16.675 &   1.049&  5790& +4.9& -0.0& +0.0 &    91 &\\
HD 199903 & 3.760& -0.07 &   42.0&  0.2 & 20050829& 2100m35b  & 034  &   56.086 &   1.128&  5924& +4.3& -0.3& +0.2 &    80 &\\ 
HD 198697 & 3.744& -0.05 &    3.0&  0.3 & 20050829& 2100m35b  & 044  &   19.458 &   0.842&  6085& +4.6& -0.1& +0.1 &    60 &\\
HD 199672 & 3.787&  0.00 &  -15.1&  0.3 & 20050829& 2100m35b  & 075  &   -4.957 &   0.735&  5806& +3.3& -0.4& +0.2 &    71 &\\
HD 200382 & 3.703&  0.16 &   16.0&  0.4 & 20050829& 2100m35b  & 108  &   11.064 &   0.928&  5062& +4.4& +0.1& +0.0 &    37 &\\
HD 201513 & 3.735& -0.17 &   20.0&  0.3 & 20050829& 2100m35b  & 113  &   16.486 &   1.698&  6277& +4.8& +0.1& +0.1 &    26 &\\
HD 200608 & 3.760&  0.10 &   13.1&  0.4 & 20050829& 2100m35b  & 139  &   18.234 &   0.901&  5762& +4.1& -0.4& +0.3 &    70 &\\
HD 200344 & 3.775& -0.08 &   49.1&  0.3 & 20050829& 2100m35b  & 144  &   62.212 &   0.981&  5833& +3.6& -0.5& +0.3 &    61 &\\
HD 199903 & 3.760& -0.07 &   42.0&  0.2 & 20050829& 2100m35   & 034  &   42.375 &   0.578&  6047& +4.6& -0.2& +0.2 &   147 &\\
HD 198697 & 3.744& -0.05 &    3.0&  0.3 & 20050829& 2100m35   & 044  &    2.443 &   0.676&  5934& +4.7& -0.3& +0.2 &   126 &\\
HD 199672 & 3.787&  0.00 &  -15.1&  0.3 & 20050829& 2100m35   & 075  &  -14.549 &   0.690&  5926& +3.5& -0.3& +0.2 &   136 &\\
HD 200382 & 3.703&  0.16 &   16.0&  0.4 & 20050829& 2100m35   & 108  &   16.506 &   0.838&  5099& +4.4& -0.0& +0.0 &    91 &\\
HD 201513 & 3.735& -0.17 &   20.0&  0.3 & 20050829& 2100m35   & 113  &   19.251 &   0.946&  5976& +4.7& -0.2& +0.1 &    79 &\\
HD 200608 & 3.760&  0.10 &   13.1&  0.4 & 20050829& 2100m35   & 139  &   10.805 &   0.663&  5777& +4.1& -0.3& +0.2 &   136 &\\
HD 200344 & 3.775& -0.08 &   49.1&  0.3 & 20050829& 2100m35   & 144  &   48.946 &   0.680&  5911& +3.8& -0.4& +0.3 &   110 &\\
HD 217221 & 3.711&  0.16 &   26.2&  0.1 & 20050829& 2305m29b  & 012  &   39.502 &   1.168&  5445& +4.5& -0.0& +0.0 &    47 &\\
HD 217123 & 3.789& -0.27 &  -10.7&  0.4 & 20050829& 2305m29b  & 049  &  -20.252 &   1.313&  6308& +4.2& -0.4& +0.3 &    59 &\\
HD 217500 & 3.729& -0.12 &    0.0&  0.3 & 20050829& 2305m29b  & 066  &   -2.920 &   1.484&  5849& +4.0& -0.2& +0.1 &    22 &cc \\
HD 218532 & 3.765&  0.09 &   35.4&  0.2 & 20050829& 2305m29b  & 085  &   38.136 &   1.140&  6029& +4.6& -0.3& +0.2 &    54 &\\
HD 219057 & 3.751& -0.22 &    0.7&  0.2 & 20050829& 2305m29b  & 132  &   -1.502 &   1.465&  6065& +4.8& -0.2& +0.2 &    43 &\\
HD 217221 & 3.711&  0.16 &   26.2&  0.1 & 20050829& 2305m29   & 012  &   26.936 &   0.728&  5224& +4.5& -0.2& +0.1 &   100 &\\
HD 217123 & 3.789& -0.27 &  -10.7&  0.4 & 20050829& 2305m29   & 049  &  -12.458 &   0.905&  6209& +4.0& -0.4& +0.3 &   127 &\\
HD 217500 & 3.729& -0.12 &    0.0&  0.3 & 20050829& 2305m29   & 066  &   -0.547 &   0.678&  5799& +4.2& -0.4& +0.1 &    51 &\\
HD 218532 & 3.765&  0.09 &   35.4&  0.2 & 20050829& 2305m29   & 085  &   36.420 &   0.701&  5657& +4.0& -0.4& +0.2 &   109 &\\
HD 219057 & 3.751& -0.22 &    0.7&  0.2 & 20050829& 2305m29   & 132  &    1.416 &   0.639&  5891& +4.5& -0.4& +0.3 &    79 &\\
\enddata
\tablecomments{Only normal stars, which have their SpectraFLAG column empty, were used to check the 
temperature values.
} 
\end{deluxetable}

\end{document}